\documentclass[conference]{IEEEtran}
\usepackage{algorithm,algorithmicx,algpseudocode,graphicx,tabulary,doi,float}
\usepackage{varwidth, multirow,amsmath,lscape}
\usepackage{booktabs,adjustbox}
\usepackage{graphicx}
\usepackage{array}
\usepackage{tabu}
\usepackage{colortbl}
\usepackage{longtable,supertabular,tikz}
\usepackage{comment}
\usepackage{xcolor,cite,caption}

\usepackage{hyperref}

\usepackage{xcolor}
\newcommand{\revision}[1]{{\textcolor{black}{#1}}}
\newcommand{\revisionB}[1]{{\textcolor{black}{#1}}}

\newskip\LTpre        \LTpre=0pt
\newskip\LTpost       \LTpost=0pt

\usepackage{fancyhdr}
\hypersetup{final}

\def\checkmark{\tikz\fill[scale=0.4](0,.35) -- (.25,0) -- (1,.7) -- (.25,.15) -- cycle;}

\IEEEoverridecommandlockouts                              
\begin{document}

\title{A Survey of Honeypots and Honeynets for \\Internet of Things, Industrial Internet of Things, and Cyber-Physical Systems}

\author{\IEEEauthorblockN{Javier Franco$^{1}$, Ahmet Aris$^{1}$, Berk Canberk$^{2}$, and A. Selcuk Uluagac$^{1}$}
\IEEEauthorblockA{ $^{1}$ Cyber-Physical Systems Security Lab., Florida International University, Florida, USA\\
$^{2}$ Department of Computer Engineering, Istanbul Technical University, Istanbul, Turkey\\ 
Emails: \{jfran243, aaris, suluagac\}@fiu.edu, canberk@itu.edu.tr}
}

\maketitle

\begin{abstract}
The Internet of Things (IoT), the Industrial Internet of Things (IIoT), and Cyber-Physical Systems (CPS) have become essential for our daily lives in contexts such as our homes, buildings, cities, health, transportation, manufacturing, infrastructure, and agriculture. However, they have become popular targets of attacks, due to their inherent limitations which create vulnerabilities. Honeypots and honeynets can prove essential to understand and defend against attacks on IoT, IIoT, and CPS environments \revision {by attracting attackers and deceiving them into thinking that they have gained access to the real systems}. Honeypots and honeynets can complement other security solutions (i.e., firewalls, Intrusion Detection Systems - IDS) to form a strong defense against malicious entities. 
This paper provides a comprehensive survey of the research that has been carried out on honeypots and honeynets for IoT, IIoT, and CPS. It provides a taxonomy and extensive analysis of the existing honeypots and honeynets, states key design factors for the state-of-the-art honeypot/honeynet research and outlines open issues for future honeypots and honeynets for IoT, IIoT, and CPS environments.
\end{abstract}


\section{Introduction}\label{sec:intro}
\revision {The Internet of Things (IoT) is a network of Internet-connected devices, such as sensors, actuators, and other embedded devices that are able to collect data and communicate. Industrial IoT (IIoT) is the application of IoT to automation applications using industrial communication technologies~\cite{Sisinni:2018}. Cyber-Physical Systems (CPS) on the other hand, are networks of devices such as sensors, actuators, Programmable Logic Controllers (PLCs), Remote Terminal Units (RTUs), Intelligent Electronic Devices (IEDs), and other embedded devices that monitor and control physical processes in critical and non-critical application areas. CPS includes, but is not limited to Industrial Control Systems (ICS), Smart Grid and other smart infrastructures (e.g., water, gas, building automation), medical devices, and smart cars~\cite{Bordel:2017, Humayed:2017}. As it can be seen from the descriptions of IoT, IIoT, and CPS, these concepts do not have explicit separation points. Border et al.~\cite{Bordel:2017} and the National Institute of Standards and Technology's (NIST) special report by Greer et al.~\cite{NIST-IoT-CPS:2019} analyzed the definitions of IoT and CPS in the literature and indicated that these concepts are viewed either as the same, or different but they have overlapping parts, or they are subsets of each other.  Greer et al.~\cite{NIST-IoT-CPS:2019} pointed out that IoT and CPS are similar as they both connect the physical world of engineered systems and the logical world of communications and information technology. These two worlds are connected by sensors that collect data about the physical elements of a system and transmit it to the logical elements, and to the actuators that respond to the logical elements and apply changes to the physical elements. At the same time, however, Greer et al.~\cite{NIST-IoT-CPS:2019} stated that IoT and CPS are different in that IoT places more emphasis on information technology and networking things in the physical world, while CPS is more of a closed system and is  focused more on the exchange of information for sensing and controlling the physical world. IIoT further connects the definitions of IoT and CPS, as it possesses characteristics from both.} 

IoT, IIoT, and CPS are converting almost every aspect of life to smart in the 21st century. Sensors, actuators, wearables, embedded devices, and many other devices are becoming ubiquitous around the world with uses in diverse contexts such as homes, buildings, cities, health, transportation, \revision {automotive}, manufacturing, critical (e.g., nuclear reactors, power plants, oil refineries) and non-critical infrastructures, and agriculture. While this promises connectivity and efficiency, the various devices in IoT, IIoT, and CPS environments have their unique properties in terms of resource limitations, network lifetimes, and application Quality-of-Service (QoS) requirements which affect the security of such applications crucially\revisionB{~\cite{Makhdoom:2019}. }

IoT devices typically have constrained power, storage, computing, and communications resources which limit the accommodation of good security mechanisms~\cite{Meneghello:2019, Neshenko:2019}. On the other hand, devices used in IIoT and CPS were not initially designed with security in mind and they had been considered secure, as they were isolated. This security by obscurity assumption was broken by the uncovering of the Stuxnet (2010), DuQu (2011), and Flame (2012) attacks~\cite{Scott:2014}. As an increasing number of industrial environments are being connected to the Internet, security updates and patches are becoming serious problems in decades-old industrial devices\revisionB{~\cite{Simoes:2013, Scott:2014, Yu:2021, systemCallsLeo}}.

In order to protect IoT, IIoT, and CPS environments from malicious entities, traditional security mechanisms such as cryptography, firewalls, Intrusion Detection and Prevention Systems (IDS, IPS), antivirus, and anti-malware solutions can be utilized. However, they do not \revision{transparently} allow security researchers to observe and analyze how attackers perform attacks and find out their behaviors~\cite{Fan:2018}. Honeypots and honeynets come to the scene as viable solutions at this point, \revision{as they} can provide actionable intelligence on the attackers. A honeypot is a tool that is used with the purpose of being attacked and possibly compromised \cite{Spitzner:2001}. Two or more honeypots implemented on a system form a honeynet \cite{Kumar:2017}. Honeypots are used to attract attackers and deceive them into thinking that they gained access to real systems. Honeypots can be integrated with firewalls and IDSs to form an IPS in order to capture all the information about attackers, study all of their actions, develop ways to improve system security and prevent attacks in the future \cite{Fan:2018}.

Although there exist a number of honeypot and honeynet works on IoT, IIoT, or CPS, no study exists in the literature which considers all of the honeypot and honeynet models, analyzes their similarities and differences, and extracts key points in the design and implementation of honeypots and honeynets for IoT, IIoT, and CPS. In order to fill this important research gap, we propose our comprehensive survey on honeypot and honeynet models that have been proposed for IoT, IIoT, and CPS environments over the period 2002-2020. To the best of our knowledge, our work is the first study in the literature that surveys the current state-of-the-art honeypot and honeynet models not only for IoT, but also for IIoT and CPS. 

\noindent\textbf{Contributions:} The contributions of our survey are as follows:
\begin{itemize}
    \item Taxonomy of honeypots and honeynets proposed for IoT, IIoT, and CPS environments,
    \item Comprehensive analysis of IoT, IIoT, and CPS honeypots and honeynets, and intriguing characteristics that are shared by studies, 
    \item Statement of the key design factors for future IoT, IIoT, and CPS honeypots and honeynets,
    \item Presentation of open research problems that still need to be addressed in honeypot and honeynet research for IoT, IIoT, and CPS.
\end{itemize}

\noindent\textbf{Organization:} The paper is organized as follows: Section~\ref{sec:related} gives the related work. Section~\ref{sec:Background} provides background information on \revision{honeypots, honeynets, }
and related terms. Section~\ref{sec:methodology} provides a methodology for the classification of honeypot and honeynet characteristics. Section~\ref{sec:honeypots} classifies and presents diverse IoT honeypot and honeynet models and research. Section~\ref{sec:taxonomy} presents a taxonomy of \revision{the proposed }IoT honeypot and honeynet models. 
Section~\ref{sec:cpshoneypots} classifies and presents diverse CPS and IIoT honeypot and honeynet models and research. Section~\ref{sec:taxonomycps} presents a taxonomy of \revision{the proposed }CPS and IIoT honeypot and honeynet models. 
Section~\ref{sec:openissues} provides lessons learned and design considerations for honeypot and honeynet implementations. In Section~\ref{sec:conclusion}, conclusions and future work are presented. 

\section{Related Work}\label{sec:related}
The security of IoT, IIoT, and CPS environments is a very broad field of research, and it is possible to find a myriad of studies. Without going into much detail, we refer the readers to the works of \revisionB{Butun et al.~\cite{ButunSurvey} and Makhdoom et al.~\cite{Makhdoom:2019} for extensive overviews of vulnerabilities, threats, and attacks, the security surveys of Lee et al.~\cite{LeeSurvey} on IoT standards and Granjal et al.~\cite{GranjalSurvey} on the existing IoT protocols,  the study of } Neshenko et al.~\cite{Neshenko:2019} for a recent comprehensive IoT security survey, \revisionB{the study of Sikder et al.~\cite{amitsurvey} for a survey of threats to IoT sensors}, the study of Humayed et al.~\cite{Humayed:2017} for an extensive survey on the threats, vulnerabilities, attacks, and defense solutions to CPS, \revisionB{the survey of Al-Garadi et al.~\cite{Al-GaradiSurvey} for machine and deep learning techniques for IoT security, the comprehensive survey of Yu et al.~\cite{Yu:2021} for CPS security and Cintuglu et al.~\cite{cintugluSurvey} for CPS testbeds}. There are also studies like that of Babun et al.~\cite{Babun:2020} which develop innovative ways to protect networks with vulnerable IoT devices. 

The honeypot and honeynet research has been a very active field. In terms of general honeypots and honeynets that are not specific to IoT, IIoT, or CPS, Fan et. al.~\cite{Fan:2015-2} proposed criteria and a methodology for the classification of honeynet solutions and analyzed the advantages and disadvantages of each criterion used in their taxonomy. In 2018, Fan et al.~\cite{Fan:2018} expanded on their earlier research and proposed a taxonomy of decoy systems with respect to decoys and captors. There also exist other survey studies on general honeypot and honeynet solutions, which include but are not limited to ~\cite{Mairh:2011, Campbell:2015}, and~\cite{Zobal:2019}. In addition, privacy and liability issues when honeypots are deployed were analyzed by Sokol et al.~\cite{Sokol:2015, Sokol:2015-2}. In terms of honeypots and honeynets for IoT, IIoT, and CPS, only a few surveys exist in the literature. Razali et al. \cite{Razali:2018} analyzed types, properties, and interaction levels of IoT honeypots and classified honeynet models based on interaction, resources, purpose, and role.   Dalamagkas et al. \cite{Dalamagkas:2019} surveyed the honeypot and honeynet frameworks for smart-grid environments. Dowling et al.~\cite{Dowling:2017-2} proposed a framework for developing data-centric, adaptive smart city honeynets that focus on the key values of data complexity, security, and criticality. Furthermore, Neshenko et al.~\cite{Neshenko:2019} discussed the IoT and CPS honeypots in their survey on IoT security. However, they did not provide a comprehensive survey on such honeypots since the focus of their study was on the security of IoT. 

In addition to proposing novel honeypot/honeynet models or surveying the existing studies, there has been research on the development of honeynet description languages and also on the detectability of honeypots. Fan et al. \cite{Fan:2015-1} presented a technology-independent, flexible honeynet description language and a tool called HoneyGen for the deployment and modification of virtual honeynets based on the VNX and Honeyd platforms. Acien et al. \cite{Acien:2018} analyzed the steps and requirements to deploy honeypots in IoT environments effectively in a way that they can look like real devices to attackers. Surnin et al.~\cite{Surnin:2019} focused on techniques for honeypot detection with SSH and Telnet, identifying issues of software architecture and implementation that make honeypots easily detected~\cite{SurninGithub:2019}. Zamiri-Gourabi et al.~\cite{Zamiri:2019} proposed a methodology to detect the ICS honeypots deployed on the Internet by means of fingerprinting methodologies.

\noindent\textbf{Differences from the existing work:} 
While the recent years have seen an increase in honeypot and honeynet research, our study is different because it is the first comprehensive study that analyzes the existing honeypot and honeynet models and research for IoT, IIoT, and CPS environments holistically, provides a taxonomy of honeypots and honeynets and identifies key design considerations and open issues for honeypots and honeynets in IoT, IIoT, and CPS.

\section{BACKGROUND INFORMATION}\label{sec:Background}

\revision {In this section, we give some brief information on honeypots, honeynets, and other related terms.}

\subsection{Honeypots and Honeynets}
A honeypot is a tool that serves as a decoy to attract attackers and deceive them into thinking that they have gained access to a real system. There exist various views of a honeynet: A honeynet can be defined simply as two or more honeypots implemented on a system~\cite{Kumar:2017}, or in a more narrow definition, a honeynet is a high interaction honeypot system of Generation I, II, or III \cite{HoneynetProject:2001}. \revision{Although honeypots and honeynets are defined in the mentioned ways, it is interesting to note that very few authors refer to their honeypot system as a honeynet, despite their research implementing multiple honeypots. For instance, as it will be reviewed in the following sections, only a few honeypots (\cite{Manzanares:2017, Evron:2017, Piggin:2016, Hilt:2020, Litchfield:2016}) in the literature were implemented with a single honeypot. For this reason, we adhered to the statements of authors about their view of their systems as honeypots or honeynets while we are reviewing the studies in this survey.}

Three main architectures/generations that are used in honeynets are described in~\cite{Oza:2018}. \emph{Generation I} was developed in 1999 and is composed of a firewall and an IDS, with honeypots behind these. Generation I can capture in-depth information and unknown attacks. However, Generation I honeynets can be 
easily detected by attackers. \emph{Generation II} was developed in 2002 and had a honeynet sensor that serves the purpose of the IDS sensor and of the firewall used in Generation I. This sensor works like a bridge, so it is much more difficult for attackers to detect that they are in a honeynet. \emph{Generation III} was developed in 2004 and had the same architecture as Generation II but has improved deployment and management capabilities. 

\begin{figure}[!t]
    \centering
    \includegraphics[scale=0.45]{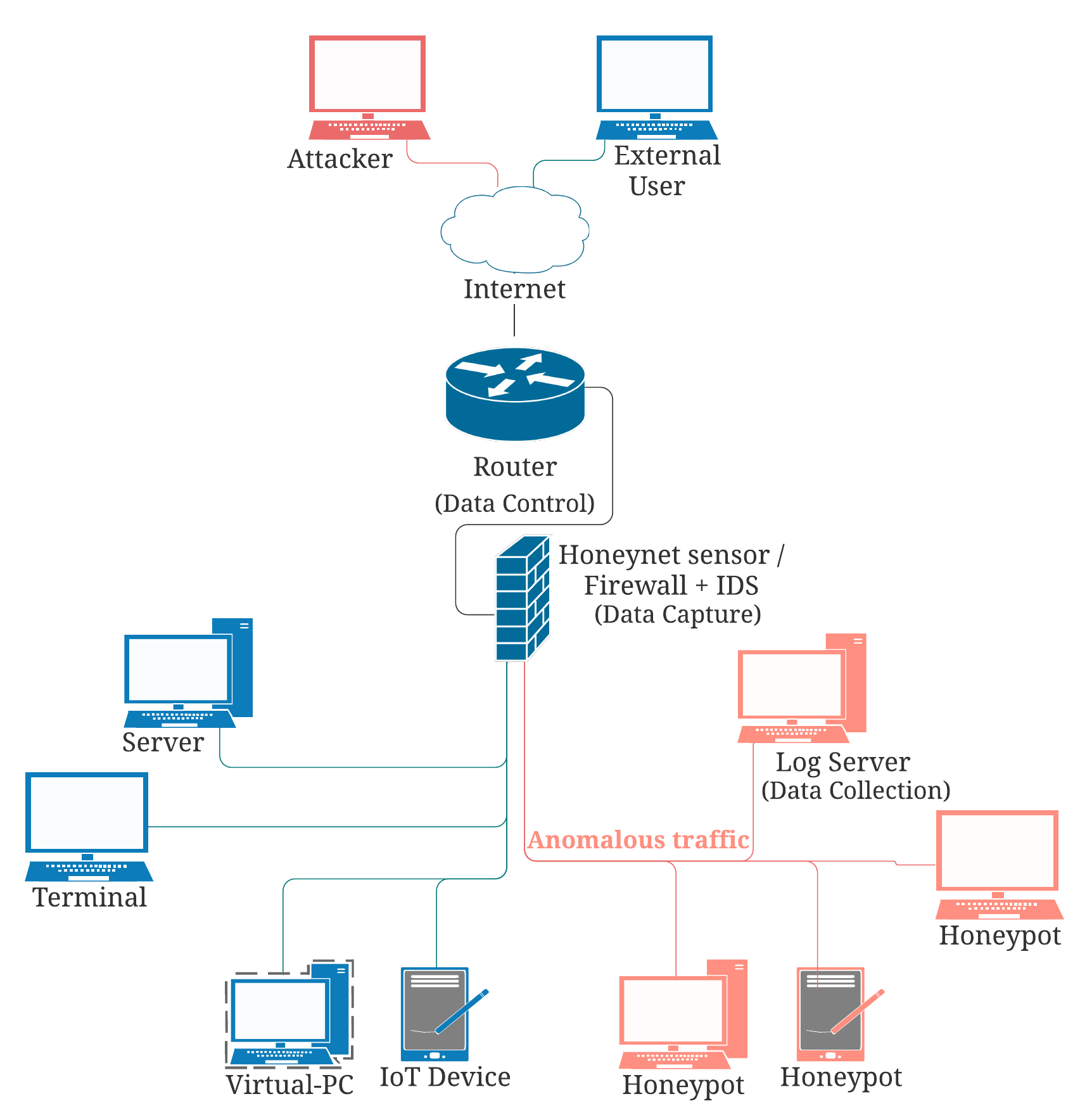}
    \vspace{0.5em}
    \caption{Basic honeynet architecture.}
    \label{fig:Basic-Honeynet-Architecture}
\end{figure}

Figure~\ref{fig:Basic-Honeynet-Architecture} depicts a basic honeynet architecture.
There are three essential elements to any honeynet: \emph{data control, data capture, and data collection}. Data control involves controlling the flow of data so that the attackers do not realize they are in a honeynet and making sure that if the honeynet is compromised, it will not be used to attack other systems. The data capture involves capturing all the data regarding movements and actions within the honeynet \cite{HoneynetProject:2001}. The data collection involves the ability to securely transfer all the captured data to a centralized place \cite{Fan:2015-2}.

Honeypots and honeynets can be deployed at various locations. They can be deployed at cloud computing environments (e.g., Amazon EC2), Demilitarized Zones (DMZ) of enterprise networks, actual application/production environments (e.g., at an IoT, IIoT, or CPS network), and private deployment environments with public IP addresses. Each of these deployment options has its own advantages and disadvantages. In addition, the decision of the deployment environment may have an effect on the choice of the most appropriate type of honeypot or honeynet. 

\subsection{Other Related Terms}
Other concepts and terms exist related to honeypots and honeynets for IoT, IIoT, and CPS applications. These are testbeds, network emulators, and simulation frameworks. Similar to honeypots and honeynets, such systems simulate or emulate devices, protocols, or even provide a physical environment where CPS devices operate and communicate using industrial protocols. However, unlike honeypots and honeynets, they do not act as decoy systems that aim to grab the attention of attackers and analyze their attacks. As we explain in the following sections, honeypot and honeynet researchers used such tools to create their decoy systems. The MiniCPS framework~\cite{MiniCPS}, the IMUNES emulator/simulator~\cite{imunes}, the GridLab-D power distribution simulator~\cite{GridLab}, the SoftGrid smart grid security toolkit~\cite{softgrid}, the PowerWorld simulator~\cite{powerworld}, and the Mininet emulator~\cite{mininet} were all used in a number of studies to simulate protocols, emulate devices and scale decoy systems. Front-end and back-end are also related terms that are used in various studies. The front-end of a honeypot/honeynet system is the part attackers interact with and gathers data, while the back-end receives data from the front-end for analysis, decryption, and storage. Self-adapting refers to the ability of a honeypot to analyze information and adapt its responses or behavior accordingly in order to accomplish its purpose better. 

\section{CLASSIFICATION METHODOLOGY}\label{sec:methodology}

\begin{figure*}[htbp]
    \centering
    \includegraphics[scale=0.38]{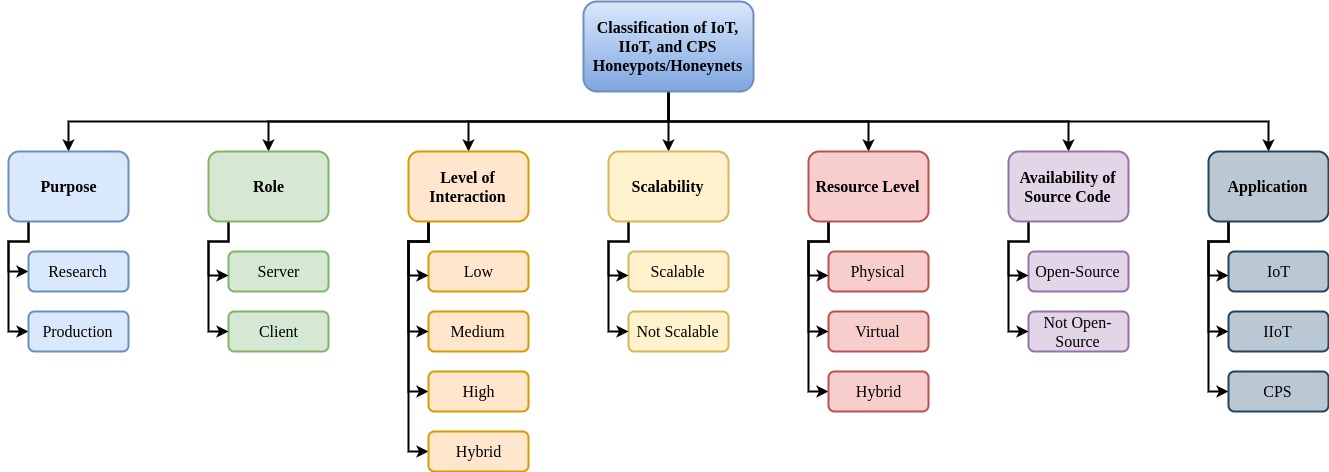}
    \vspace{1em}
    \caption{Classification \revision{categories} of honeypots and honeynets for IoT, IIoT, and CPS \revision{in which some of the items in the categorization build upon~\cite{Fan:2018,Campbell:2015,Zobal:2019,Razali:2018}}. Details of works corresponding to each category are tabulated in Tables I, II, and III. }
    \label{fig:classification}
\end{figure*}

Honeypots and honeynets can be classified in various ways. \revision {In order to classify the honeypots and honeynets for IoT, IIoT, and CPS in this survey, we build upon prior surveys \cite{Fan:2018,Campbell:2015,Zobal:2019,Razali:2018}. However, our classification in this work improves the existing works by identifying some of the recurring key characteristics of the surveyed works.}   
Specifically, we classify the honeypots and honeynets for IoT, IIoT, and CPS \revision{with respect to} their \emph{purpose, role, level of interaction, scalability, resource level, availability of the source code,} and their \emph{application} as shown in Fig.~\ref{fig:classification}. We also consider the simulated services, the inheritance relationships between the honeypots and honeynets, the platforms they were built on, and  the programming languages they used.

\textit{Classification by Purpose: }
Honeypots can be categorized into two classes based on the purpose for which they were created: \emph{research} and \emph{production} honeypots.
Research honeypots are used to gather and analyze information about attacks in order to develop better protection against those attacks. Production honeypots are more defense-focused. They are usually implemented to keep an attacker from accessing the actual system of the organization that implements it~\cite{Spitzner:2001}. 

\textit{Classification by Role: }
Role refers to whether a honeypot actively detects or passively captures traffic. A \emph{client} honeypot can actively initiate a request to a server to investigate a malicious program while a \emph{server} honeypot waits for attacks. The great majority of honeypots are server honeypots~\cite{Fan:2018}.

\textit{Classification by Level of Interaction: }
Honeypots can be classified by the level of interaction that they allow to the attacker: \emph{low interaction, medium interaction, high interaction,} and \emph{hybrid}.
Low interaction honeypots emulate one or more services with simple functions and do not give access to an operating system. The benefits of low interaction honeypots are ease of setup, low risk, low cost, and low maintenance. However, low interaction honeypots are identified much more easily by attackers because of their limitations, and the information they gather is limited and has low fidelity~\cite{Razali:2018}.

High interaction honeypots provide much more interaction, not only emulating services but also allowing access to an operating system~\cite{Razali:2018}. While some of the research refers to high interaction when a honeypot is created using real devices, other works also include virtual environments that emulate complete devices and services as high interaction. High interaction honeypots collect information about all of the attacker’s movements and actions, which is an advantage of high interaction honeypots because the information gathered has high fidelity. However, they come with high risk because everything they allow attackers to access is on real resources to gather more information. Moreover, they are more complex to set up, they collect much more data, and they are more difficult to maintain and run~\cite{Razali:2018}. Once they are compromised, rebuilding them becomes necessary. Also, attackers can compromise them to attack other targets, 
which creates liability issues.

As the name indicates, medium interaction honeypots provide a level of interaction in-between a low and a high interaction honeypot. Although there are different perspectives on whether they have a real operating system or an emulated operating system, they do emulate more services than a low-interaction honeypot, providing for more interaction which increases risk, and makes them more difficult to detect compared to low interaction honeypots.

\begin{figure}[!b]
    \centering
    \includegraphics[scale=0.23]{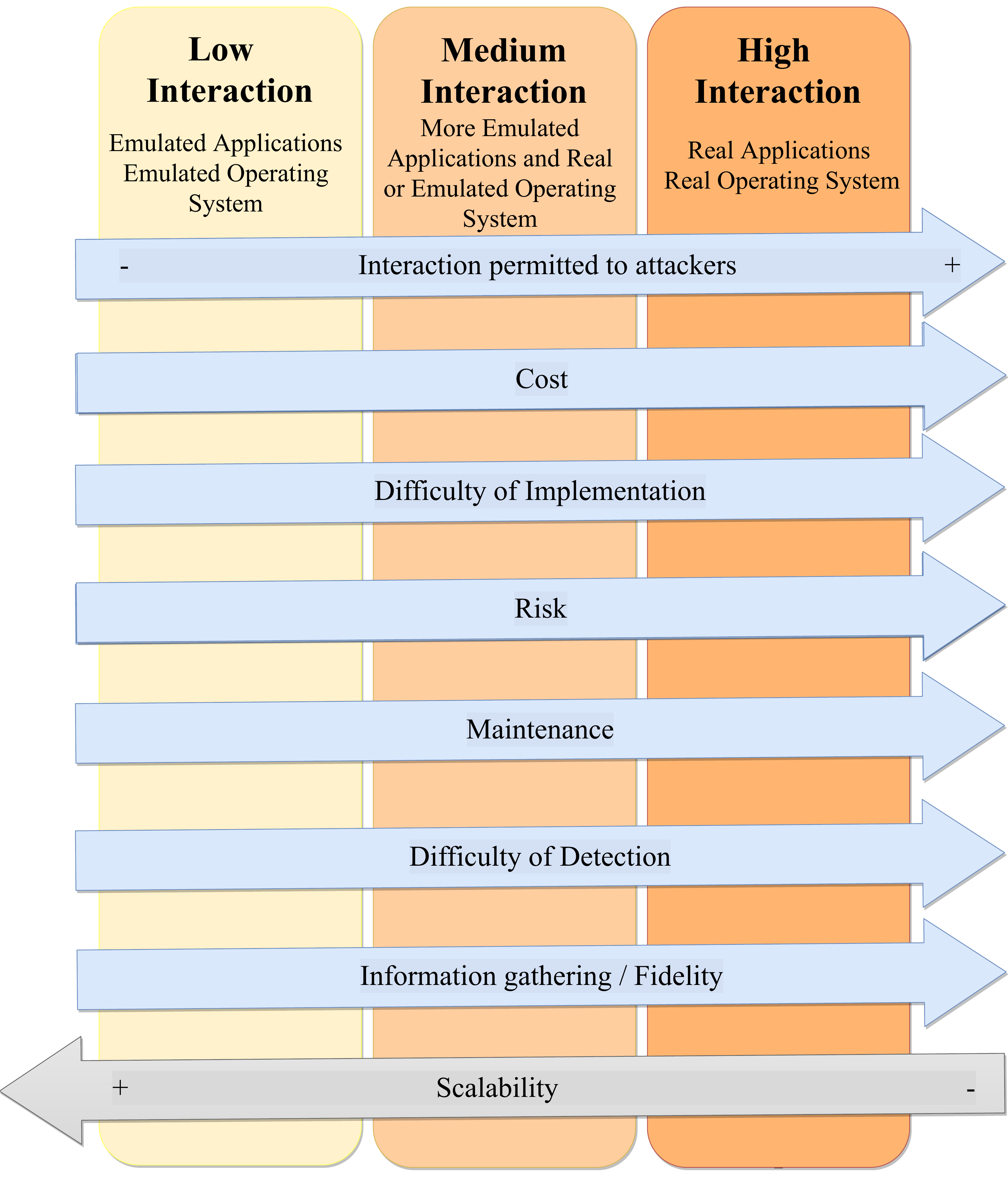}
    \caption{Characteristics by level of interaction.}
    \label{fig:Level-interaction}
\end{figure}

Figure~\ref{fig:Level-interaction} shows how the level of interaction varies in relation to the different characteristics. This should be seen as more of a fluid continuum rather than set characteristics. 

A mix of honeypots with different levels of interaction implemented in the same system is called a hybrid honeynet. Hybrid honeynets are able to provide a better balance by providing the benefits of each type of honeypot~\cite{ Dalamagkas:2019}.

\textit{Classification by Scalability: }
Scalability refers to the ability of a honeypot to grow and provide more decoys. An unscalable honeypot has only a certain number of decoys and cannot be changed. A scalable honeypot can expand the number of decoys it deploys and monitors \cite{Fan:2018}. Scalability is important because various honeypots implemented together in a honeynet provide greater protection, services, data collection, and variety of data compared to a single honeypot. Physical honeypots are usually harder to scale because of the resources needed. High-interaction honeypots also tend to have lower scalability because of their complexity.

\textit{Classification by Resource Level: }
The type of resources used to create the honeypot system
can be physical or virtual. A \emph{physical} honeypot system is composed of several honeypots running on physical machines, while a \emph{virtual} honeypot system is made up of virtual honeypots that are hosted on one or more physical machines. Physical honeypots have high interaction and have more data capture fidelity than virtual honeypots. However, they are more costly and require more resources to implement. Virtual honeypots require fewer resources to implement and are therefore less costly. A \emph{hybrid} honeynet that uses both physical and virtual honeypots is able to better balance cost and data capture fidelity \cite{Fan:2015-2}.

\textit{Classification by Availability of Source Code: }
Open-source refers to a software's source code being released in a way that anyone can have access to it, modify it, and/or distribute it. Open-source software allows for collaborative development. Not all of the honeypot and honeynet authors provide the source code of their decoy systems. Making the source code available allows other researchers and developers to understand and improve the existing honeypots and honeynets.

\textit{Classification by Application: }
Application refers to the intended application for which the honeypot system is created. \revision{In this survey, we classify the} IoT honeypot systems 
as general use, IoT, or Smart Home IoT. General use honeypots are those that were not originally 
created for IoT. However, these are relevant because they have subsequently been used in research with IoT honeypots. IoT honeypots target general IoT applications. IoT Smart Home honeypots are honeypots with a specific focus on applications for Smart Home uses. \revision{We classify }the CPS and IIoT honeypot systems 
as ICS, Smart Grid, Water System, Gas System, Building Automation System, and IIoT applications. Although the boundary between ICS and other smart infrastructures are not very obvious, we adhere to the authors' statements about their honeypots in this paper for the classification by application purposes.

\section{HONEYPOTS AND HONEYNETS FOR INTERNET OF THINGS}\label{sec:honeypots}
In this section, we give a brief overview of honeypot and honeynet studies for IoT. First, we identify some general application honeypots available. Next, we present the research with IoT honeypots and honeynets with full device emulation. Finally, we present the IoT honeypot and honeynet research focused on the type of attack. \revision{We would like to note that, unless otherwise stated, honeypots reviewed in this section are in the role of server honeypots.}

\subsection{General Application Honeypots}

There are various general application honeypots that have an inheritance relationship with later research and honeypots for IoT applications. In other words, while these \revision{honeypots and honeynets were not specifically created for IoT, they are being used in research for IoT honeypots and honeynets. It is important to note that all of these are open-source, except for the Adaptive Honeypot Alternative (AHA) with Rootkit Detection~\cite{Pauna:2012}.} \revisionB{Table~\ref{tab:generalApp} provides a list of the considered general IoT honeypots. } 

\begin{table}[H]
\caption{List of General IoT Honeypots.}
\label{tab:generalApp}
\begin{tabular}{>{}p{2cm} p{1.2cm}p{4.3cm}}
\hline
\cellcolor[HTML]{3465A4}{\color[HTML]{FFFFFF} \textbf{Honeypot}} &
  \cellcolor[HTML]{3465A4}{\color[HTML]{FFFFFF} \textbf{Interaction Level}} &
  \cellcolor[HTML]{3465A4}{\color[HTML]{FFFFFF} \textbf{Simulated Services}} \\ \hline
HoneyD \cite{Provos:2007}                                     & Low         & FTP, SMTP, Telnet, IIS, POP \\ \hline
Dionaea \cite{DinoTools:2019} &
  Medium &
  Black hole, EPMAP, FTP, HTTP, Memcache, MongoDB, MQTT, MySQL, Nfq, PPTP, SIP, SMB, TFTP, UPnP \\ \hline
Kippo \cite{Kippo:2016}                                       & Medium      & SSH                        \\ \hline
Cowrie \cite{Cowrie:2019}                                     & Medium/High & SSH, Telnet, SFTP, SCP     \\ \hline
HoneyPy \cite{HONEYPY:2015}                                   & Low/Medium  & Created as required        \\ \hline
AHA \cite{Wagener:2011}      & Low/High    & SSH                        \\ \hline
AHA with Rootkit Detection \cite{Pauna:2012}                 & Medium      & SSH                        \\ \hline
RASSH \cite{Pauna:2014}   & Medium      & SSH                        \\ \hline
QRASSH \cite{Pauna:2018} & Medium      & SSH                        \\ \hline
\end{tabular}
\end{table}

\textit{Honeyd}: \revision{HoneyD \cite{Provos:2007} is an open-source software for the creation of low interaction, scalable 
honeypots. Honeyd creates virtual honeypots, but it also allows physical machine integration. It can simulate UDP, TCP, FTP, SMTP, Telnet, IIS, and POP services. Stafira \cite{Stafira:2019} examined} 
\revision{if HoneyD is able to create effective honeypots to attract attackers. They compared honeypots 
simulating 
IoT devices with 
real devices. The results showed that, although the content served by both honeypots and real devices were similar, there are significant differences between} 
\revision{average times for query responses and Nmap scans.} 

\textit{Dionaea}: \revision{Dionaea \cite{DinoTools:2019} is an open-source software for the creation of medium interaction 
honeypots that can simulate several services (e.g., 
FTP, HTTP, 
MongoDB, MQTT, MySQL, 
SIP, SMB, TFTP, UPnP, etc.)~\cite{Dionaea:2015}. It targets adversaries that attack hosts on the Internet with vulnerable services. Since adversaries try to install malware on the infected hosts, Dionaea aims to obtain a copy of malware and help researchers to analyze it. Dionaea has a static configuration, which makes it 
difficult to adapt the configuration as needed to respond to events~\cite{Fan:2018}. Metognon et al. \cite{Metongnon:2018} used Dionaea in their IoT honeypot research. Kaur and Pateriya~\cite{Kaur:2018} proposed the setup of a cost-effective honeypot for IoT using Dionaea on Raspberry Pi and analyzed captured data using VirusTotal tool and Shodan search engine to identify the characteristics and vulnerabilities of devices in order to improve their security.}

\textit{Kippo}: Kippo \cite{Kippo:2016} is an open-source, medium interaction, scalable 
honeypot. It focuses on SSH, and it logs brute force attacks, as well as interactions from automated or individual attacks~\cite{Dowling:2017}. Dowling et al.~\cite{Dowling:2017} modified Kippo in order to implement a ZigBee IoT honeypot. Kippo is chosen because of the high number of attacks that SSH receives. Pauna~\cite{Pauna:2014} used Kippo for the creation of Reinforced  Adaptive  SSH (RASSH) honeypot. 

\textit{Adaptive Honeypot Alternative (AHA)}: \revision{Wagener~\cite{Wagener:2011} used both a low- 
and a high-interaction honeypot to gather data from attackers. With this data, he applied game-theory 
and Machine Learning (ML) techniques to develop a self-adaptive SSH honeypot called Adaptive Honeypot Alternative (AHA)~\cite{AHA:2018}. 
While Wagener does not implement his honeypot in an IoT environment, his honeypot serves as the basis for Pauna's works \cite{Pauna:2012, Pauna:2014, Pauna:2018, Pauna:2019}. 
Wagener 
reported that attackers carried out three times more interactions when they were responding to the customized tools of an adaptive honeypot, which shows 
the important role that adaptive honeypots 
can 
play in honeypot research.} 

\textit{AHA with Rootkit Detection}: \revision{In 2012, Pauna~\cite{Pauna:2012} improved on Wagener’s adaptive honeypot, creating a medium interaction, scalable, virtual honeypot with the ability to detect rootkit malware installed by attackers. Pauna's honeypot resides on the Argos emulator as a guest OS, and 
utilizes Argos to detect 
rootkit malware. 
This research was followed by \cite{Pauna:2014, Pauna:2018, Pauna:2019}.
}

\textit{RASSH}: \revision{In 2014, Pauna et al. presented an adaptive honeypot, RASSH~\cite{Pauna:2014}, which uses a medium-interaction Kippo honeypot integrated with two modules: Actions module and Reinforcement Learning module. RASSH interacts with attackers and takes dynamic actions (e.g., allowing, blocking, delaying, etc.) using the Reinforcement Learning module.  
This research was followed by \cite{Pauna:2018} \cite{Pauna:2019}, which led to the creation of IRASSH-T \cite{IRASSH:2018} self-adaptive IoT honeypot.}

\textit{Cowrie}: \revision{Cowrie~\cite{Cowrie:2019} is a software for the creation of medium to high interaction, scalable, 
virtual honeypots. As a medium interaction honeypot, it logs an attacker's shell interaction on a simulated UNIX system via emulating several commands. As a high interaction honeypot, it is a proxy for SSH and Telnet to observe an attacker's interaction on another system. To be more specific, it can act as a proxy between an attacker and a pool of virtual machines configured in a backend site which allows flexibility. Cowrie was forked from Kippo honeypot and simulates SSH, Telnet, SFTP, SCP, and TCP/IP services. It supports integration to ElasticSearch, LogStash, and Kibana for logging, storage, and visualization. It has been used in the IoT honeypot research for Metognon et al.~\cite{Metongnon:2018}, IRASSH-T~\cite{Pauna:2019}, ML-Enhanced Cowrie~\cite{Shrivastava:2019}, and Lingenfelter et al.~\cite{Lingenfelter:2020}.}

\textit{HoneyPy}: \revision{HoneyPy~\cite{HONEYPY:2015} is a software for the creation of low to medium interaction honeypots, depending on services that are simulated. HoneyPy comes with a large range of plugins that can be used for simulating services such as DNS, NTP, SIP, SMTP, web, etc. It can also be configured to run with custom configurations as needed. HoneyPy provides researchers several options for logging, which include but are not limited to ElasticSearch, Logstash, RabbitMQ, Slack, Splunk, Twitter. In this way, external services can be used to analyze HoneyPy logs. Metognon et al.~\cite{Metongnon:2018} used HoneyPy in their IoT honeypot research.}

\textit{QRASSH}: \revision{In 2018, Pauna et al.~\cite{Pauna:2018} proposed another SSH honeypot, namely Q Reinforced Adaptive SSH (QRASSH) ~\cite{QRASSH:2018} honeypot, which uses Cowrie and Deep Q-learning. However, Pauna et al. identified that the reward functions in the algorithms used in QRASSH were subjective. For this reason, they proposed further research for being able to generate optimal reward functions for the desired behavior. This study was further advanced in \cite{Pauna:2019} and led to the creation of IoT Reinforced Adaptive SSH (IRASSH-T) \cite{IRASSH:2018} honeypot.}

\textit{Metongnon and Sadre}: \revision {Metongnon and Sadre \cite{Metongnon:2018} carried out a measurement study to observe attacks against protocols that are commonly used by IoT devices. They used a large /15 network telescope to observe large-scale events/traffic on the dark address-space of the Internet. They deployed three honeypots: Cowrie~\cite{Cowrie:2019}, HoneyPy~\cite{HONEYPY:2015}, and Dionaea~\cite{Dionaea:2015} to get more details about specific attacks. The top three most attacked protocols observed via telescope were Telnet (Ports 23 and 2323), SSH (Port 22), and HTTP(S) (Ports 80, 81, 8080, 443). The most attacked protocols observed on the honeypots were Telnet, SMB, and SSH.}

\subsection{Research with IoT Honeypots and Honeynets with Full Device Emulation}

IoT honeypots and honeynets that provide full device emulation provide the most versatility. Full device emulation allows for greater realism and increases the difficulty for attackers to detect it as a honeypot. In this section, only those honeypots/honeynets which have the ability to fully emulate all kinds of devices are included. It is important to note that five of the six IoT honeypot/honeynet 
studies which are identified as providing full device emulation are also self-adaptive. \revisionB{Table~\ref{tab:deviceEmulation} provides a list of the considered IoT honeypots that perform full IoT device emulation. }

\begin{table}[H]
\caption{List of IoT Honeypots for Full Device Emulation.}
\label{tab:deviceEmulation}
\begin{tabular}{>{}p{2.8cm} p{1cm}p{3.6cm}}
\hline
\cellcolor[HTML]{3465A4}{\color[HTML]{FFFFFF} \textbf{Honeypot}} &
  \cellcolor[HTML]{3465A4}{\color[HTML]{FFFFFF} \textbf{Interaction Level}} &
  \cellcolor[HTML]{3465A4}{\color[HTML]{FFFFFF} \textbf{Emulated Devices}} \\ \hline
FIRMADYNE \cite{Chen:2016}                   & High        & COTS network-enabled IoT devices   \\ \hline
ThingPot \cite{Wang:2018}                     & Medium      & Philips Hue, Belkin, Wemo, Tplink \\ \hline
ML-Enhanced ThingPot~\cite{Vishwakarma:2019} & Medium      & General IoT devices       \\ \hline
IoTCandyJar \cite{Luo:2017}                  & Intelligent & General IoT devices      \\ \hline
Chameleon \cite{Zhou:2019}                   & Hybrid      & Any real IoT device                   \\ \hline
Honware \cite{Vetterl:2019}                  & High        & CPE devices                        \\ \hline
\end{tabular}
\end{table}

\textit{FIRMADYNE}: Chen et al.~\cite{Chen:2016} presented FIRMADYNE~\cite{FIRMADYNE:2016}, an open-source, extensible, self-adaptive automated framework for discovering vulnerabilities in commercial-off-the-shelf network-enabled devices. FIRMADYNE works by emulating the full system with an instrumented kernel. It has a web crawler component to download firmware images and their metadata, an extract firmware filesystem, an initial emulation component, and a dynamic analysis component. FIRMADYNE was evaluated using a real-world dataset of more than 23,000 firmware images from 42 device vendors and 74 exploits. Out of 9,486 firmware images that were successfully extracted, 887 prove vulnerable to at least one exploit, and 14 previously unknown vulnerabilities were discovered. 

\textit{ThingPot and ML-Enhanced ThingPot}: \revision{Wang et al.~\cite{Wang:2018} proposed 
ThingPot~\cite{ThingPot:2017}, 
a medium-interaction, scalable, 
virtual open-source honeypot that simulates the complete IoT platform and all supported application layer protocols. 
ThingPot was tested for 
45 days with Extensible Messaging and Presence Protocol (XMPP) and REST API, and most of the captured requests were HTTP REST requests. The authors noted that 
the attackers were looking for certain devices like Philips Hue, Belkin, Wemo, and TPlink, scanning to get information about the devices, and then using more targeted attacks such as brute force or fuzzing to control them. They also noted that the attackers were using The Onion Router (TOR) network~\cite{TOR:2020} 
to stay anonymous. Vishwakarma and Jain~\cite{Vishwakarma:2019} used ThingPot to propose ML-Enhanced ThingPot, a self-adaptive honeypot solution for the detection of DDoS attacks through the Telnet port that uses unsupervised machine learning (ML) techniques in real-time. 
} 

\textit{IoTCandyJar}: \revision{Luo et al.~\cite{Luo:2017} proposed a new type of honeypot which they define as \emph{intelligent interaction}, 
and has the benefits of both low and high interaction honeypots, simulating the behaviors of IoT devices without the risk of the honeypot being compromised. The honeypot uses ML with Markov Decision Process to automatically learn the behaviors of IoT devices that are publicly available on the Internet and learn which has the best response to extend the session with attackers. IoTCandyjar captured 18 million raw requests during the time of the study, including about 1 million IoT related requests. Ports 80, 7547, 8443, 81, 8080, and 88 were the most scanned, with the majority of requests being HTTP.
} 

\textit{Chameleon}: \revision{Zhou~\cite{Zhou:2019} proposed a self-adaptive IoT honeypot that can emulate all kinds of IoT devices. Chameleon has front-end responder, evaluator, and back-end interactor modules. The front-end responder 
processes requests and responds accordingly. 
If the request is new, the responder sends the request to the evaluator. The evaluator evaluates the security of the request with the IP whitelist. If the source is untrusted, Chameleon responds with a default response and the request is stored for manual study. The back-end interactor establishes a connection with the target IoT device and detects the open ports and services to open/start them on Chameleon. 
As the honeypot receives more requests, 
Chameleon's characteristics become more like those of the target device. Chameleon is evaluated by simulating a variety of 100 IoT devices on the Internet, and comparing this to 100 traditional honeypots using Shodan Honeyscore~\cite{Honeyscore:2020} fingerprinting tool. The honeypots simulated by Chameleon were not fingerprinted while all the traditional honeypots were.} 

\textit{Honware}: \revision{Vetterl and Clayton~\cite{Vetterl:2019} presented a high interaction virtual self-adaptive honeypot that emulates diverse IoT and Customer Premise Equipment (CPE) devices by processing a standard firmware image and extracting and adapting the filesystem. Honware uses Quick Emulator (QEMU) to be able to fully emulate devices, and runs this with a customized pre-built kernel and the filesystem on a host OS. 
}

\subsection{Research with IoT Honeypots and Honeynets Focused on Type of Attack}

This section contains all of the remaining research with IoT honeypots and honeynets, organized by their focus on attack type. \revisionB{Table ~\ref{tab:attackType} provides a list of the considered IoT honeypots by their target attack types. }
\begin{table}[]
\caption{List of IoT Honeypots that Focus on Specific Attacks.}
\label{tab:attackType}
\begin{tabular}{
>{}p{2.5cm}p{3.5cm}p{1.2cm}}
\hline
\cellcolor[HTML]{3465A4}{\color[HTML]{FFFFFF} \textbf{Target Attack(s)}} &
  \cellcolor[HTML]{3465A4}{\color[HTML]{FFFFFF} \textbf{Honeypots}} &
  \cellcolor[HTML]{3465A4}{\color[HTML]{FFFFFF} \textbf{Interaction Level}} \\ \hline
  
  \multirow{3}{*}{Telnet}  & IoTPOT~\cite{Pa:2016} &  Hybrid \\ \cline{2-3}
                                        & MTPot~\cite{Cymmetria:2017}, \newline Semic and Mrdovic~\cite{Semic:2017} & Low \\ \cline{2-3}
                                        &  Phype~\cite{Phype:2019} &  Medium\\ \hline
SSH and Telnet &
  Shrivastava et al.~\cite{Shrivastava:2019}, IRASSH-T~\cite{IRASSH:2018}, Lingenfelter et al.~\cite{Lingenfelter:2020} &
  Medium \\ \hline
Telnet, SSH, HTTP, and CPE WAN Management & Krishnaprasad~\cite{Krishnaprasad:2017}     & Hybrid         \\ \hline
Man-in-the-Middle                        & Oza et al.~\cite{Oza:2019}                  & High           \\ \hline

\multirow{2}{*}{D/DoS}                  & Anirudh et al.~\cite{Anirudh:2017}, \newline Vishwakarma and Jain~\cite{Vishwakarma:2019} & Medium \\ \cline{2-3}
                                        & Tambe et al.~\cite{Tambe:2019}, Molina et al.~\cite{Molina:2020} & High \\ \hline
Fileless attacks                                & HoneyCloud~\cite{Dang:2019}       & High           \\ \hline
SSH on Zigbee networks                           & Dowling et al.~\cite{Dowling:2017} & Medium         \\ \hline
UPnP                                                      & U-Pot~\cite{HakimGithub:2019}               & Medium         \\ \hline
Attacks on \newline Authentication                                   & HioTPot~\cite{Gandhi:2018}                  & Not identified \\ \hline
Reconnaissance                                  & HoneyIo4~\cite{Manzanares:2017}             & Low            \\ \hline
Attacks on home\newline networks                                   & Pot2DPI~\cite{Martin:2017}                 & Medium         \\ \hline

\multirow{3}{2cm}{Attacks on device characteristics} & Siphon~\cite{Guarnizo:2017} &  High \\\cline{2-3}
                                        & Metongnon and Sadre~\cite{Metongnon:2018} &  Low/Medium  \\\cline{2-3}
                                        & Zhang et al.~\cite{Zhang:2019} & Hybrid  \\\hline
\end{tabular}
\end{table}

\textit{Only Telnet Attacks: }\revision{
 IoTPOT \cite{Pa:2016} is a hybrid honeypot proposed by Pa et al.~\cite{Pa:2015} that simulates Telnet services for different IoT devices and focuses on Telnet intrusions. IoTPOT uses a front-end low-interaction responder that simulates IoT devices by responding to TCP requests, banner interactions, authentication, and command interactions. It is proposed to work on the back-end with a high-interaction virtual environment called IoTBOX running a Linux OS to analyze the attacks and the captured malware, and run the malware on 
 multiple CPU architectures. 
 }
 
\revision{MTPot~\cite{Cymmetria:2017} is a low-interaction, unscalable, 
virtual IoT honeypot that was designed 
specifically for 
Mirai attacks. According to Evron~\cite{Evron:2017}, it detects connections on ports using Telnet, identifies 
Mirai based on the commands requested, alters parameters to identify Mirai attacks, and reports to a syslog server. 
Evron notes that while the tool can be easily fingerprinted, it is simple and can also prove useful.} 

\revision{Semic and Mrdovic~\cite{Semic:2017} presented a multi-component low-interaction honeypot with a focus on Telnet Mirai attacks. The front-end of their 
honeypot is designed to attract and interact with attackers by using a weak, generic password. Instead of using an emulation file, the front-end is programmed to generate responses based on the input from the attacker, with the logic defined in the code. The back-end is protected by a firewall and receives the information from the front-end for decryption, reporting, and storage. 
}

Phype Telnet IoT Honeypot~\cite{Phype:2019} is an open-source software for the creation of medium interaction, scalable, virtual honeypots with a focus on IoT malware. According to the Phype GitHub repository~\cite{Phype:2019}, Phype simulates a UNIX system shell environment. It tracks and analyses botnet connections, mapping together connections and networks. The application includes a client honeypot that accepts Telnet connections and a server to receive and analyze the information gathered about these connections. 

\textit{Telnet and SSH Attacks: }  
\revision{Shrivastava et al.~\cite{Shrivastava:2019} focused on the use of Cowrie Honeypot to detect attacks on IoT devices and created a Machine Learning (ML)-Enhanced Cowrie. They opened the Telnet and SSH ports, and classified requests as malicious payload, SSH attack, XOR DDoS, suspicious, spying, or clean (non-malicious). They 
evaluated various ML algorithms to analyze and classify data, 
and concluded that Support Vector Machine (SVM) gives the best results with an accuracy of 97.39 \%. 
}

\revision{
Based on the their prior QRASSH honeypot, Pauna et al.~\cite{Pauna:2019} proposed a self-adaptive IoT honeypot named 
IRASSH-T that focuses on SSH/Telnet. 
IRASSH-T 
uses 
reinforcement learning algorithms to identify optimal reward functions for self-adaptive honeypots to communicate with attackers and capture more information about target malware. Their evaluation shows that IRASSH-T improves on previously identified reward functions for self-adaptive honeypots and will be able to attract more attacks and enable collection of more malware from attackers. 
} 

\revision{Lingenfelter et al.~\cite{Lingenfelter:2020} focused on capturing data on IoT botnets using three Cowrie SSH/Telnet 
honeypots to emulate an IoT system. Their system 
sets the prefab command outputs to match those of actual IoT devices and uses sequence matching connections on ports 
to facilitate as much traffic as possible. They analyzed remote login sessions that created or downloaded files. They also used a clustering method with edit distance between command sequences to find identical attack patterns. 
During their study, two Mirai attack patterns accounted for 97.7 \% of the attacks received on the honeypot. They concluded that botnet attacks on Telnet ports are the most common attack to download or create files, 
and many attacks on IoT devices are carried out with Mirai.
}

\textit{Telnet, SSH, HTTP, and CWMP Attacks: } \revision{ Krishnaprasad~\cite{Krishnaprasad:2017} used IoTPOT~\cite{Pa:2015} as a model in creating a honeypot with a low interaction front-end. The front-end 
has a proxy for Telnet, SSH, HTTP, and CPE WAN Management (CWMP) protocols and gathers attack data. The high interaction backend on Krishnaprasad's model can be physical or virtual, a single machine or a network of machines, and has a module for each of the protocols. 
The honeypot uses Twisted~\cite{Twisted:2014} event-driven networking engine, and employs Logstash~\cite{Logstash:2020} to collect log data. The log data is pushed to 
Elasticsearch~\cite{Elasticsearch:2017} 
for 
storage and Kibana~\cite{Kibana:2020} is used for visualization. 
For evaluation, 
Docker containers were setup to simulate IoT devices, and the honeypot was deployed in seven locations around the world. 
In seven days, the honeypot was reported to have received attacks from 6774 distinct IPs. More than half of these were Telnet attacks, followed by CWMP and SSH, with HTTP receiving significantly less attacks than the others. 
}

\textit{Man-in-the-middle Attacks: } Oza et al.~\cite{Oza:2019} addressed the issue of Man-in-the-Middle (MitM) attacks and presented a deception and authorization mechanism called OAuth to mitigate these attacks. When a user sends a request to an IoT device in the system, if the user information is not stored in the database, it is sent to an Authenticator that sends a message to the valid user. If the request is not authenticated by the user, it is sent to the honeynet instead of sending to the IoT device. 

\textit{DoS Attacks: } \revision{Anirudh et al.~\cite{Anirudh:2017} investigated 
how a DoS attack in an IoT network can be blocked by a medium-high interaction honeypot. 
Their system employs an IDS which passes malicious requests to the honeypot for further analysis. In order to evaluate their system, they simulated IoT data, and compared the performance of their system in blocking DoS attacks with and without the honeypot. 
} 

\textit{DDoS and Other Large Scale Attacks: } \revision{Using ThingPot~\cite{Wang:2018}, Vishwakarma and Jain~\cite{Vishwakarma:2019} 
proposed a self-adaptive honeypot 
to detect malware and identify unknown malware like those used in zero-day DDoS attacks. The proposed solution collects logs of attacks received by ThingPot honeypots and uses the logs to train ML classifiers. 
The authors considered 
deploying virtual box images of ThingPot on 
the IoT devices in a network, and placing the ML classifier on the router. 
}

\revision{Tambe et al.~\cite{Tambe:2019} proposed a scalable 
high interaction honeypot to attract and detect large scale botnet attacks. In order to solve the scalability problem of high interaction honeypots using real devices, Tambe et al. used VPN tunnels which allowed a small number of real IoT devices to appear as multiple IoT devices with different IP addresses around the world. 
Their evaluations using commercial-off-the-shelf IoT devices 
showed that the devices were being detected as honeypots 
by Shodan Honeyscore~\cite{Honeyscore:2020}. The authors also proposed two live traffic analysis methods for the detection of large scale attacks.} 

\revision{Molina et al.~\cite{Molina:2020} presented a self-adaptive high interaction IoT honeynet as part of a full cyber-security framework. Their framework uses Network Function Virtualization (NFV) and Software Defined Networks (SDN) to emulate a network of physical devices and allow IoT systems to self-protect and self-heal from DDoS botnet attacks. The honeynet uses NFV to allow for the autonomic deployment of virtual high interaction honeypots with dynamic configuration and reconfiguration. They used SDN for connectivity, data control, traffic filtering, forwarding, and redirecting 
between the honeynet and the real IoT environment. This allowed them to deploy honeynets both pro-actively and reactively. 
}

\textit{Fileless Malware Attacks: } \revision{ Dang et al.~\cite{Dang:2019} presented HoneyCloud for 
fileless attacks on Linux-based IoT devices. HoneyCloud was implemented using both physical and virtual honeypots. The 
virtual honeypots provided full device emulation for the six IoT device types. 
They used four 
physical IoT honeypots (a Raspberry Pi, a Beaglebone, a Netgear R6100, and a Linksys WRT54GS) and 108 
virtual IoT honeypots to attract 
and closely analyze the fileless attacks and to propose defense strategies. Their research revealed that approximately 9.7\% of malware-based attacks on IoT devices are fileless and these attacks can be powerful. 
They also identified the top ten most used shell commands in fileless attacks, 
65.7\% of which are launched through rm, kill, ps, and psswd commands, enabled by default on Linux-based IoT devices. 
} 

\textit{Only SSH Protocol Attacks:} \revision{Dowling et al.~\cite{Dowling:2017} focused on SSH protocol attacks on Zigbee networks. 
A Wireless Sensor Network (WSN) was created with Arduino and XBee modules to transmit medical information in pcap files that serve as honeytokens to catch the attention of attackers. A Kippo medium interaction SSH honeypot was modified 
to simulate a Zigbee Gateway available through SSH to attract the maximum amount of traffic. The attacks were analyzed to see which ones were directed at Zigbee. Of all the attacks documented, only individual attacks demonstrated interest in the honeytokens, or the files, leading to the conclusion that the attacks were not geared toward Zigbee in particular. On the other hand, 94\% of honeypot activity was dictionary attacks that continuously tried to access the network by sequentially trying different username and password combinations.
}

\textit {Attacks on UPnP Devices: } \revision{U-Pot~\cite{HakimGithub:2019} is an open-source medium-interaction, 
virtual honeypot platform for Universal Plug and Play-based (UPnP) IoT devices. 
U-Pot can be used to emulate real 
IoT devices, and can be scaled to mimic multiple instances at once. A honeypot can even be automatically created using UPnP device description documents for a UPnP IoT device. The main benefits of U-Pot are its flexibility, scalability, and low cost.}

\textit { Authentication Attacks:} \revision{HIoTPot~\cite{Gandhi:2018} is a virtual IoT honeypot created on a Raspberry Pi for both research and production. 
Using Raspberry Pi 3 as a server, HIoTPoT maintains a database of authenticated users. When any user attempts to gain access to the IoT network, it compares the user with the MySQL database. Unidentified users are sent to the honeypot, where their attack patterns, logs, and chat details are tracked, while the system sends an alert to notify all devices in the network of the attempted intrusion. }
 
\textit{ Reconaissance Attacks: } \revision{HoneyIo4~\cite{Manzanares:2017} is a low-interaction virtual production honeypot that simulates four IoT devices (
a camera, a printer, a video game console, and a cash register
). 
HoneyIo4 fools network scanners conducting reconnaissance attacks by simulating IoT OS fingerprints. With this fake 
OS information, the attack is redirected and becomes unsuccessful. }

\textit{ Attacks on Home Networks: } \revision{Martin et al.~\cite{Martin:2017} presented a comprehensive system for home network defense with four major components: a local honeypot to interact with attackers and collect data, a module to capture packet patterns and recognize malicious traffic, a deep packet inspection (DPI) for signature-based filtering, and a port manager for port re-mapping between the router and IoT devices. HoneyD~\cite{HoneyD:2002} low interaction honeypot is used to monitor the ports that are supposed to be inactive and Pot2DPI serves as a connection between the port manager and honeypot to inform the honeypot when packet forwarding port mapping has happened. 
Evaluation of the proposed system was carried out using 
Alman-trojan, Cerber, Fereit, and Torrentlocker pcap traces and the system was able to detect the first three with 99.84 \% 
accuracy, while it was only 48.84 \% 
accurate in detecting Torrentlocker. 
}

\textit{ Attacks Focused on Device Characteristics: } 
\revision{Guarnizo et. al~\cite{Guarnizo:2017} proposed Siphon, a high-interaction, scalable, physical honeypot. 
Siphon was implemented on seven IoT devices (IP cameras, a network video recorder (NVR), and an IP printer). The devices were made visible as 85 geographically distributed unique services on the Internet by connecting them to Amazon, LiNode, and Digital Ocean cloud servers in different cities via 
creating wormholes. 
}

Zhang et al. \cite{Zhang:2019} focused on attacks aimed at the Huawei CVE-2017-17215 vulnerability that can be exploited for remote code execution. They implemented a medium-high interaction honeypot to simulate UPnP services, a high interaction honeypot using IoT device firmware, and a hybrid multi-port honeypot using Simple Object Access Protocol (SOAP) service ports to increase the honeynet capacity and simulate honeypots. They used a Docker image to package and rapidly deploy the honeynet to capture IoT attacks.

\section{TAXONOMY OF HONEYPOTS AND HONEYNETS FOR INTERNET OF THINGS}\label{sec:taxonomy}

Honeypots and honeynets proposed for IoT are listed in Table~\ref{tab:iot-honeypots} and the tools, implementation and attack type details of the corresponding honeypots and honeynets are also outlined in Table~\ref{tab:iot-honeypots-tools}. In this section, we consider all of the proposals for IoT and provide an overview of these studies based on the development of research over time, common characteristics, level of interaction, application, scalability, resource level, simulated services, most commonly used tools, availability of the source codes, \revision{and the most common attacks}.


\subsection{Development of Research Over Time}
\begin{figure}[htbp]
    \centering
    \includegraphics[scale=0.29]{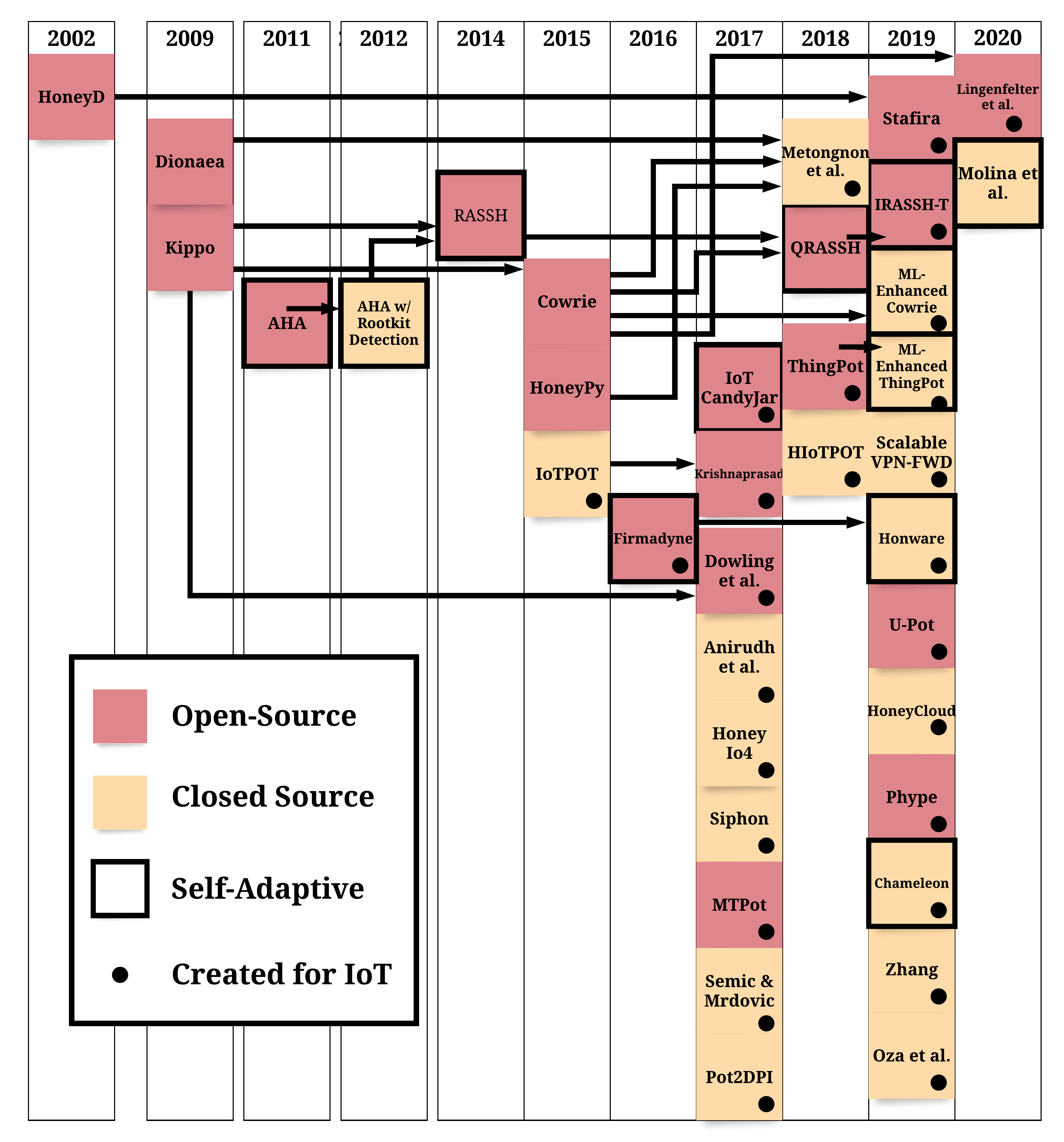}
    \caption{Evolution of Inheritance for the IoT Honeypot and Honeynet Models and Research}
    \label{fig:Parent-Child-models}
    
\end{figure}

Research of honeypots specifically created for IoT begins in 2015 with the creation of IoTPOT \cite{Pa:2015}. Previous research included in this survey was originally created for general application and later built upon for IoT applications. As shown in Figure~\ref{fig:Parent-Child-models}, about half of IoT honeypot models have a form of inheritance with each other, where a honeypot is built based on another. Cowrie \cite{Cowrie:2019} is the open-source honeypot with the greatest number of IoT honeypots which have been built directly from it.  This could be in part because Cowrie continues to be actively maintained. In 2016, Firmadyne honeypot was the second IoT specific honeypot, and the first self-adaptive IoT honeypot. After the worldwide effects of Mirai malware in 2016, including attacks on IoT devices, it is interesting to note there was a large increase in IoT honeypot and honeynet research in 2017. Of the nine studies published in 2017, seven studies explicitly refer to Mirai \cite{Luo:2017} \cite{Manzanares:2017} \cite{Guarnizo:2017} \cite{Martin:2017} \cite{Krishnaprasad:2017} \cite{Evron:2017} \cite{Semic:2017}. Also, as the number of IoT devices has increased rapidly in recent years, so has the research. 2019 saw a noticeable increase in the development of self-adaptive IoT honeypots. We can also see that more than half of the studies proposed independent honeypots, which may be due to shortcomings of existing honeypots to meet their needs. 

\smallskip
\subsection{Common Characteristics}
All of the honeypot/honeynet models 
surveyed were created for research purposes, except for HoneyIo4~\cite{Manzanares:2017} and the IoT honeynet presented by Molina~\cite{Molina:2020}, which are production, and HIoTPot~\cite{Gandhi:2018} which is identified as both research and production. All of the decoys use Linux, and all can be classified as having a server role, except for Phype Telnet IoT Honeypot~\cite{Phype:2019}, which has both a client and a server role. In addition, all of the open-source models were written in Python programming language. 

\smallskip
\subsection{Level of Interaction} In this study, classification based on level of interaction proved to be the most fluid of all the classifications regarding honeypots. Although most research can agree when a honeypot is low-interaction, definitions for the other levels can vary. For the purposes of this paper, level of interaction identified by the authors was used.  Most research seeks to leverage the benefits of both low-interaction and high-interaction honeypots, many times calling this medium interaction. In other cases, this is done using hybrid honeynet systems.

\smallskip
\subsection{Resource Level}
The great majority of available research on IoT honeynets has been carried out with virtual resources rather than physical resources. Only Siphon~\cite{Guarnizo:2017}, Stafira~\cite{Stafira:2019}, and Scalable VPN forwarded Honeypots~\cite{Razali:2018} were carried out with physical resources.

\smallskip
\subsection{Scalability} Most of the honeypot and honeynet research was carried out using scalable honeypot systems, except for Dionaea~\cite{Dionaea:2015} and HoneyIo4~\cite{Manzanares:2017}, which can only deploy one simulation at a time. It is interesting to note that despite using virtual resources rather than physical resources, these two systems cannot be expanded to provide more decoys. 

\smallskip
\subsection{Application}
Considering the application areas of honeypots and honeynets for IoT, nine of the models and research studies considered were for general use, 22 were for IoT application, and four were created for Smart Home applications. 

\smallskip
\subsection{Simulated Services}  The most commonly simulated services in the research coincide with the top three most attacked protocols identified by Metongnon and Sadre~\cite{Metongnon:2018}: Telnet, SSH, and HTTP(S). These are standard TCP/IP protocols, none of which are IoT specific. Two reasons for this may be that these common application protocols are targeted because they are in the most exposed and vulnerable layer and 75\% of attacks on IoT devices were carried out through a router \cite{Symantec:2019}. Each of the models and studies considered in this survey have their focus and specific purpose. However, there are five research models that stand out as the most versatile as they emulate full devices and are self-adaptive: IoTCandyJar~\cite{Luo:2017}, Chameleon~\cite{Zhou:2019}, Firmadyne~\cite{Chen:2016}, Honware~\cite{Vetterl:2019}, and ML-enhanced ThingPot~\cite{Vishwakarma:2019}. However, of these, only IoTCandyJar and Firmadyne are open-source. 

\smallskip
\subsection{Availability of Open-Source Honeypot and Honeynet Solutions} 
 Approximately half of the IoT honeypot and honeynet models considered in this survey are open-source. This highlights the importance of open-source software in contributing to the development of improved models.


\smallskip
\subsection{Most Commonly Used Tools} The three tools that were most commonly used in the honeypot research studies included in this survey are Shodan~\cite{Shodan:2020}, Nmap~\cite{Nmap:2020}, and MASSCAN~\cite{MASSCAN:2019}. 

Shodan~\cite{Shodan:2020} is a search engine for Internet-connected devices, which includes everything from web cams, to medical devices, appliances, and water treatment facilities. Shodan indexes everything that is somehow connected to the Internet, their location, and their users, providing valuable information about the vulnerabilities of today’s interconnected world. Shodan is used around the world, especially by corporations, researchers, security professionals, and law enforcement. 

Nmap~\cite{Nmap:2020} is an open-source, free tool for exploring networks and security auditing. It works by sending packets and then analyzing the responses. It is especially used by network administrators, auditors, and hackers to scan and determine what hosts are available on a network, the services they are offering, their operating systems, and other valuable information. The Nmap suite also has an advanced GUI, a data transfer and debugging tool called Ncat, a tool to compare scan results called Ndiff, and a packet generation and response analysis tool.  Nmap is a flexible, easy, and powerful tool. Nmap is used by honeypot and honeynet developers as a tool to gain valuable information including checking for network connectivity, scanning for open ports on real or simulated devices, comparing scan results of real vs. simulated devices, and testing the fingerprintability of honeypots. 

MASSCAN~\cite{MASSCAN:2019} is another open-source, free tool, which is very similar to Nmap and has many similar functionalities. It is a TCP port scanner and its speed sets it apart from similar tools because it transmits 10 million packets per second, which allows it to scan the entire Internet in less than six minutes. 

Although there are many other diverse tools (e.g. Wireshark, VirusTotal, Pcap, Zmap, Censys, Scapy, etc.) that have been used in IoT honeynet research, these three are by far the most commonly used. This can be attributed to their availability, their low cost, their ease of use, and their effectiveness. Nmap is the most widely recognized and used network and security auditing tool and Shodan is the first and largest search engine for Internet connected devices.

  \clearpage
\onecolumn
  \LTcapwidth=\textwidth
  \begin{longtable}{|>{\footnotesize}p{2.6cm}|>{\footnotesize}c|>{\footnotesize}p{1.25cm}|>{\footnotesize}p{1.1cm}|>{\footnotesize}p{1.5cm}|>{\footnotesize}p{4.1cm}|>{\footnotesize}p{0.7cm}|>{\footnotesize}p{0.7cm}|>{\footnotesize}p{1.5cm}|}\caption{\revision{Classification of IoT Honeypots and Honeynets}}\label{tab:iot-honeypots}\\\hline
   \rowcolor[rgb]{ .141,  .251,  .384} \textcolor[rgb]{ 1,  1,  1}{\textbf{Work}} & {\textcolor[rgb]{ 1,  1,  1}{\textbf{Year}}} & \textcolor[rgb]{ 1,  1,  1}{\textbf{Level of Interaction}} & \textcolor[rgb]{ 1,  1,  1}{\textbf{Scalability}} & \textcolor[rgb]{ 1,  1,  1}{\textbf{Resource level}} & \textcolor[rgb]{ 1,  1,  1}{\textbf{Simulated services}} & \textcolor[rgb]{ 1,  1,  1}{\textbf{Role}} & \textcolor[rgb]{ 1,  1,  1}{\textbf{Open-source}} & \textcolor[rgb]{ 1,  1,  1}{\textbf{Application}} \\\endhead

    \rowcolor[rgb]{ .863,  .902,  .945} \textbf{HoneyD \cite{HoneyD:2002}} & \cellcolor[rgb]{ 1,  1,  1}2002 & \cellcolor[rgb]{ 1,  1,  1}Low & \cellcolor[rgb]{ 1,  1,  1} \checkmark & \cellcolor[rgb]{ 1,  1,  1}Virtual& \cellcolor[rgb]{ 1,  1,  1}FTP, SMTP, Telnet,IIS, POP & \cellcolor[rgb]{ 1,  1,  1}Server & \cellcolor[rgb]{ 1,  1,  1}Yes & \cellcolor[rgb]{ 1,  1,  1}General \\\hline
    
    \rowcolor[rgb]{ .863,  .902,  .945} \textbf{Dionaea \cite{Dionaea:2015}} & \cellcolor[rgb]{ 1,  1,  1}2009 & \cellcolor[rgb]{ 1,  1,  1}Medium & \cellcolor[rgb]{ 1,  1,  1}X & \cellcolor[rgb]{ 1,  1,  1}Virtual & \cellcolor[rgb]{ 1,  1,  1}Black hole, EPMAP, FTP, HTTP, Memache, Mirror, MongoDB, MQTT, MSSQL, MySQL, nfq, PPTP, SIP, SMB, TFTP, UPnP & \cellcolor[rgb]{ 1,  1,  1}Server & \cellcolor[rgb]{ 1,  1,  1}Yes & \cellcolor[rgb]{ 1,  1,  1}General \\\hline
    
    \rowcolor[rgb]{ .863,  .902,  .945} \textbf{Kippo \cite{Kippo:2016}} & \cellcolor[rgb]{ 1,  1,  1}2009 & \cellcolor[rgb]{ 1,  1,  1}Medium & \cellcolor[rgb]{ 1,  1,  1} \checkmark & \cellcolor[rgb]{ 1,  1,  1}Virtual & \cellcolor[rgb]{ 1,  1,  1}SSH & \cellcolor[rgb]{ 1,  1,  1}Server & \cellcolor[rgb]{ 1,  1,  1}Yes & \cellcolor[rgb]{ 1,  1,  1}General \\\hline
    
    \rowcolor[rgb]{ .863,  .902,  .945} \textbf{Adaptive Honeypot Alternative~\cite{Wagener:2011}} & \cellcolor[rgb]{ 1,  1,  1}2011 & \cellcolor[rgb]{ 1,  1,  1}Low and High & \cellcolor[rgb]{ 1,  1,  1} \checkmark & \cellcolor[rgb]{ 1,  1,  1}Virtual & \cellcolor[rgb]{ 1,  1,  1}SSH & \cellcolor[rgb]{ 1,  1,  1}Server & \cellcolor[rgb]{ 1,  1,  1}Yes & \cellcolor[rgb]{ 1,  1,  1}General \\\hline
   
    \rowcolor[rgb]{ .863,  .902,  .945} \textbf{AHA with Rootkit Detection \cite{Pauna:2012}} & \cellcolor[rgb]{ 1,  1,  1}2012 & \cellcolor[rgb]{ 1,  1,  1}Medium & \cellcolor[rgb]{ 1,  1,  1} \checkmark & \cellcolor[rgb]{ 1,  1,  1}Virtual & \cellcolor[rgb]{ 1,  1,  1}SSH & \cellcolor[rgb]{ 1,  1,  1}Server & \cellcolor[rgb]{ 1,  1,  1}No & \cellcolor[rgb]{ 1,  1,  1}General \\\hline
    
    \rowcolor[rgb]{ .863,  .902,  .945} \textbf{RASSH \cite{Pauna:2014}} & \cellcolor[rgb]{ 1,  1,  1}2014 & \cellcolor[rgb]{ 1,  1,  1}Medium & \cellcolor[rgb]{ 1,  1,  1} \checkmark & \cellcolor[rgb]{ 1,  1,  1}Virtual & \cellcolor[rgb]{ 1,  1,  1}SSH & \cellcolor[rgb]{ 1,  1,  1}Server & \cellcolor[rgb]{ 1,  1,  1}Yes & \cellcolor[rgb]{ 1,  1,  1}General \\\hline
    
    \rowcolor[rgb]{ .863,  .902,  .945} \textbf{Cowrie \cite{Cowrie:2019}} & \cellcolor[rgb]{ 1,  1,  1}2015 & \cellcolor[rgb]{ 1,  1,  1}Medium/High & \cellcolor[rgb]{ 1,  1,  1} \checkmark  & \cellcolor[rgb]{ 1,  1,  1}Virtual & \cellcolor[rgb]{ 1,  1,  1}SSH, Telnet, SFTP, SCP & \cellcolor[rgb]{ 1,  1,  1}Server & \cellcolor[rgb]{ 1,  1,  1}Yes & \cellcolor[rgb]{ 1,  1,  1}General \\\hline
    
     \rowcolor[rgb]{ .863,  .902,  .945} \textbf{HoneyPy \cite{HONEYPY:2015}} & \cellcolor[rgb]{ 1,  1,  1}2015 & \cellcolor[rgb]{ 1,  1,  1}Low/Medium & \cellcolor[rgb]{ 1,  1,  1} \checkmark & \cellcolor[rgb]{ 1,  1,  1}Virtual & \cellcolor[rgb]{ 1,  1,  1}Created as required & \cellcolor[rgb]{ 1,  1,  1}Server & \cellcolor[rgb]{ 1,  1,  1}Yes & \cellcolor[rgb]{ 1,  1,  1}General \\\hline
     
     \rowcolor[rgb]{ .863,  .902,  .945} \textbf{IoTPOT \cite{Pa:2015}} & \cellcolor[rgb]{ 1,  1,  1}2015 & \cellcolor[rgb]{ 1,  1,  1}Hybrid & \cellcolor[rgb]{ 1,  1,  1} \checkmark & \cellcolor[rgb]{ 1,  1,  1}Virtual & \cellcolor[rgb]{ 1,  1,  1}Telnet & \cellcolor[rgb]{ 1,  1,  1}Server & \cellcolor[rgb]{ 1,  1,  1}No & \cellcolor[rgb]{ 1,  1,  1}IoT \\\hline
    
    \rowcolor[rgb]{ .863,  .902,  .945} \textbf{Firmadyne \cite{Chen:2016}} & \cellcolor[rgb]{ 1,  1,  1}2016 & {\cellcolor[rgb]{ 1,  1,  1}High} & {\cellcolor[rgb]{ 1,  1,  1} \checkmark} & {\cellcolor[rgb]{ 1,  1,  1}Virtual} & \cellcolor[rgb]{ 1,  1,  1}Full device emulation & {\cellcolor[rgb]{ 1,  1,  1}Server} & {\cellcolor[rgb]{ 1,  1,  1}Yes} & \cellcolor[rgb]{ 1,  1,  1}IoT \\\hline
    
    \rowcolor[rgb]{ .863,  .902,  .945} \textbf{Dowling et al. \cite{Dowling:2017}} & \cellcolor[rgb]{ 1,  1,  1}2017 & \cellcolor[rgb]{ 1,  1,  1}Medium & \cellcolor[rgb]{ 1,  1,  1} \checkmark & \cellcolor[rgb]{ 1,  1,  1}Virtual & \cellcolor[rgb]{ 1,  1,  1}Zigbee, SSH, HTTP  & \cellcolor[rgb]{ 1,  1,  1}Server & \cellcolor[rgb]{ 1,  1,  1}Yes & \cellcolor[rgb]{ 1,  1,  1}IoT \\\hline
  
    \rowcolor[rgb]{ .863,  .902,  .945} \textbf{IoT CandyJar \cite{Luo:2017}} & \cellcolor[rgb]{ 1,  1,  1}2017 & \cellcolor[rgb]{ 1,  1,  1}Intelligent & \cellcolor[rgb]{ 1,  1,  1} \checkmark & \cellcolor[rgb]{ 1,  1,  1}Virtual & \cellcolor[rgb]{ 1,  1,  1}Full device emulation  & \cellcolor[rgb]{ 1,  1,  1}Server & \cellcolor[rgb]{ 1,  1,  1}Yes & \cellcolor[rgb]{ 1,  1,  1}IoT \\\hline
   
    \rowcolor[rgb]{ .863,  .902,  .945} \textbf{Krishnaprasad \cite{Krishnaprasad:2017}} & \cellcolor[rgb]{ 1,  1,  1}2017 & \cellcolor[rgb]{ 1,  1,  1}Hybrid & \cellcolor[rgb]{ 1,  1,  1} \checkmark & \cellcolor[rgb]{ 1,  1,  1}Virtual  & \cellcolor[rgb]{ 1,  1,  1}Telnet, SSH, HTTP, CWMP & \cellcolor[rgb]{ 1,  1,  1}Server & \cellcolor[rgb]{ 1,  1,  1}Yes & \cellcolor[rgb]{ 1,  1,  1}IoT \\\hline
    
    \rowcolor[rgb]{ .863,  .902,  .945} \textbf{Anirudh et al. \cite{Anirudh:2017}} & \cellcolor[rgb]{ 1,  1,  1}2017 & \cellcolor[rgb]{ 1,  1,  1}Medium/High & \cellcolor[rgb]{ 1,  1,  1} \checkmark & \cellcolor[rgb]{ 1,  1,  1}Virtual & \cellcolor[rgb]{ 1,  1,  1}Not identified & \cellcolor[rgb]{ 1,  1,  1}Server & \cellcolor[rgb]{ 1,  1,  1}No & \cellcolor[rgb]{ 1,  1,  1}IoT \\\hline
   
    \rowcolor[rgb]{ .863,  .902,  .945} \textbf{HoneyIo4 Production Honeypot \cite{Manzanares:2017}} & \cellcolor[rgb]{ 1,  1,  1}2017 & \cellcolor[rgb]{ 1,  1,  1}Low & \cellcolor[rgb]{ 1,  1,  1}X & \cellcolor[rgb]{ 1,  1,  1}Virtual & \cellcolor[rgb]{ 1,  1,  1}SNMP, SSH, SMTP, DNS, HTTP & \cellcolor[rgb]{ 1,  1,  1}Server & \cellcolor[rgb]{ 1,  1,  1}No & \cellcolor[rgb]{ 1,  1,  1}IoT \\\hline
   
    \rowcolor[rgb]{ .863,  .902,  .945} \textbf{Siphon \cite{Guarnizo:2017}} & \cellcolor[rgb]{ 1,  1,  1}2017 & \cellcolor[rgb]{ 1,  1,  1}High & \cellcolor[rgb]{ 1,  1,  1} \checkmark & \cellcolor[rgb]{ 1,  1,  1}Physical & \cellcolor[rgb]{ 1,  1,  1}HTTP, Telnet, SSH, RTSP & \cellcolor[rgb]{ 1,  1,  1}Server & \cellcolor[rgb]{ 1,  1,  1}No & \cellcolor[rgb]{ 1,  1,  1}IoT \\\hline
    
    \rowcolor[rgb]{ .863,  .902,  .945} \textbf{MTPot \cite{Evron:2017}} & \cellcolor[rgb]{ 1,  1,  1}2017 & \cellcolor[rgb]{ 1,  1,  1}Low & \cellcolor[rgb]{ 1,  1,  1}X & \cellcolor[rgb]{ 1,  1,  1}Virtual & \cellcolor[rgb]{ 1,  1,  1}Telnet & \cellcolor[rgb]{ 1,  1,  1}Server & \cellcolor[rgb]{ 1,  1,  1}Yes & \cellcolor[rgb]{ 1,  1,  1}IoT \\\hline
    
    \rowcolor[rgb]{ .863,  .902,  .945} \textbf{Semic and \newline Mrdovic~\cite{Semic:2017}} & \cellcolor[rgb]{ 1,  1,  1}2017 & \cellcolor[rgb]{ 1,  1,  1}Low & \cellcolor[rgb]{ 1,  1,  1} \checkmark & \cellcolor[rgb]{ 1,  1,  1}Virtual & \cellcolor[rgb]{ 1,  1,  1}Telnet & \cellcolor[rgb]{ 1,  1,  1}Server & \cellcolor[rgb]{ 1,  1,  1}No & \cellcolor[rgb]{ 1,  1,  1}IoT \\\hline
    
    \rowcolor[rgb]{ .863,  .902,  .945} \textbf{Pot2DPI \cite{Martin:2017}} & \cellcolor[rgb]{ 1,  1,  1}2017 & {\cellcolor[rgb]{ 1,  1,  1}Medium} & {\cellcolor[rgb]{ 1,  1,  1} \checkmark} & {\cellcolor[rgb]{ 1,  1,  1}Virtual} & {\cellcolor[rgb]{ 1,  1,  1}Telnet, UPnP} & {\cellcolor[rgb]{ 1,  1,  1}Server} & {\cellcolor[rgb]{ 1,  1,  1}No} & \cellcolor[rgb]{ 1,  1,  1}Smart Home \\\hline
    
     \rowcolor[rgb]{ .863,  .902,  .945} \textbf{Metongnon et al. \cite{Metongnon:2018}} & \cellcolor[rgb]{ 1,  1,  1}2018 & \cellcolor[rgb]{ 1,  1,  1}Low/Medium & \cellcolor[rgb]{ 1,  1,  1} \checkmark & \cellcolor[rgb]{ 1,  1,  1}Virtual & \cellcolor[rgb]{ 1,  1,  1}SSH, Telnet, EPMAP, FTP, HTTP, Memcache, MQTT, MSSQL, MySQL, PPTP, SIP, SMB, UPnP, TFTP, TR-069.1,TR-069.2, CoAP & \cellcolor[rgb]{ 1,  1,  1}Server & \cellcolor[rgb]{ 1,  1,  1}No & \cellcolor[rgb]{ 1,  1,  1}IoT \\\hline
    
    \rowcolor[rgb]{ .863,  .902,  .945} \textbf{QRASSH \cite{Pauna:2018}} & \cellcolor[rgb]{ 1,  1,  1}2018 & \cellcolor[rgb]{ 1,  1,  1}Medium & \cellcolor[rgb]{ 1,  1,  1} \checkmark & \cellcolor[rgb]{ 1,  1,  1}Virtual & \cellcolor[rgb]{ 1,  1,  1}SSH & \cellcolor[rgb]{ 1,  1,  1}Server & \cellcolor[rgb]{ 1,  1,  1}Yes & \cellcolor[rgb]{ 1,  1,  1}General \\\hline
    
    \rowcolor[rgb]{ .863,  .902,  .945} \textbf{ThingPot et al. \cite{Wang:2018}} & \cellcolor[rgb]{ 1,  1,  1}2018 & \cellcolor[rgb]{ 1,  1,  1}Medium & \cellcolor[rgb]{ 1,  1,  1} \checkmark & \cellcolor[rgb]{ 1,  1,  1}Virtual & \cellcolor[rgb]{ 1,  1,  1}Full device emulation & \cellcolor[rgb]{ 1,  1,  1}Server & \cellcolor[rgb]{ 1,  1,  1}Yes & \cellcolor[rgb]{ 1,  1,  1}Smart Home \\\hline
    
    \rowcolor[rgb]{ .863,  .902,  .945} \textbf{HIoTPOT \cite{Gandhi:2018}} & \cellcolor[rgb]{ 1,  1,  1}2018 & \cellcolor[rgb]{ 1,  1,  1}Not identified & \cellcolor[rgb]{ 1,  1,  1} \checkmark & \cellcolor[rgb]{ 1,  1,  1}Virtual & \cellcolor[rgb]{ 1,  1,  1}Not identified & \cellcolor[rgb]{ 1,  1,  1}Server & \cellcolor[rgb]{ 1,  1,  1}No & \cellcolor[rgb]{ 1,  1,  1}IoT \\\hline

    \rowcolor[rgb]{ .863,  .902,  .945} \textbf{Stafira \cite{Stafira:2019}} & \cellcolor[rgb]{ 1,  1,  1}2019 & \cellcolor[rgb]{ 1,  1,  1}Low & \cellcolor[rgb]{ 1,  1,  1} \checkmark & \cellcolor[rgb]{ 1,  1,  1}Physical & \cellcolor[rgb]{ 1,  1,  1}TCP/IP, HTTP & \cellcolor[rgb]{ 1,  1,  1}Server & \cellcolor[rgb]{ 1,  1,  1}Yes & \cellcolor[rgb]{ 1,  1,  1}Smart Home \\\hline
   
    \rowcolor[rgb]{ .863,  .902,  .945} \textbf{IRASSH-T \cite{Pauna:2019}} & \cellcolor[rgb]{ 1,  1,  1}2019 & \cellcolor[rgb]{ 1,  1,  1}Medium & \cellcolor[rgb]{ 1,  1,  1} \checkmark & \cellcolor[rgb]{ 1,  1,  1}Virtual & \cellcolor[rgb]{ 1,  1,  1}SSH & \cellcolor[rgb]{ 1,  1,  1}Server & \cellcolor[rgb]{ 1,  1,  1}Yes & \cellcolor[rgb]{ 1,  1,  1}IoT \\\hline
    
    \rowcolor[rgb]{ .863,  .902,  .945} \textbf{ML enhanced \newline Cowrie~\cite{Shrivastava:2019}} & \cellcolor[rgb]{ 1,  1,  1}2019 & \cellcolor[rgb]{ 1,  1,  1}Medium & \cellcolor[rgb]{ 1,  1,  1} \checkmark & \cellcolor[rgb]{ 1,  1,  1}Virtual & \cellcolor[rgb]{ 1,  1,  1}SSH, Telnet & \cellcolor[rgb]{ 1,  1,  1}Server & \cellcolor[rgb]{ 1,  1,  1}Yes & \cellcolor[rgb]{ 1,  1,  1}IoT \\\hline
    
    \rowcolor[rgb]{ .863,  .902,  .945} \textbf{ML enhanced \newline ThingPot~\cite{Vishwakarma:2019}} & \cellcolor[rgb]{ 1,  1,  1}2019 & \cellcolor[rgb]{ 1,  1,  1}Medium & \cellcolor[rgb]{ 1,  1,  1} \checkmark & \cellcolor[rgb]{ 1,  1,  1}Virtual & \cellcolor[rgb]{ 1,  1,  1} Full device emulation & \cellcolor[rgb]{ 1,  1,  1}Server & \cellcolor[rgb]{ 1,  1,  1}No & \cellcolor[rgb]{ 1,  1,  1}IoT \\\hline
    
    \rowcolor[rgb]{ .863,  .902,  .945} \textbf{Scalable \newline VPN-forwarded Honeypots \cite{Tambe:2019}} & \cellcolor[rgb]{ 1,  1,  1}2019 & \cellcolor[rgb]{ 1,  1,  1}High & \cellcolor[rgb]{ 1,  1,  1} \checkmark & \cellcolor[rgb]{ 1,  1,  1}Physical & \cellcolor[rgb]{ 1,  1,  1}HTTP, TFTP, Telnet, others not specified & \cellcolor[rgb]{ 1,  1,  1}Server & \cellcolor[rgb]{ 1,  1,  1}No & \cellcolor[rgb]{ 1,  1,  1}IoT \\\hline
    
    \rowcolor[rgb]{ .863,  .902,  .945} \textbf{Zhang \cite{Zhang:2019}} & \cellcolor[rgb]{ 1,  1,  1}2019 & \cellcolor[rgb]{ 1,  1,  1}Hybrid & \cellcolor[rgb]{ 1,  1,  1} \checkmark & \cellcolor[rgb]{ 1,  1,  1}Physical/Virtual & \cellcolor[rgb]{ 1,  1,  1}UPnP, SOAP & \cellcolor[rgb]{ 1,  1,  1}Server & \cellcolor[rgb]{ 1,  1,  1}No & \cellcolor[rgb]{ 1,  1,  1}IoT \\\hline
    
     \rowcolor[rgb]{ .863,  .902,  .945} \textbf{U-Pot \cite{HakimGithub:2019}} & \cellcolor[rgb]{ 1,  1,  1}2019 & \cellcolor[rgb]{ 1,  1,  1}Medium & \cellcolor[rgb]{ 1,  1,  1} \checkmark & \cellcolor[rgb]{ 1,  1,  1}Virtual & \cellcolor[rgb]{ 1,  1,  1}UPnP & \cellcolor[rgb]{ 1,  1,  1}Server & \cellcolor[rgb]{ 1,  1,  1}Yes & \cellcolor[rgb]{ 1,  1,  1}IoT \\\hline
    
    \rowcolor[rgb]{ .863,  .902,  .945} \textbf{HoneyCloud \cite{Dang:2019}} & \cellcolor[rgb]{ 1,  1,  1}2019 & \cellcolor[rgb]{ 1,  1,  1}High & \cellcolor[rgb]{ 1,  1,  1} \checkmark & \cellcolor[rgb]{ 1,  1,  1}Physical/Virtual & \cellcolor[rgb]{ 1,  1,  1}SSH, Telnet, SMB, HTTP, HTTPS, RDP, MySQL, SQL Server & \cellcolor[rgb]{ 1,  1,  1}Server & \cellcolor[rgb]{ 1,  1,  1}No & \cellcolor[rgb]{ 1,  1,  1}Smart Home \\\hline
    
    \rowcolor[rgb]{ .863,  .902,  .945} \textbf{Phype \cite{Phype:2019}} & \cellcolor[rgb]{ 1,  1,  1}2019 & \cellcolor[rgb]{ 1,  1,  1}Medium & \cellcolor[rgb]{ 1,  1,  1} \checkmark & \cellcolor[rgb]{ 1,  1,  1}Virtual & \cellcolor[rgb]{ 1,  1,  1}Telnet & \cellcolor[rgb]{ 1,  1,  1}Server & \cellcolor[rgb]{ 1,  1,  1}Yes & \cellcolor[rgb]{ 1,  1,  1}IoT \\\hline
   
    \rowcolor[rgb]{ .863,  .902,  .945} \textbf{Oza et al. \cite{Oza:2019}} & \cellcolor[rgb]{ 1,  1,  1}2019 & {\cellcolor[rgb]{ 1,  1,  1}High} & {\cellcolor[rgb]{ 1,  1,  1} \checkmark} & \cellcolor[rgb]{ 1,  1,  1}Virtual & {\cellcolor[rgb]{ 1,  1,  1}Not identified } & {\cellcolor[rgb]{ 1,  1,  1}Server} & {\cellcolor[rgb]{ 1,  1,  1}No} & \cellcolor[rgb]{ 1,  1,  1}IoT \\\hline
  
    \rowcolor[rgb]{ .863,  .902,  .945} \textbf{Honware~\cite{Vetterl:2019}} & \cellcolor[rgb]{ 1,  1,  1}2019 & {\cellcolor[rgb]{ 1,  1,  1}High} & {\cellcolor[rgb]{ 1,  1,  1} \checkmark} & {\cellcolor[rgb]{ 1,  1,  1}Virtual} & \cellcolor[rgb]{ 1,  1,  1}Full device emulation & {\cellcolor[rgb]{ 1,  1,  1}Server} & {\cellcolor[rgb]{ 1,  1,  1}No} & \cellcolor[rgb]{ 1,  1,  1}IoT \\\hline
    
    \rowcolor[rgb]{ .863,  .902,  .945} \textbf{Chameleon~\cite{Zhou:2019}} & \cellcolor[rgb]{ 1,  1,  1}2019 & {\cellcolor[rgb]{ 1,  1,  1}Hybrid} & {\cellcolor[rgb]{ 1,  1,  1} \checkmark} & {\cellcolor[rgb]{ 1,  1,  1}Virtual} & \cellcolor[rgb]{ 1,  1,  1}Full device emulation  & {\cellcolor[rgb]{ 1,  1,  1}Server} & {\cellcolor[rgb]{ 1,  1,  1}Yes} & \cellcolor[rgb]{ 1,  1,  1}IoT \\\hline
        
    \rowcolor[rgb]{ .863,  .902,  .945} \textbf{Lingenfelter \newline et al. \cite{Lingenfelter:2020}} & \cellcolor[rgb]{ 1,  1,  1}2020 & {\cellcolor[rgb]{ 1,  1,  1}Medium} & {\cellcolor[rgb]{ 1,  1,  1} \checkmark} & {\cellcolor[rgb]{ 1,  1,  1}Virtual} & \cellcolor[rgb]{ 1,  1,  1}SSH, Telnet, SMTP, HTTP & {\cellcolor[rgb]{ 1,  1,  1}Server} & {\cellcolor[rgb]{ 1,  1,  1}Yes} & \cellcolor[rgb]{ 1,  1,  1}IoT \\\hline
    
    \rowcolor[rgb]{ .863,  .902,  .945} \textbf{Molina et al. \cite{Molina:2020}} & \cellcolor[rgb]{ 1,  1,  1}2020 & {\cellcolor[rgb]{ 1,  1,  1}High} & {\cellcolor[rgb]{ 1,  1,  1} \checkmark} & {\cellcolor[rgb]{ 1,  1,  1}Virtual} & \cellcolor[rgb]{ 1,  1,  1}Not identified & {\cellcolor[rgb]{ 1,  1,  1}Server} & {\cellcolor[rgb]{ 1,  1,  1}No} & \cellcolor[rgb]{ 1,  1,  1}IoT \\\hline
  %
\end{longtable}%
\clearpage
\twocolumn

	\clearpage
\onecolumn
	\LTcapwidth=\textwidth
	\renewcommand\arraystretch{1.5}
	\begin{longtable} {|>{\footnotesize}p{1.8cm}|>{\footnotesize}p{2.5cm}|>{\footnotesize}p{3.75cm}|>{\footnotesize}p{3.15cm}|>{\footnotesize}p{3.7cm}|>{\footnotesize}p{0.9cm}|}\caption{Tools, Implementation and Attack Types of Honeypots and Honeynets for IoT}\label{tab:iot-honeypots-tools} \\\hline
	\rowcolor[rgb]{ .141,  .251,  .384} \textcolor[rgb]{ 1,  1,  1}{\textbf{Work}} & \textcolor[rgb]{ 1,  1,  1}{\textbf{Tools}} & \textcolor[rgb]{ 1,  1,  1}{\textbf{Simulated services}} & \textcolor[rgb]{ 1,  1,  1}{\textbf{Attack Types}} & \textcolor[rgb]{ 1,  1,  1}{\textbf{Data Analyzed}} & \textcolor[rgb]{ 1,  1,  1}{\textbf{Length of the Study}} \\
    
    \endfirsthead
\multicolumn{6}{c}%
{\tablename\ \thetable: Tools, Implementation and Attack Types of Honeypots and Honeynets for IoT (Cont.)} \\
\hline
\rowcolor[rgb]{ .141,  .251,  .384} \textcolor[rgb]{ 1,  1,  1}{\textbf{Work}} & \textcolor[rgb]{ 1,  1,  1}{\textbf{Tools}} & \textcolor[rgb]{ 1,  1,  1}{\textbf{Simulated services}} & \textcolor[rgb]{ 1,  1,  1}{\textbf{Attack Types}} & \textcolor[rgb]{ 1,  1,  1}{\textbf{Data Analyzed}} & \textcolor[rgb]{ 1,  1,  1}{\textbf{Length of the Study}} \\\endhead

    \rowcolor[rgb]{ .863,  .902,  .945} \textbf{HoneyD \cite{HoneyD:2002}} & {\cellcolor[rgb]{ 1,  1,  1}N/A} & \cellcolor[rgb]{ 1,  1,  1}FTP, SMTP, Telnet,IIS, POP & \cellcolor[rgb]{ 1,  1,  1}N/A & \cellcolor[rgb]{ 1,  1,  1}N/A & \cellcolor[rgb]{ 1,  1,  1}N/A \\\hline
    
    \rowcolor[rgb]{ .863,  .902,  .945} \textbf{Dionaea \cite{Dionaea:2015}} & {\cellcolor[rgb]{ 1,  1,  1}N/A} & \cellcolor[rgb]{ 1,  1,  1}Black TTP, Memache, Mirror, MongoDB, MQTT, MSSQL, MySQL, Nfq, PPTP, SIP, SMB, TFTP, UPnP & \cellcolor[rgb]{ 1,  1,  1}N/A & \cellcolor[rgb]{ 1,  1,  1}N/A & \cellcolor[rgb]{ 1,  1,  1}N/A \\\hline
    
    \rowcolor[rgb]{ .863,  .902,  .945} \textbf{Kippo \cite{Kippo:2016}} & {\cellcolor[rgb]{ 1,  1,  1}N/A} & \cellcolor[rgb]{ 1,  1,  1}SSH & \cellcolor[rgb]{ 1,  1,  1}N/A & \cellcolor[rgb]{ 1,  1,  1}N/A & \cellcolor[rgb]{ 1,  1,  1}N/A \\\hline
    
    \rowcolor[rgb]{ .863,  .902,  .945} \textbf{Adaptive \newline Honeypot \newline Alternative~\cite{Wagener:2011}} & \cellcolor[rgb]{ 1,  1,  1}AHA Daemon & \cellcolor[rgb]{ 1,  1,  1}SSH & \cellcolor[rgb]{ 1,  1,  1}SSH-brute force & \cellcolor[rgb]{ 1,  1,  1}User/passwords, TTY buffer, TCP/UDP packets & \cellcolor[rgb]{ 1,  1,  1}8 hours \\\hline
    
    \rowcolor[rgb]{ .863,  .902,  .945} \textbf{AHA with Rootkit Detection \cite{Pauna:2012}} & \cellcolor[rgb]{ 1,  1,  1}AHA Daemon, Kernel rootkit Kbeast, Argos& \cellcolor[rgb]{ 1,  1,  1}SSH & \cellcolor[rgb]{ 1,  1,  1}Rootkit malware & \cellcolor[rgb]{ 1,  1,  1}Keystroke logging, Rootkit malware & \cellcolor[rgb]{ 1,  1,  1}7 days \\\hline
    
    \rowcolor[rgb]{ .863,  .902,  .945} \textbf{RASSH \cite{Pauna:2014}} & \cellcolor[rgb]{ 1,  1,  1}Pybrain RL, SARSA, Markov & \cellcolor[rgb]{ 1,  1,  1}SSH & \cellcolor[rgb]{ 1,  1,  1}SSH attack & \cellcolor[rgb]{ 1,  1,  1}Logs, commands offering downloading & \cellcolor[rgb]{ 1,  1,  1}N/A \\\hline
   
    \rowcolor[rgb]{ .863,  .902,  .945} \textbf{Cowrie \cite{Cowrie:2019}} & \cellcolor[rgb]{ 1,  1,  1}N/A & \cellcolor[rgb]{ 1,  1,  1}SSH, Telnet, SFTP, SCP & \cellcolor[rgb]{ 1,  1,  1}N/A & \cellcolor[rgb]{ 1,  1,  1}N/A & \cellcolor[rgb]{ 1,  1,  1}N/A \\\hline
    
    \rowcolor[rgb]{ .863,  .902,  .945} \textbf{HoneyPy \cite{HONEYPY:2015}} & {\cellcolor[rgb]{ 1,  1,  1}N/A} & \cellcolor[rgb]{ 1,  1,  1}Created as required & \cellcolor[rgb]{ 1,  1,  1}N/A& \cellcolor[rgb]{ 1,  1,  1}N/A & \cellcolor[rgb]{ 1,  1,  1}N/A \\\hline
    
    \rowcolor[rgb]{ .863,  .902,  .945} \textbf{IoTPOT \cite{Pa:2015}} & \cellcolor[rgb]{ 1,  1,  1}Masscan, pcap& \cellcolor[rgb]{ 1,  1,  1}Telnet & \cellcolor[rgb]{ 1,  1,  1}DNS Water Torture, SSL attack, DoS, DDoS, UDP Flood, SYN Flood, ACk Flood, SynAck Flood, Null Flood, Telnet Scan, DNS attacks, Fake Web Hosting & \cellcolor[rgb]{ 1,  1,  1}PCAP analysis includes total \# of packets, start/end time of packet captures, data byte/bit rate, average packet size and rate, number of victim IP address for each attack & \cellcolor[rgb]{ 1,  1,  1}39 days \\\hline
    
    \rowcolor[rgb]{ .863,  .902,  .945} {\textbf{Firmadyne \cite{Chen:2016}}} & \cellcolor[rgb]{ 1,  1,  1}Nmap, Metasploit framework, Binwalk, Scrapy, QEMU, Sasquatch, Firmware-mod-kit & \cellcolor[rgb]{ 1,  1,  1}HTTP, Telnet, DNS, dec-notes, HTTPS, UPnP, RIPD, Freeciv & \cellcolor[rgb]{ 1,  1,  1}Reconnaissance attacks, buffer overflow & \cellcolor[rgb]{ 1,  1,  1}Firmwares, results from web analysis, MIB files & \cellcolor[rgb]{ 1,  1,  1}N/A \\\hline
    
    \rowcolor[rgb]{ .863,  .902,  .945} \textbf{IoT CandyJar \cite{Luo:2017}} & \cellcolor[rgb]{ 1,  1,  1}pyLDAvis, Digital Ocean VM, Amazon AWS, MDP, Censys, ZoomEye, Shodan, MASSCAN & \cellcolor[rgb]{ 1,  1,  1}HTTP, RTSP, SOAP\_Envelope & \cellcolor[rgb]{ 1,  1,  1}HTTP, HTTP\_HEAD, UDP, HTTP\_OPTIONS, TCP, SOAP\_Envelope, RTSP, HTTP\_CONNECT & \cellcolor[rgb]{ 1,  1,  1}Attack types and characteristics & \cellcolor[rgb]{ 1,  1,  1}1 month \\\hline
    
    \rowcolor[rgb]{ .863,  .902,  .945} \textbf{Krishnaprasad \cite{Krishnaprasad:2017}} & \cellcolor[rgb]{ 1,  1,  1}Twisted, ELK Stack, Docker,  & \cellcolor[rgb]{ 1,  1,  1}Telnet, SSH, HTTP, CWMP & \cellcolor[rgb]{ 1,  1,  1}Brute-force attack, Hajime, ZmEu attacks & \cellcolor[rgb]{ 1,  1,  1}Attack types and characteristics & \cellcolor[rgb]{ 1,  1,  1}7 days \\\hline
    
    \rowcolor[rgb]{ .863,  .902,  .945} \textbf{Anirudh et al. \cite{Anirudh:2017}} & {\cellcolor[rgb]{ 1,  1,  1}IDS, logs} & \cellcolor[rgb]{ 1,  1,  1}N/A & \cellcolor[rgb]{ 1,  1,  1}DoS attacks & \cellcolor[rgb]{ 1,  1,  1}IP Address, MAC Address & \cellcolor[rgb]{ 1,  1,  1}N/A \\\hline
    
    \rowcolor[rgb]{ .863,  .902,  .945} \textbf{HoneyIo4 \newline Production \newline Honeypot \cite{Manzanares:2017}} & \cellcolor[rgb]{ 1,  1,  1}Shodan, Nmap, Wireshark, Scapy, VM. & \cellcolor[rgb]{ 1,  1,  1}SNMP, SSH, SMTP, DNS, HTTP & \cellcolor[rgb]{ 1,  1,  1}Reconnaissance attacks & \cellcolor[rgb]{ 1,  1,  1}TCP, UDP and ICMP packets& \cellcolor[rgb]{ 1,  1,  1}N/A \\\hline
    
    \rowcolor[rgb]{ .863,  .902,  .945} \textbf{Siphon \cite{Guarnizo:2017}} & \cellcolor[rgb]{ 1,  1,  1}Shodan, Tcpdump, Nmap & \cellcolor[rgb]{ 1,  1,  1}HTTP, Telnet, SSH, RTSP & \cellcolor[rgb]{ 1,  1,  1}Brute-force login attempts & \cellcolor[rgb]{ 1,  1,  1}TCP connections per wormhole, services consulted, access gained, movements statistics  & \cellcolor[rgb]{ 1,  1,  1}60 days \\\hline
    
    \rowcolor[rgb]{ .863,  .902,  .945} \textbf{MTPot \cite{Evron:2017}} & {\cellcolor[rgb]{ 1,  1,  1}N/A} & \cellcolor[rgb]{ 1,  1,  1}Telnet & \cellcolor[rgb]{ 1,  1,  1}N/A & \cellcolor[rgb]{ 1,  1,  1}Incoming connections on any port using telnet & \cellcolor[rgb]{ 1,  1,  1}N/A \\\hline
    
    \rowcolor[rgb]{ .863,  .902,  .945} \textbf{Semic and \newline Mrdovic  \cite{Semic:2017}} & {\cellcolor[rgb]{ 1,  1,  1}N/A} & \cellcolor[rgb]{ 1,  1,  1}Telnet & \cellcolor[rgb]{ 1,  1,  1}Telnet attack & \cellcolor[rgb]{ 1,  1,  1}Protocols, IP addresses, logs & \cellcolor[rgb]{ 1,  1,  1}N/A \\\hline
    
    \rowcolor[rgb]{ .863,  .902,  .945} \textbf{Pot2DPI \cite{Martin:2017}} & {\cellcolor[rgb]{ 1,  1,  1}N/A} & {\cellcolor[rgb]{ 1,  1,  1}Telnet, UPnP} & \cellcolor[rgb]{ 1,  1,  1}Mirai and Persirai attacks protocols, ports scans  & \cellcolor[rgb]{ 1,  1,  1}Packet traces, attack signatures, protocols, ports & \cellcolor[rgb]{ 1,  1,  1}N/A \\\hline
    
    \rowcolor[rgb]{ .863,  .902,  .945} \textbf{Metongnon et al. \cite{Metongnon:2018}} & \cellcolor[rgb]{ 1,  1,  1}Eemo,  Shodan  & \cellcolor[rgb]{ 1,  1,  1}SSH, Telnet, EPMAP, FTP, HTTP, Memcache, MQTT, MSSQL, MySQL, PPTP, SIP, SMB, UPnP, TFTP, TR-069.1, TR-069.2, CoAP & \cellcolor[rgb]{ 1,  1,  1}Attack URL, SYN packet, Mirai and Mirai-like attacks, Harvest cryptocurrencies, Login attempts, Reconaissance & \cellcolor[rgb]{ 1,  1,  1}Protocols, packets per port, packets characteristics & \cellcolor[rgb]{ 1,  1,  1}5 months \\\hline
    
    \rowcolor[rgb]{ .863,  .902,  .945} \textbf{QRASSH \cite{Pauna:2018}} & \cellcolor[rgb]{ 1,  1,  1}Deep Q-learning, Keras with Theano backend, Nmap & \cellcolor[rgb]{ 1,  1,  1}SSH & \cellcolor[rgb]{ 1,  1,  1}SSH attack & \cellcolor[rgb]{ 1,  1,  1}Commands(downloading, hacking, linux) & \cellcolor[rgb]{ 1,  1,  1}N/A \\\hline
    
    \rowcolor[rgb]{ .863,  .902,  .945} \textbf{ThingPot et al. \cite{Wang:2018}} & \cellcolor[rgb]{ 1,  1,  1}Skipfish, Nikto, Masscan & \cellcolor[rgb]{ 1,  1,  1}HTTP, XMPP, ZigBee & \cellcolor[rgb]{ 1,  1,  1}HTTP POST request, HTTP GET with URLs, scanning tools, SQL malware & \cellcolor[rgb]{ 1,  1,  1}HTTP request, SQL access request, scanning network & \cellcolor[rgb]{ 1,  1,  1}1.5 months \\\hline
    
    \rowcolor[rgb]{ .863,  .902,  .945} \textbf{Stafira \cite{Stafira:2019}} & \cellcolor[rgb]{ 1,  1,  1}Nmap, Wireshark, VMWare Workstation  & \cellcolor[rgb]{ 1,  1,  1}TCP/ IP, HTTP & \cellcolor[rgb]{ 1,  1,  1}Only user testing & \cellcolor[rgb]{ 1,  1,  1}Access time, HTML code, network headers and Nmap scan & \cellcolor[rgb]{ 1,  1,  1}N/A \\\hline
    
    \rowcolor[rgb]{ .863,  .902,  .945} \textbf{IRASSH-T \cite{Pauna:2019}} & \cellcolor[rgb]{ 1,  1,  1}Apprenticeship Learning & \cellcolor[rgb]{ 1,  1,  1}SSH & \cellcolor[rgb]{ 1,  1,  1}SSH attack & \cellcolor[rgb]{ 1,  1,  1}N/A & \cellcolor[rgb]{ 1,  1,  1}N/A \\\hline
    
    \rowcolor[rgb]{ .863,  .902,  .945} \textbf{ML enhanced Cowrie \cite{Shrivastava:2019}} & \cellcolor[rgb]{ 1,  1,  1}Support Vector Machine (SVM), Random Forest, Naive Bayes, J48 decision tree, VirusTotal website, Weka, machine learning algorithms & \cellcolor[rgb]{ 1,  1,  1}SSH, Telnet & \cellcolor[rgb]{ 1,  1,  1}Malicious payload, SSH attack, XOR DDoS, Spying, Suspicious, Clean & \cellcolor[rgb]{ 1,  1,  1}System logs, IP, attack types and characteristics, commands executed, behavior analysis & \cellcolor[rgb]{ 1,  1,  1}40 days \\\hline
    
    \rowcolor[rgb]{ .863,  .902,  .945} \textbf{ML enhanced ThingPot \cite{Vishwakarma:2019}} & \cellcolor[rgb]{ 1,  1,  1}Linux bash scripts, Microsoft Azure, MATLAB & \cellcolor[rgb]{ 1,  1,  1}Telnet, MQTT, XMPP, AMQP, CoAP, UPnP, HTTP, REST & \cellcolor[rgb]{ 1,  1,  1}DDoS, malware, TCP SYN flood, UDP flood, HTTP GET flood & \cellcolor[rgb]{ 1,  1,  1}Network traffic, payload, malware samples, the toolkit by attacker & \cellcolor[rgb]{ 1,  1,  1}N/A \\\hline
    
    \rowcolor[rgb]{ .863,  .902,  .945} \textbf{Scalable VPN- forwarded Honeypots \cite{Tambe:2019}} & \cellcolor[rgb]{ 1,  1,  1}VPN, TShark, HONAN, pcap, VM, MySQL, own script & \cellcolor[rgb]{ 1,  1,  1}HTTP, TFTP, Telnet, others not specified & \cellcolor[rgb]{ 1,  1,  1}DDoS style attacks & \cellcolor[rgb]{ 1,  1,  1}Protocols, packets per port, packets characteristics & \cellcolor[rgb]{ 1,  1,  1}16 months \\\hline
    
    \rowcolor[rgb]{ .863,  .902,  .945} \textbf{Zhang \cite{Zhang:2019}} & \cellcolor[rgb]{ 1,  1,  1}Tc, own script & \cellcolor[rgb]{ 1,  1,  1}UPnP, SOAP & \cellcolor[rgb]{ 1,  1,  1}UPnP & \cellcolor[rgb]{ 1,  1,  1}Protocols, packets per port, timestamp, inject behaviors & \cellcolor[rgb]{ 1,  1,  1}7 days \\\hline
    
     \rowcolor[rgb]{ .863,  .902,  .945} \textbf{U-Pot \cite{HakimGithub:2019}} & \cellcolor[rgb]{ 1,  1,  1}Shodan, Zmap, U-Pot & \cellcolor[rgb]{ 1,  1,  1}UPnP & \cellcolor[rgb]{ 1,  1,  1}N/A & \cellcolor[rgb]{ 1,  1,  1}N/A & \cellcolor[rgb]{ 1,  1,  1}N/A \\\hline
   
    \rowcolor[rgb]{ .863,  .902,  .945} \textbf{HoneyCloud \cite{Dang:2019}} & \cellcolor[rgb]{ 1,  1,  1}VM, Cloud storage, antivirus communities, Honeycomb, VirusTotal website & \cellcolor[rgb]{ 1,  1,  1}SSH, Telnet, SMB, HTTP, HTTPS, RDP, MySQL, SQL Server & \cellcolor[rgb]{ 1,  1,  1}Fileless attacks, malware-based attacks & \cellcolor[rgb]{ 1,  1,  1}Symmetry/asymmetry of data flows, packets analysis, attack types and characteristics, keystrokes, trace of network activities, CPU usage, Process list. & \cellcolor[rgb]{ 1,  1,  1}1 year \\\hline
   
     \rowcolor[rgb]{ .863,  .902,  .945} \textbf{Phype \cite{Phype:2019}} & \cellcolor[rgb]{ 1,  1,  1}Phype Telnet & \cellcolor[rgb]{ 1,  1,  1}Telnet & \cellcolor[rgb]{ 1,  1,  1} N/A & \cellcolor[rgb]{ 1,  1,  1}N/A & \cellcolor[rgb]{ 1,  1,  1}N/A \\\hline
    
    \rowcolor[rgb]{ .863,  .902,  .945} {\textbf{Chameleon \cite{Zhou:2019}}} & {\cellcolor[rgb]{ 1,  1,  1}Nmap, Shodan} & {\cellcolor[rgb]{ 1,  1,  1}N/A} & \cellcolor[rgb]{ 1,  1,  1}Reconnaissance attacks & \cellcolor[rgb]{ 1,  1,  1}IP whitelist, received requests & \cellcolor[rgb]{ 1,  1,  1}N/A \\\hline
    
    \rowcolor[rgb]{ .863,  .902,  .945}{\textbf{Honware \cite{Vetterl:2019}}} & \cellcolor[rgb]{ 1,  1,  1}QEMU, Binwalk, Wireshark, Ping, Nmap, Firmadyne, Shodan & \cellcolor[rgb]{ 1,  1,  1}SSH, Telnet, HTTP, UPnP, DHCP, DNS, dec-notes, freciv, netbios, HTTPS, MDNS, TFTP & \cellcolor[rgb]{ 1,  1,  1}Reconnaissance attacks, Zero days, capture attacks traffic & \cellcolor[rgb]{ 1,  1,  1}Kernel logs, firmwares, & {\cellcolor[rgb]{ 1,  1,  1}$<$ 2 months} \\\hline
    
    \rowcolor[rgb]{ .863,  .902,  .945} {\textbf{Oza et al. \cite{Oza:2019}}} & {\cellcolor[rgb]{ 1,  1,  1}OAuth2, MySQL, QEMU} & {\cellcolor[rgb]{ 1,  1,  1}N/A} & \cellcolor[rgb]{ 1,  1,  1}Man in the Middle attacks & \cellcolor[rgb]{ 1,  1,  1}MAC address, Unauthorized access  & \cellcolor[rgb]{ 1,  1,  1}N/A \\\hline
   
    \rowcolor[rgb]{ .863,  .902,  .945} \textbf{Lingenfelter \cite{Lingenfelter:2020}} & \cellcolor[rgb]{ 1,  1,  1}Filebeat, ELK stack, Logstash, VirusTotal & \cellcolor[rgb]{ 1,  1, 1}SSH, Telnet & \cellcolor[rgb]{ 1,  1, 1}IoT botnet malware & \cellcolor[rgb]{ 1,  1,  1}Packets per port, System logs, IPs, Brute-force scan, file hash  & \cellcolor[rgb]{ 1,  1,  1}40 days\\\hline

\end{longtable}%
\clearpage
\twocolumn

\subsection{The Most Common Attacks}\revision{
The most commonly detected/tested attacks in IoT honeypots/honeynets are Telnet, SSH, DoS/DDoS, and HTTP(S) attacks. In addition, reconnaissance attacks, brute-force attacks, malware, and Mirai attacks were also detected/tested in the proposed honeypots and honeynets. Although less common than the mentioned attacks, botnet, Man-in-the-Middle, malicious cryptocurrency mining, and buffer overflow attacks were also detected/tested in the proposed systems.
}

\section{HONEYPOTS AND HONEYNETS FOR IIOT AND CPS}\label{sec:cpshoneypots}
In this section, we give brief overview of honeypots and honeynets proposed for IIoT and CPS applications. We group the IIoT and CPS honeypots and honeynets based on the application types as follows: ICS, Smart Grid, Water Systems, Gas Pipeline, Building Automation Systems, and IIoT.

\subsection{Honeypots and Honeynets for Industrial Control Systems}\label{subsec:ics-honeypots}

\revisionB{In this subsection, we give brief overview of honeypots and honeynets ICS. Table~\ref{tab:ics-general} provides a list of the considered general ICS honeypots. }

\textit{CISCO SCADA HoneyNet Project: }
The first honeynet for SCADA ICS was proposed by Pothamsetty and Franz in Cisco Systems' SCADA HoneyNet Project~\cite{CISCO-SCADA-Honeynet} in 2004. SCADA HoneyNet is based on the Honeyd~\cite{Provos:2007} open-source honeypot framework and is a low-interaction honeynet that supports the simulation of Modbus/TCP, FTP, Telnet, and HTTP services running on a \revision{programmable logic controller} (PLC).

\textit{Digital Bond SCADA Honeynet: }
The second honeynet for SCADA ICS was introduced by Digital Bond in 2006 under the name of SCADA Honeynet~\cite{DB-SCADA-Honeynet, DB-SCADA-Honeynet2}. It consists of two virtual machines: one of them simulates a PLC with Modbus/TCP, FTP, Telnet, HTTP, and SNMP services while the other one is a Generation III Honeywall. The Honeywall is a modified version of SCADA HoneyNet~\cite{CISCO-SCADA-Honeynet} that aims to monitor and control the honeypot's traffic and attacker interactions.

Wade~\cite{Wade:2011} used Digital Bond's SCADA honeynet in her thesis to analyze the attractiveness of honeypots in ICS systems. Her honeypot simulated a Schneider Modicon PLC with Modbus TCP, FTP, Telnet, and SNMP services.

\textit{Conpot and Conpot-based ICS Honeypots: }
One of the most popular ICS honeypots that has been used by researchers is Conpot~\cite{Conpot}. It is an open-source low-interaction honeypot that was developed under the Honeynet Project~\cite{HoneynetProject} and is still being maintained. Conpot supports various industrial protocols including IEC 60870-5-104, \revision{Building Automation and Control Network} (BACnet), EtherNet/IP, Guardian AST, Kamstrup, Modbus, S7comm, and other protocols such as HTTP, FTP, SNMP, \revision{Intelligent Platform Management Interface} (IPMI), and TFTP. It provides templates for Siemens S7 class PLCs, Guardian AST tank monitoring systems, and Kamstrup 382 smart meters.

\begin{table}[!t]
\caption{List of General ICS Honeypots.}
\label{tab:ics-general}
\begin{tabular}{>{}p{2.1cm}p{1cm}p{4.4cm}}
\rowcolor[HTML]{3465A4} 
{\color[HTML]{FFFFFF} \textbf{Honeypots}} &
  {\color[HTML]{FFFFFF} \textbf{Interaction Level}} &
  {\color[HTML]{FFFFFF} \textbf{Simulated Services}} \\
CISCO~\cite{CISCO-SCADA-Honeynet} & Low    & Modbus/TCP, Telnet, HTTP, FTP                          \\\hline
Digital Bond~\cite{DB-SCADA-Honeynet} &
  Low &
  Modbus/TCP, Telnet, HTTP, FTP, SNMP \\\hline
  
Conpot~\cite{Conpot} &
  Low & IEC 60870-5-104, BACnet, EtherNet/IP, Guardian AST, Kamstrup, Modbus, S7comm, HTTP, FTP, SNMP, IPMI, TFTP \\\hline
Zhao and Qin~\cite{Zhao:2017}                   & Medium & S7comm, Modbus, SNMP, HTTP                             \\\hline
DiPot~\cite{Cao:2018} &
  Low &  HTTP, Modbus, Kamstrup, SNMP, IMPI, BACnet, Guardian AST, S7comm \\\hline
XPOT~\cite{Lau:2016}            & Medium & S7comm, SNMP                                           \\\hline
HosTaGe~\cite{Vasilomanolakis:2016} &  Low &  Modbus, S7comm, HTTPS, FTP, MySQL, SIP,SSH, SNMP, HTTP, Telnet, SMB, and SMT \\\hline

S7CommTrace~\cite{Xiao:2018}                    & Medium & S7comm                                                 \\\hline
Honeyd+ ~\cite{Winn:2015}                       & High   & EtherNet/IP, HTTP                                      \\\hline

Gallenstein~\cite{Gallenstein:2017}             & Low    & EtherNet/IP. ISO-TSAP, HTTP                            \\\hline

Abe et al.~\cite{ABE:2018}                      & Low    & Modbus, S7comm, BACNet, IPMI, Guardian AST, HTTP, SNMP \\\hline
\end{tabular}
\end{table}

Jicha et al.~\cite{Jicha:2016} deployed Conpot honeypots at six different locations around the world via Amazon Web Services platform. The authors configured and deployed two Conpot instances at every location, one with the default configuration and the other one with gas tank level SCADA Conpot. Honeypots ran for 15 days and they analyzed the behavior of simulated protocols against Nmap scanning tool and Shodan search engine scan data. They realized that the ports identified by Shodan and Nmap may differ.

Zhao and Qin~\cite{Zhao:2017} improved Conpot honeypots with additional Siemens S7comm protocol functions and sub-functions support and a dynamic Human Machine Interface (HMI) for evaluation of threats to ICS environments. The authors state that their study improved the interaction level of Conpot and provided better support for the simulation of Siemens S7 class PLCs. Their 43-day long deployment received traffic from 244 valid IP addresses from 34 different countries.

Cao et al. proposed a distributed ICS honeypot called DiPot~\cite{Cao:2018}. DiPot is based on Conpot honeypot framework. It enhances Conpot framework by adding higher-fidelity ICS protocol simulations, data capture and analysis with K-means clustering, and visualization and statistics support. The authors indicated that deployed DiPot honeypots in cloud virtual machines around the world successfully deceived Shodan search engine and were recognized as real ICS devices. 

Lu et al.~\cite{Lu2019} deployed Conpot on a Raspberry Pi to simulate Siemens S7 class PLCs. In addition, they used an Arduino board to simulate a PLC and another Arduino for sensor simulations. However, they did not give much detail about PLC simulation on Arduino and the supported industrial protocols. The authors did not perform any deployments or tests against the proposed honeypot architecture.

\revision {Ferretti et al.~\cite{Ferretti:2019} aimed to analyze the scanning traffic on the Internet that is targeting ICS. To analyze the scanners and their behaviors, the authors deployed several low interaction Conpot honeypots. 
Each Conpot honeypot was configured to simulate a specific ICS device with a specific communication protocol (i.e., S7comm, Modbus/TCP, IEC-61850-104, EtherNet/IP, BACnet, HTTP, FTP, and SSH). Their analysis, which covered four months of operation, showed that the majority of the scanners were legitimate (e.g., Shodan, Censys, etc.) and showed certain scanning patterns. The authors pointed out that the usage of legitimate scanner patterns could give a clue in detecting malicious scanning and attack activities targeting ICS environments.}

\textit{CamouflageNet: }
\revision{Naruoka et al.~\cite{Naruoka:2015} proposed CamouflageNet, which is a honeypot system for ICS environments. CamouflageNet creates a set of Honeyd honeypots in the ICS network with the same fingerprints (i.e., services, ports. vendors) and changes the IP addresses of the devices in the network dynamically when an intrusion attempt is caught by one of the honeypots. 
In terms of ICS protocols and devices, CamouflageNet does not emulate any ICS-specific protocols or simulate any ICS device.}

\textit{XPOT: }
Lau et al.~\cite{Lau:2016} proposed an ongoing study of a medium-interaction honeypot for ICS, namely XPOT. XPOT simulates Siemens S7-300 series PLCs, and allows the attacker to compile, interpret and load PLC program onto XPOT. It supports S7comm and SNMP protocols.

\textit{HosTaGe: } \revision{
Vasilomanolakis et al.~\cite{Vasilomanolakis:2016} proposed HosTaGe ICS honeypot which is an ICS protocols-extended version of their earlier mobile honeypot~\cite{Vasilomanolakis:2013}. The proposed honeypot system consists of three parts: protocol emulation, multi-stage attack detection, and signature generation for IDS. Protocol emulation supports several protocols  
(e.g., Modbus, S7comm, HTTP(S), FTP, SIP, SSH, SNMP, etc.).  
The attack detection module employs an Extended Finite State Machine model to detect multiple-stage attacks that consist of attacks applied by the same source in a serial manner within a specified time window. HosTaGe ICS honeypot 
can generate signatures that open-source Bro IDS can use. In addition, the honeypot system uploads attacker-injected files to VirusTotal to 
determine if they are malicious or not. The authors deployed HosTaGe along with a Conpot honeypot. 
}

\textit{S7CommTrace: }\revision{
S7CommTrace~\cite{Xiao:2018} is a honeypot for ICS that uses Siemens S7comm protocol. It consists of TCP communication, S7 communications protocol, user template, and data storage modules. 
Compared to Conpot, S7CommTrace supports more S7comm functions and sub-fuctions. 
Both S7CommTrace and Conpot instances were deployed to four different locations around the world on cloud environments and analyzed 
for 60 days. The analysis showed that S7CommTrace was able to receive more connections and provide more attack data compared to Conpot. 
In addition, while all Conpot instances were fingerprinted 
by Shodan, 
S7CommTrace instances were not fingerprinted. 
}

\textit{ICS Honeypots based on Honeyd: }
Disso et al.~\cite{Disso:2013} researched the security of SCADA systems from the honeypots' point of view. They created a testbed which consists of a real PLC device as a high-interaction honeypot and a Honeyd-based low-interaction honeypot. They placed Honeynet Project's Roo honeywall~\cite{honeywallcdrom} in front of the honeypots. They conducted latency, network traffic counter and background traffic level analysis (i.e., anti-honeypot techniques) to compare the high and low interaction honeypots.

Winn et al.~\cite{Winn:2015} proposed Honeyd+, which aims to construct several high-interaction ICS honeypots using a proxying technique with a single physical PLC device. Honeyd+ uses Honeyd honeypots with templates that mimic ICS PLCs. The performance analysis of the authors indicated that Honeyd+ is able to simulate 75 ICS honeypots using a single PLC device with a Rapsberry Pi board. However, the analysis showed that Honeyd+ starts to see serious performance drops starting from five simultaneous connections from attackers.

Gallenstein conducted research on the automated creation and configuration of ICS PLC honeypots that would emulate PLCs from different vendors with minimum effort. In his thesis~\cite{Gallenstein:2017}, he integrated Honeyd with ScriptGenE framework proposed by Warner~\cite{Warner:2015}. ScriptGenE is an extended version of ScriptGen which is automated protocol replay framework proposed by Leita et al.~\cite{Leita:2005}. Gallenstein emulated a prison ICS environment that had three PLCs from Allen-Bradley and Siemens. He tested the legitimacy of his honeypots with Shodan Honeyscore, Nmap and vendor tools (i.e., RSLinkx and STEP7).

\revision{Abe et al.~\cite{ABE:2018} proposed an ICS honeypot system 
that employs Honeyd and Conpot 
frameworks and adds a traceback capability to gain more information about 
attackers. The proposed system is able to emulate the ICS protocols and devices by means of Conpot framework, and performs basic honeypot functions by means of Honeyd. The authors implemented Nmap in the Honeyd to perform a reverse scan to the attackers and obtain useful information regarding the attack. 
}

\textit{ICS Honeypots based on Network Simulators/Emulators: }

\revisionB{In this category, we give brief overview of honeypots and honeynets for ICS that are based on network simulators/emulators. Table~\ref{tab:ics-emulators} provides a list of the ICS honeypots that are based on network emulators/simulators. }

\begin{table}[htbp]
\caption{List of ICS Honeypots Based on Network Simulators/Emulators.}
\label{tab:ics-emulators}
\begin{tabular}{>{}p{1.9cm}p{1.8cm}p{3.8cm}}
\rowcolor[HTML]{3465A4} 
{\color[HTML]{FFFFFF} \textbf{Honeypots}} &
  {\color[HTML]{FFFFFF} \textbf{Interaction Level}} &
  {\color[HTML]{FFFFFF} \textbf{Simulator/Emulator}} \\
Haney et al.~\cite{Haney:2014}      & High   & IMUNES Simulator, JAMOD Library                    \\\hline
Kuman et al.~\cite{Kuman:2017}                  & Low    & IMUNES Simulator                                             \\\hline
Ding et al.~\cite{Ding:2018}                    & Medium & IMUNES Simulator                                           \\\hline
\end{tabular}
\end{table}

\revision{Haney and Mauricio proposed a SCADA honeynet framework~\cite{Haney:2014} 
based on the IMUNES Network Simulator~\cite{imunes}. The proposed honeynet employs a honeywall to control the activities of attackers and IMUNES-based honeypots that simulate PLCs and RTUs in SCADA systems. The authors used Java Modbus library (JAMOD) to simulate Modbus/TCP and developed Java code for PLC state variables, control logic, and process to be controlled by the simulated PLC. They also utilized Honeyd to provide Telnet, SSH, and HTTP(S) services for the virtual PLC. 
}

Kuman et al.~\cite{Kuman:2017} utilized Conpot honeypot on top of IMUNES network emulator to analyze attacks targeting ICS networks. They modified IMUNES emulated nodes so that they were able to install Conpot instances simulating Siemens S7-300 PLCs. They also employed OSSEC host-based IDS to monitor the activities with the Conpot instances. Two weeks long experiments showed that Siemens S7-300 PLC honeypots received port scanning activities to Modbus and web server ports.

Ding et al.~\cite{Ding:2018} proposed an ICS honeypot that used and modified Siemens S7comm protocol on top of IMUNES network emulator. The authors paid attention to configure SNAP7 tool to provide a fingerprint like real Siemens PLC devices. They also added SNMP service support. IMUNES gives the proposed honeypot chance to use lightweight Docker container technology to quickly start up. The authors used Nmap and PLCscan tools to verify the fingerprinting results of the proposed honeypot.

\textit{ Honeypots and Honeynets Using Real ICS Devices: }

\revisionB{In this category, we give brief overview of honeypots and honeynets that use real ICS devices. Table~\ref{tab:ics-real} provides a list of the ICS honeypots that are based on network emulators/simulators. }

\begin{table}[htbp]
\caption{List of ICS Honeypots that Use Real Devices.}
\label{tab:ics-real}
\begin{tabular}{>{}p{1.7cm}p{1cm}p{4.9cm}}
\rowcolor[HTML]{3465A4} 
{\color[HTML]{FFFFFF} \textbf{Honeypots}} &
  {\color[HTML]{FFFFFF} \textbf{Interaction Level}} &
  {\color[HTML]{FFFFFF} \textbf{Used Device(s)}} \\

Bodenheim~\cite{Bodenheim:2014}                 & High   & Allen-Bradley PLCs                                \\\hline
Piggin et al.~\cite{Piggin:2016}            & High   & Not Specified                                         \\\hline
Haney~\cite{Haney:2019}                         & Low    & Direct Logic PLC                      \\\hline
Hilt et al.~\cite{Hilt:2020}                    & High   &  Siemens  S7-1200,  Allen-BradleyMicroLogix 1100, Omron CP1L\\\hline
\end{tabular}
\end{table}

\revision{
Bodenheim deployed four high-interaction ICS honeypots to an integrator's site in his thesis~\cite{Bodenheim:2014}. He aimed 
to analyze the impact of Shodan search engine on identification of the Internet-connected ICS devices and 
understand if being indexed by Shodan increases the number of attacks. He used real PLCs of Allen-Bradley. He configured two honeypots with standard settings, one honeypot with obfuscated banner information and the last one with a banner that advertised itself clearly as an Allen-Bradley PLC. A 55 day deployment period showed no evidence that Shodan's indexing increased the number of attacks.
}

\revision{Piggin and Buffey~\cite{Piggin:2016} presented a high-interaction 
honeypot 
for ICS production environments. 
They used real PLC hardware 
to obtain high fidelity data and prevent being detected as a honeypot. They also implemented a process simulation for the PLC hardware. 
However, they did not give 
details in regards to the PLC used, 
supported 
protocols, and the process simulation. 
They also implemented a Situational Awareness and Forensics platform for honeypot traffic analysis. 
Their results 
show that a majority 
of the attacks were scanning activities. In addition, they realized some targeted attacks, including a password attack with default credentials, a dictionary attack, an SSH brute-force attack, an attack to the industrial protocols of the PLC, and a knowledgeable attack originating from the 
TOR network.}

Haney proposed a hybrid high interaction honeynet framework for ICS in~\cite{Haney:2019}. The proposed honeynet consists of a honeywall system, a SCADA HMI on a virtual machine, physical PLC devices, virtual nodes emulating PLC devices and services (e.g., HTTP, SNMP, SSH, Telnet), and a physical process simulation. Haney paid attention to not only the requirements of Gen III honeynets (i.e., data control, data capture and information sharing), but also other requirements of realism, scalability, and detection resistance.

\revision{Hilt et al.~\cite{Hilt:2020} 
constructed the most realistic ICS honeypot to date. They set up a smart-factory honeypot using four real PLC devices (i.e., Siemens S7-1200, Allen-Bradley MicroLogix 1100, and Omron CP1L) with corresponding role implementations (i.e., agitator, burner control, belt control, palletizer). They implemented HMIs for each PLC on virtual machines. In addition to PLCs and HMIs, they also accommodated robotics and engineering workstations and installed corresponding software to them. Moreover, they set up a file server and placed fake files on it. In order to monitor the honeypots, they used Ethernet taps connected to a Raspberry Pi to collect monitoring information. Their honeypot system opened Siemens S7comm, Omron FINS, EtherNet/IP, and VNC protocols and services to the Internet. To convince the attackers that they are not a honeypot, they created a fake company profile with employees, Artificial Intelligence (AI)-generated pictures, profiles, website, e-mail addresses, and phone numbers. They tried to attract the attackers using messages posted on Pastebin. Analysis of honeypot deployments 
for seven months shows that the honeypots received scanning activities from unknown sources as well as from legitimate scanners. They were attacked by three ransomware attacks, one malicious cryptomining operation 
and robotic workstation beaconing attempt. However, the PLC honeypots did not receive any targeted attacks but only scanning activities. 
}

\textit{ Other ICS Honeypots and Honeynets: }

\revisionB{In this category, we give brief overview of the other ICS honeypots and honeynets. Table~\ref{tab:ics-other} provides a list of rest of the ICS honeypots. }

\begin{table}[htbp]
\caption{Other ICS honeypots.}
\label{tab:ics-other}
\begin{tabular}{>{}p{2.1cm}p{1cm}p{4.4cm}}
\rowcolor[HTML]{3465A4} 
{\color[HTML]{FFFFFF} \textbf{Honeypots}} &
  {\color[HTML]{FFFFFF} \textbf{Interaction Level}} &
  {\color[HTML]{FFFFFF} \textbf{Simulated Services}} \\
Berman~\cite{Berman:2012}                       & Low    & Modbus/TCP                                             \\\hline
Jaromin~\cite{Jaromin:2013}                     & Low    & Modbus/TCP, HAP, HTTP                                  \\\hline
Holczer et al.~\cite{Holczer:2015}              & High   & S7comm, SNMP, HTTP(S)                                  \\\hline
Serbanescu \newline et al.~\cite{SerbanescuHoneynet:2015} &
  Low &
  DNP3, IEC-104, Modbus, ICCP, SNMP, TFTP, XMPP \\\hline
Simões et al.~\cite{Simoes:2015}                & Low    & Modbus, SNMP, FTP                                      \\\hline
Ahn et al.~\cite{Ahn:2019}                      & Low    & Modbus                                                 \\\hline
Belqruch et al.~\cite{Belqruch:2019}         & Medium & SSH  \\\hline                                                 
\end{tabular}
\end{table}
Berman~\cite{Berman:2012} designed and implemented a PLC emulator on a Gumstix board that could be used as an ICS honeypot. The proposed emulator was developed in Python using Scapy library and acts like an Allen-Bradley PLC running Modbus/TCP protocol. Berman evaluated the performance of the emulator via fingerprinting and standard traffic response tests.

\revision{Jaromin proposed an ICS honeypot~\cite{Jaromin:2013} that emulated Koyo DirectLogic 405 PLC devices. His honeypot used Berman's Modbus/TCP emulator~\cite{Berman:2012} and added Host Automation Products (HAP) protocol emulation and a web server implementation. He performed several tests 
to compare the Gumstix deployment, laptop Personal Computer (PC) deployment, and the 
real Koyo DirectLogic PLC device. }

\revision{Holczer et al.~\cite{Holczer:2015} proposed a high interaction PLC honeypot for ICS 
which simulates the main functionalities of Siemens ET 200S PLCs. It can support Siemens STEP7 PLC management service, HTTP(S) and SNMP services. 
The authors deployed the honeypot within the IP range of their university and realized that they did not receive much traffic from the attackers.}

\revision{Serbanescu et al.~\cite{SerbanescuHoneynet:2015} proposed a low-interaction honeynet architecture for ICS environments. Their work 
extends their earlier 
honeypot~\cite{Serbanescu:2015} 
with ICS protocols and creates a honeynet. Deployment of the honeynet was made on the Amazon EC2 platform 
with six honeypots supporting combinations of different protocols (i.e.,  Distributed Network Protocol 3 (DNP3), IEC-104, Modbus, ICCP, SNMP, TFTP, XMPP). 
Analysis of the collected data for 28 days showed that 
the 
honeypots received 
only the reconnaissance activities. In addition, they indicated that the popularity of the ICS protocols based on the received traffic is as follows: Modbus, ICCP, DNP3, and IEC-104 in descending order. They drew two important inferences: i) Attackers are using the Shodan results to determine which ICS systems to attack, ii) Attackers send non-ICS-specific traffic to the standard ports of ICS protocols.}

\revision{Simões et al.~\cite{Simoes:2015} proposed a SCADA honeypot architecture for ICS 
which is extremely similar to the honeypot architecture proposed in their previous work~\cite{Simoes:2013}. Their architecture 
simulates ICS devices and supports both ICS-specific protocols and other protocols (i.e., SNMP, FTP). It also 
employs a port scan module to detect the reconnaissance activities, a firewall to prevent the honeypot from being used as an attack tool, an event monitoring module to observe the traffic and attacker interactions, 
and a management and watchdog module 
for configuration of the honeypot. As a Proof-of-Concept (PoC), 
they implemented two low-interaction Modbus honeypots on a Rapsberry Pi, one placed 
in an ICS network, and the second placed in a 
datacenter.}

Ahn et al.~\cite{Ahn:2019} proposed a security architecture for SCADA ICS systems that uses a low interaction honeypot to detect a possible intruder and performs ARP poisoning attack to poison the ARP table of the attacker later on. The authors did not provide any implementation details or any performance evaluation results.

Belqruch and Maach \cite{Belqruch:2019} implemented a Kippo~\cite{Kippo:2016} honeypot for brute-force SSH attacks on an SCADA ICS. They simulated attacks via Kali Linux tools.

\subsection{Honeypots and Honeynets for Smart Grid}\label{subsec:smart-grid-honeypots}

\revisionB{In this category, we give brief overview of the smart grid honeypots and honeynets. Table~\ref{tab:smartGrid} provides a list of the smart grid honeypots. }

\textit{CryPLH: }
Buza et al.~\cite{Buza:2014} proposed CryPLH, a low interaction and a virtual Smart-Grid ICS honeypot simulating Siemens Simatic 300 PLC devices. CryPLH uses NGINX and miniweb web servers to simulate HTTP(S), a Python script to simulate Step 7 ISO-TSAP protocol and a custom SNMP implementation. The authors deployed the honeypot within the university's IP range and observed scanning, pinging, and SSH login attempts.

\textit{SHaPe: }
Kołtyś and Gajewski proposed a low-interaction honeypot, namely  SHaPe~\cite{Koltys2015}, for electric power substations. SHaPe is capable of emulating any IEDs in an electric power substation that is compliant with IEC 61850 standard. The proposed honeypot extended the general purpose open-source Dionaea honeypot by means of libiec61850 library.

\begin{table}[!t]
\caption{List of Smart Grid Honeypots and Honeynets.}
\label{tab:smartGrid}
\begin{tabular}{>{}p{2.1cm}p{1cm}p{4.5cm}}
\rowcolor[HTML]{3465A4} 
{\color[HTML]{FFFFFF} \textbf{Honeypots}} &
  {\color[HTML]{FFFFFF} \textbf{Interaction Level}} &
  {\color[HTML]{FFFFFF} \textbf{Simulated Services}} \\\hline
CryPLH~\cite{Buza:2014}         & Low    & HTTP(S), SNMP, Step7 ISO-TSAP                             \\\hline
SHaPe~\cite{Koltys2015}                  & Low    & IEC 61850 MMS, HTTP, FTP, SMB                             \\\hline
GridPot~\cite{Redwood:2015}              & Hybrid & IEC 61850 GOOSE/MMS, Modbus, HTTP                         \\\hline
Scott~\cite{Scott:2014}                  & Low    & Modbus/TCP, HTTP, SNMP                                    \\\hline
Mashima et al.~\cite{Mashima:2017}       & Low    & IEC 60870-5-104, IEC 61850, SSH                           \\\hline
Pliatsios et al.~\cite{Pliatsios:2019}   & Low    & Modbus/TCP                                                \\\hline
Mashima et al.~\cite{Mashima:2019} &
  Low &
  TCP port listener on IEC 61850 MMS, S7comm, Modbus/TCP, Niagara Fox, EtherNet/IP, IEC 60870-5-104, DNP3, BACnet\\\hline
\end{tabular}
\end{table}

\textit{GridPot: }
\revision{Redwood et al.~\cite{Redwood:2015} proposed a symbolic honeynet framework, namely SCyPH, for SCADA systems. 
The proposed framework aims to incorporate emulated SCADA system components with physics simulations and employ anomaly detection systems based on the changes on the data obtained from the physics simulation. In their demonstration, namely GridPot, 
the authors utilized GridLab-D simulator~\cite{GridLab} 
for electric substation simulations and IEC 61850-based communication, and implemented Newton-Raphson power flow solver algorithm for the voltage and current flow between the actors. They utilized Conpot to emulate IEDs and also implemented GOOSE/MMS and Modbus protocols for the interactions between the devices.} 

Kendrick and Rucker~\cite{Kendrick:2019} deployed GridPot in their thesis to analyze the threats to smart energy grids. Their honeypot deployment emulated Modbus TCP, S7comm, HTTP, and SNMP services. Although Shodan Honeyscore detected their deployment as a honeypot, a 19-day period of data collection showed that, GridPot received heavy HTTP scanning activities, over 600 Modbus, and 102 S7comm connections.

Scott~\cite{Scott:2014} implemented a SCADA honeypot that uses the open-source Conpot honeypot to simulate a Scheider Electric PowerLogic ION6200 smart meter. They deployed the honeypot in a facilities network beside other SCADA components. They configured the honeypot to send its logs to a logging server, which alerts the network administrators based on the severity of the interactions that attackers are performing. Their honeypot supports Modbus, HTTP (for HMI), and SNMP.

\revision{Mashima et al.~\cite{Mashima:2017} proposed a scalable high-fidelity honeynet system for electrical substations in smart-grid environments. The proposed honeynet consists of a virtual substation gateway that supports the standardised smart-grid communication protocols (i.e., IEC 60870-5-104 and IEC 61850) and opens the entry point to the external attackers; virtual IEDs that are represented by Mininet~\cite{mininet} virtual hosts and SoftGrid~\cite{softgrid} IED simulations; and simulation of smart grid components (e.g., circuit breakers, transformers, etc.) via POWERWORLD~\cite{powerworld} power simulator. The proposed honeynet is highly scalable and resistant to fingerprinting against Shodan and attacker tools such as Nmap. 
}

Hyun~\cite{Hyun:2018} used Conpot honeypot to discover the compromise attempt indicators for ICS environments. She configured Conpot to simulate a Siemens S7-200 PLC in an electric power plant. The simulated instance supported HTTP, Modbus/TCP, S7comm, SNMP, BACnet, IMPI, and EtherNet/IP services. The deployment of the honeypot outside of the university's network for four months revealed that popular choices for compromise attempts were HTTP, Modbus, and S7comm services.

Pliatsios et al.~\cite{Pliatsios:2019} proposed a honeypot system for Smart-Grid which is based on the Conpot honeypot framework. The proposed honeypot consists of real \revision{Human-Machine Interface} HMI and real \revision{Remote Terminal Unit} RTU devices, and two virtual machines, one for virtual HMI and the other for a Conpot-based honeypot emulating an RTU device. The Conpot honeypot uses the real traffic generated by the real RTU device in order to make the attackers believe that they are interacting with a real ICS device. 

\revision{Mashima et al.~\cite{Mashima:2019} deployed low interaction smart-grid honeypots in five geographic regions 
via 
Amazon cloud platform and analyzed the traffic coming to the honeypots for six months. 
They did not use open-source honeypot frameworks in order to avoid fingerprinting by attackers. Instead, they set up TCP listeners on several ports for ICS protocols. 
They realized that their honeypot instances received SYN-flooding DoS attack on IEC 61850 and S7comm protocols' port and also scanning activity for DNP3 and Modbus/TCP protocols. Their analysis showed that the same group of attackers, using the same IP addresses, was targeting smart grid devices on their honeypot instances around the world and sometimes an attack targeting a specific honeypot instance was applied to another instance the following week.}

\subsection{Honeypots and Honeynets for Water Systems}\label{subsec:water-honeypots}

\revisionB{In this category, we give brief overview of the  honeypots and honeynets for water systems. Table~\ref{tab:waterSystems} provides a list of the water system honeypots. }
\begin{table}[!b]
\caption{List of ICS Honeypots for Water Systems.}
\label{tab:waterSystems}
\begin{tabular}{>{}p{2.3cm}p{1cm}p{4.2cm}}
\rowcolor[HTML]{3465A4} 
{\color[HTML]{FFFFFF} \textbf{Honeypots}} & {\color[HTML]{FFFFFF} \textbf{Interaction Level}} & {\color[HTML]{FFFFFF} \textbf{Simulated Services}} \\\hline
Wilhoit~\cite{Wilhoit:2013-1} & Hybrid & Modbus/TCP, HTTP, FTP         \\\hline
Antonioli et al.~\cite{Antonioli:2016}       & High   & EtherNet/IP, SSH, Telnet, VPN \\\hline
Murillo et al.~\cite{Murillo:2018}           & Low    & EtherNet/IP                   \\\hline
Petre and Korodi~\cite{Petre:2019}           & Medium & Modbus                        \\\hline
MimePot~\cite{Bernieri:2019}         & High   & Modbus/TCP  \\\hline                 
\end{tabular}
\end{table}

Wilhoit~\cite{Wilhoit:2013-1,Wilhoit:2013-2} deployed high and low interaction honeypots to understand the sources and motivations of attacks targeting ICS environments. His honeypot system mimicked a water pressure station. For high interaction honeypots, he used Nano-10 PLC and Siemens Simatic PLC. As low interaction honeypots, he created virtual HMI instances which look like controlling PLCs of an ICS. The low-interaction honeypots were deployed on Amazon EC2 cloud environments around the world.

\revision{Antonioli et al.~\cite{Antonioli:2016} proposed a virtual high interaction 
honeypot for ICS 
that is based on the MiniCPS ICS simulation framework~\cite{MiniCPS}. The proposed design separates the honeypot system from the real ICS, and places virtual VPN, Telnet and SSH servers as the entry points for attackers to the honeypot. Network, ICS devices and physical process simulations/emulations are performed utilizing MiniCPS framework. In addition, the authors considered to manage the bandwidth, delay, and packet loss of the emulated links in the honeypot via Tc program, and enabled EtherNet/IP communication via cpppo Python library. 
As a PoC
, the authors implemented a water treatment ICS.}

\revision{A virtual testbed environment for ICS which can be used to deploy ICS honeypots was proposed by Murillo et al.~\cite{Murillo:2018}. The presented virtual testbed environment which uses MiniCPS~\cite{MiniCPS} pays attention to realistic mathematical modeling of the ICS plants and the response time of the simulated ICS devices. 
The authors added a nonlinear plant model to MiniCPS to create a realistic ICS plant. An emulated network of a nonlinear control system which represents three water tanks, sensors, actuators and PLC devices was developed. In addition, the authors simulated a bias injection attack on the control system and proposed a mitigation mechanism.
}

Petre and Korodi~\cite{Petre:2019} proposed a solution for protecting water pumping stations from threats using a honeypot inside an Object Linking and Embedding Process Control (OPC) Unified Architecture (UA) wrapping structure. OPC UA~\cite{OPCUA} is a middleware that can be used to interface standard ICS protocols (e.g., Modbus) to Service Oriented Architecture (SOA) systems and web services. The proposed honeypot uses Node-RED library to simulate a system consisting of two water pumps and two water tanks and runs in an OPC UA Wrapper. 

Bernieri et al.~\cite{Bernieri:2019} presented a model-based ICS honeypot, namely MimePot, that utilizes \revision{Software Defined Network} SDN for traffic redirection and network address camouflage for the real devices. The proposed honeypot simulates the ICS components and control routines based on the Linear Time Invariant model. The authors provided a water distribution PoC implementation which used a simulated attacker that injects and modifies the communication between honeypot elements.

\subsection{Honeypots and Honeynets for Gas Pipelines}\label{subsec:gas-honeypots}

Wilhoit and Hilt developed a low-interaction virtual honeypot, namely GasPot~\cite{Wilhoit:2015}, for gas-tank-monitoring systems. Their honeypot represented a virtual Guardian AST gas-tank-monitoring system. Based on their deployments with physical IP addresses in seven countries around the world, they realized reconnaissance attempts and DDoS attacks were performed by attackers.

Zamiri-Gourabi et al.~\cite{Zamiri:2019} proposed an enhanced version of GasPot honeypot for ICS. Their upgrade applied patches to GasPot so that it will not be detected as a honeypot on the Internet. They fixed the incomplete command support for ATG protocol, made response times more realistic, and patched the problem of responding with static inventory values and the output formatting issue which can help an attacker to understand that it is a honeypot.

\subsection{Honeypots and Honeynets for Building Automation Systems}\label{subsec:building-automation-honeypots}

\revision{Litchfield et al.~\cite{Litchfield:2016} stated that  high interaction honeypots are unsuitable for CPS due to safety risks, costs, and limitations with the usefulness of the honeypot without the physical part of the CPS. Therefore, they suggested the use of hybrid interaction honeypots 
in which real CPS devices interact with the simulation of the physical part of the CPS, and 
proposed HoneyPhy. HoneyPhy 
consists of three modules: Internet interface(s), process model(s), and device model(s). 
A PoC implementation of HoneyPhy was given where a Heating, Ventilation and Air Conditioning (HVAC) honeypot is constructed by means of a physical SEL-751A relay, a black-box simulation model of a physical relay and a heating and cooling process simulation model. The extendability of the proposed honeypot framework for other CPS applications is limited since 
device and process models for the corresponding CPS application are needed.
}

\subsection{Honeypots and Honeynets for IIoT}\label{subsec:iiot-honeypots}

Ammar and AlSharif~\cite{Ammar:2018} proposed a model called HoneyIo3, composed of three honeynets carried out with three Raspberry Pi devices with Linux OS and Honeeepi sensor, that mimic IIoT/ICS services. Services/Protocols used in HoneyIo3 model are IPMI, S7comm, HTTP, Kamstrup, SNMP and SSH.

\revision{Du and Wang~\cite{Du:2020} focused on DDoS attacks on SDNs in IIoT environments. They identified a new kind of attack that could identify a honeypot being used in an SDN and disable it. 
Analyzing attacker strategies, they presented a pseudo-honeypot game 
strategy to dynamically protect SDNs. The 
evaluation was performed on a testbed using servers and hybrid honeypots, and showed that the proposed strategy can 
protect against 
DDoS attacks. }

\section{TAXONOMY OF HONEYPOTS AND HONEYNETS FOR IIOT AND CPS}\label{sec:taxonomycps}
Honeypots and honeynets proposed for IIoT and CPS are listed in Table~\ref{tab:iiot-cps-honeypots} and the tools, implementation, and attack type details of the corresponding honeypots and honeynets are also outlined in Table~\ref{tab:iiot-cps-honeypots-tools}. In this section, we consider all of the proposals for IIoT and CPS and provide an overview of these studies based on the development of research over time, common characteristics, scalability, simulated services, most commonly used tools, availability of the source codes, \revision{and the most common attacks}.

\begin{figure}[!t]
    \centering
    \includegraphics[scale = 0.19]{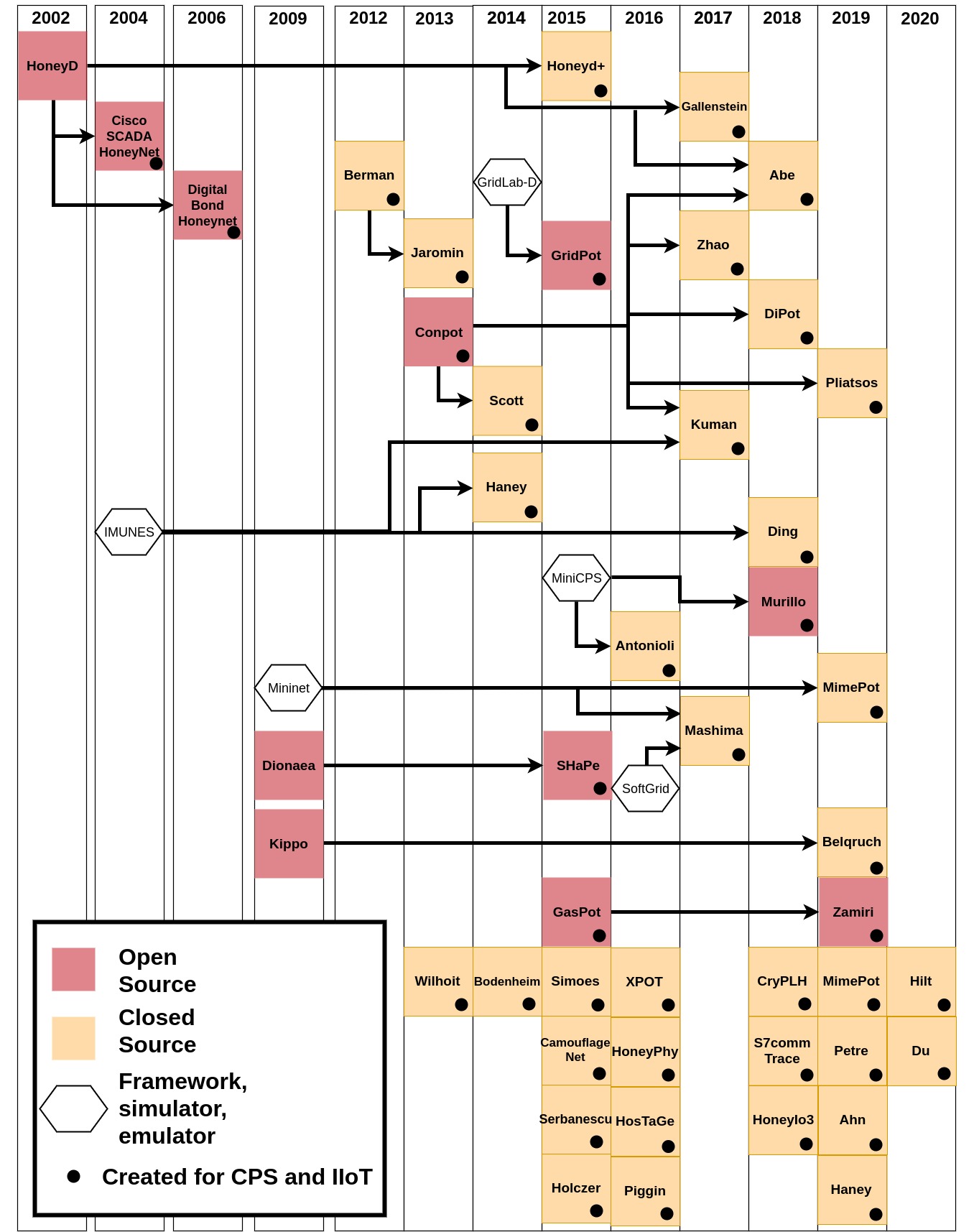}
    \caption{Evolution of Inheritance for the honeypots and honeynets of IIoT and CPS.}
    \label{fig:inheritance-iiot-cps}
\end{figure}

\subsection{Development of research over time}
We analyzed the studies and depicted the development of research over time and also the inheritance relationship between the honeypots and honeynets for IIoT and CPS in Figure~\ref{fig:inheritance-iiot-cps}. As shown in the figure, honeypots and honeynets for IIoT and CPS started with the SCADA HoneyNet~\cite{CISCO-SCADA-Honeynet} project of Pothamsetty and Franz from Cisco Systems in 2004. This project was followed by Digital Bond's SCADA Honeynet~\cite{DB-SCADA-Honeynet} in 2006. In terms of the honeypot and honeynet research for IIoT and CPS systems in the literature, we can see that Berman's thesis~\cite{Berman:2012} in 2012 was the first study. His thesis was followed by another thesis conducted by Jaromin~\cite{Jaromin:2013} the following year. It is interesting to note that both studies were performed in the US Air Force Institute of Technology. This also corresponds to a time in which notorious malware (i.e., Stuxnet (2010), DuQu (2011), Night Dragon (2011) and Flame (2012)) appeared in the wild against nations' critical infrastructure environments, and quickly grabbed the attention of military/defense organisations. In the same year, 2013, the most popular ICS honeypot - Conpot~\cite{Conpot} started and Wilhoit from Trend Micro Research published the white paper of their low interaction ICS honeypots~\cite{Wilhoit:2013-1}. After these works, honeypot and honeynet research and practice in IIoT and CPS gained a momentum.

As shown in Figure~\ref{fig:inheritance-iiot-cps}, more than one-third of works have a form of inheritance relationship with each other, where a honeypot is built based on another. In this respect, Conpot~\cite{Conpot} is the leading honeypot, as six honeypots were developed based on Conpot (this number does not include the studies that do not extend Conpot but only use it). The underlying reasons can be manifold. Conpot is open-source and is still being actively maintained. It supports several industrial and non-industrial protocols. In addition, it is being developed under the umbrella of Honeynet Project~\cite{HoneynetProject}, which has a significant background with honeypots such as Honeyd, Honeywall CDROM, Dionaea and Kippo.

In addition to extending the existing honeypots, researchers also considered to employ simulators, emulators, or frameworks as the main building block for their studies. As Figure~\ref{fig:inheritance-iiot-cps} shows, Mininet and IMUNES emulators, GridLab-D, SoftGrid and POWERWORLD simulators, and MiniCPS framework were utilized in a number of honeypot/honeynet studies.

Apart from extending honeypots or using simulators, emulators and frameworks, we can see that half of the studies proposed independent honeypots. This may be due to the shortcomings of existing honeypots to support CPS and IIoT environments or fingerprintability of them from attackers' point of view.

\subsection{Common characteristics}
Honeypots and honeynets proposed for IIoT and CPS have several characteristics in common. 

In terms of purpose of the honeypots, we can see that the majority of the honeypots and honeynets outlined in Table \ref{tab:iiot-cps-honeypots} and Table \ref{tab:iiot-cps-honeypots-tools} have  research purposes. The only studies which have production purposes are Antonioli et al.~\cite{Antonioli:2016}, Piggin et al.~\cite{Piggin:2016}, and Scott~\cite{Scott:2014}. This is understandable since IIoT and CPS environments have unique features that make it hard for security tools including honeypots to be actively deployed in such areas. Equipments in SCADA environments work continuously, and interruptions and downtimes are highly refrained from~\cite{Simoes:2013, Scott:2014}. In addition to this, industrial devices typically have real-time constraints with guaranteed response times~\cite{Holczer:2015}. For these reasons, it is very difficult to insert a honeypot in an ICS production environment which may affect the ICS communication and has the danger of being compromised (if it is a high-interaction honeypot).

Considering the roles of honeypots, we see that the overwhelming majority of the proposals have server roles. The honeypots and honeynets that have components which act like clients are Haney et al.~\cite{Haney:2019}, Pliatsos et al.~\cite{Pliatsios:2019}, and MimePot~\cite{Bernieri:2019}.

Linux is by far the most popular operating system environment choice of honeypot and honeynet developers. Apart from Linux, we see that only Haney et al.~\cite{Haney:2014} used FreeBSD.

In terms of the programming languages used for the development of honeypots and honeynets for IIoT and CPS, we note that Python is the most favored one. Aside from Python, C/C++ and Java are also used by the studies. We believe that this has a relation with the library support that these languages have for industrial protocols. In this regard, Modbus-tk, pymodbus and cpppo EtherNet/IP libraries of Python; libiec61850 and OpenDNP3 libraries of C/C++ and JAMOD Modbus library of Java are utilized by the developers in the studies. In addition, Conpot - the most popular open-source honeypot for IIoT and CPS is also written in Python. 

\vspace{0.4em}
\subsection{Level of interaction}
Honeypots and honeynets proposed for IIoT and CPS environments exhibit all possible interaction levels. In this respect, as Table~\ref{tab:iiot-cps-honeypots} shows, half of the works allow low interaction capabilities to an attacker. On the other hand, numbers of medium, high and hybrid interaction honeypots are almost equal to each other. We had to make a decision on setting the interaction level for some of the studies since not every author explicitly stated that information in their proposals. Low interaction honeypots in IIoT and CPS systems can provide valuable information in terms of scanning, target protocol, attack origin and brute-force attempts. On the other hand, it is possible to see other more advanced attacks and industrial protocol and process specific attacks only with medium and high interaction honeypots. However, one has to be extremely careful when deploying a high interaction honeypot especially in IIoT and CPS environments since they allow attackers to compromise the system and then apply other operations using the honeypot (e.g., downloading malware, trying to exploit other devices on the same network, performing attacks on behalf of the attacker).

\vspace{0.4em}
\subsection{Resource Level}
In terms of resource levels of honeypots and honeynets for IIoT and CPS, we can see that most of the decoy systems use virtual resources. However, honeypots and honeynets utilizing real industrial devices and a combination of real and virtual devices also exist. One of the biggest driving factor for researchers to propose virtual honeypots may be the high cost of actual industrial devices. As several researchers (\cite{Winn:2015, Gallenstein:2017, Mashima:2017} and \cite{Green:2017}) highlighted, components of an industrial system such as PLCs have high costs in the order of tens of thousands of dollars. 

\vspace{0.4em}
\subsection{Scalability}
The majority of the honeypots and honeynets for IIoT and CPS have scalable designs. This is also related to these honeypots having virtual resources. As we explained in Section~\ref{sec:methodology}, physical honeypots are difficult to scale as they need more physical resources, and real industrial environments can have several industrial devices. For instance, Mashima et al.~\cite{Mashima:2017} noted the number of substations in a power grid in Hong Kong as 200. In order to propose a realistic decoy system, scalable honeypot design gains importance.

\vspace{0.4em}
\subsection{Target IIoT and CPS Application}
As target application areas of the existing honeypots, we can state that more than half of the works targeted ICS environments. However, considerably fewer decoys exist for specific CPS and IIoT applications such as smart grid, water, gas, and building automation systems. Although the majority of the studies are for ICS, we would like to note the fact that the similar industrial devices (e.g., PLCs) can be used both by ICS and smart infrastructures (e.g., grid, water, gas).

\vspace{0.4em}
\subsection{Industrial process simulations}
In terms of industrial process simulations, we see that only five studies considered to employ some form of simulations. For water management CPS environments, Antonioli et al.~\cite{Antonioli:2016} used equation of continuity from hydraulics and Bernoulli's principle for the trajectories (for drain orefice), Murillo et al.~\cite{Murillo:2018} utilized a nonlinear model with Luenberger observer, and Bernieri et al.~\cite{Bernieri:2019} employed linear time invariant model for plant simulation. GridPot~\cite{Redwood:2015} made use of Newton-Raphson power flow solver for electrical grid process. Lastly, for building automation systems, Litchfield et al.~\cite{Litchfield:2016} considered Newton's Law of Cooling for the building process model.

\clearpage
\onecolumn
  \LTcapwidth=\textwidth
  \renewcommand\arraystretch{1}
  \begin{longtable}{|>{\footnotesize}p{2.5cm}|>{\footnotesize}p{0.5cm}|>{\footnotesize}p{1.4cm}|>{\footnotesize}p{1cm}|>{\footnotesize}p{1cm}|>{\footnotesize}p{5.55cm}|>{\footnotesize}p{0.7cm}|>{\footnotesize}p{2cm}|}\caption{\revision{Classification of Honeypots and Honeynets for IIoT and CPS}}\label{tab:iiot-cps-honeypots}\\\hline

    \rowcolor[rgb]{ .141,  .251,  .384} \textcolor[rgb]{ 1,  1,  1}{\textbf{Work}} & {\textcolor[rgb]{ 1,  1,  1}{\textbf{Year}}} & \textcolor[rgb]{ 1,  1,  1}{\textbf{Level of Interaction}} & \textcolor[rgb]{ 1,  1,  1}{\textbf{Scalability}} & \textcolor[rgb]{ 1,  1,  1}{\textbf{Resource level}} & \textcolor[rgb]{ 1,  1,  1}{\textbf{Simulated services}} & \textcolor[rgb]{ 1,  1,  1}{\textbf{Role}} & \textcolor[rgb]{ 1,  1,  1}{\textbf{Application}} \\\endhead

    \rowcolor[rgb]{ .863,  .902,  .945} \textbf{CISCO  \cite{CISCO-SCADA-Honeynet}} & \cellcolor[rgb]{ 1,  1,  1}2004 & \cellcolor[rgb]{ 1,  1,  1}Low & \cellcolor[rgb]{ 1,  1,  1} \checkmark & \cellcolor[rgb]{ 1,  1,  1}Virtual & \cellcolor[rgb]{ 1,  1,  1}Modbus/TCP, Telnet, HTTP, FTP & \cellcolor[rgb]{ 1,  1,  1}Server & \cellcolor[rgb]{ 1,  1,  1}ICS \\\hline
    
    \rowcolor[rgb]{ .863,  .902,  .945} \textbf{Digital Bond \cite{DB-SCADA-Honeynet}} & \cellcolor[rgb]{ 1,  1,  1}2006 & \cellcolor[rgb]{ 1,  1,  1}Low & \cellcolor[rgb]{ 1,  1,  1} \checkmark & \cellcolor[rgb]{ 1,  1,  1}Virtual & \cellcolor[rgb]{ 1,  1,  1}Modbus/TCP, Telnet, HTTP, FTP, SNMP& \cellcolor[rgb]{ 1,  1,  1}Server & \cellcolor[rgb]{ 1,  1,  1}ICS \\\hline
    
    \rowcolor[rgb]{ .863,  .902,  .945} \textbf{Conpot~\cite{Conpot}} & \cellcolor[rgb]{ 1,  1,  1}2013 & \cellcolor[rgb]{ 1,  1,  1}Low & \cellcolor[rgb]{ 1,  1,  1} \checkmark & \cellcolor[rgb]{ 1,  1,  1}Virtual & \cellcolor[rgb]{ 1,  1,  1}IEC 60870-5-104, BACnet, EtherNet/IP, Guardian AST, Kamstrup, Modbus, S7comm, HTTP, FTP, SNMP, IPMI, TFTP& \cellcolor[rgb]{ 1,  1,  1}Server & \cellcolor[rgb]{ 1,  1,  1}ICS \\\hline
    
     \rowcolor[rgb]{ .863,  .902,  .945} \textbf{Zhao and Qin~\cite{Zhao:2017}} & \cellcolor[rgb]{ 1,  1,  1}2017 & \cellcolor[rgb]{ 1,  1,  1}Medium & \cellcolor[rgb]{ 1,  1,  1} \checkmark & \cellcolor[rgb]{ 1,  1,  1}Virtual & \cellcolor[rgb]{ 1,  1,  1}S7comm, Modbus, SNMP, HTTP& \cellcolor[rgb]{ 1,  1,  1}Server & \cellcolor[rgb]{ 1,  1,  1}ICS \\\hline
     
     \rowcolor[rgb]{ .863,  .902,  .945} \textbf{DiPot~\cite{Cao:2018}} & \cellcolor[rgb]{ 1,  1,  1}2018 & \cellcolor[rgb]{ 1,  1,  1}Low & \cellcolor[rgb]{ 1,  1,  1} \checkmark & \cellcolor[rgb]{ 1,  1,  1}Virtual & \cellcolor[rgb]{ 1,  1,  1}HTTP, Modbus, Kamstrup, SNMP, IMPI, BACnet, Guardian AST, S7comm& \cellcolor[rgb]{ 1,  1,  1}Server & \cellcolor[rgb]{ 1,  1,  1}ICS \\\hline
     
    \rowcolor[rgb]{ .863,  .902,  .945} \textbf{CamouflageNet~\cite{Naruoka:2015}} & \cellcolor[rgb]{ 1,  1,  1}2015 & \cellcolor[rgb]{ 1,  1,  1}Low & \cellcolor[rgb]{ 1,  1,  1} \checkmark & \cellcolor[rgb]{ 1,  1,  1}Virtual & \cellcolor[rgb]{ 1,  1,  1}N/A& \cellcolor[rgb]{ 1,  1,  1}Server & \cellcolor[rgb]{ 1,  1,  1}ICS \\\hline
    
    \rowcolor[rgb]{ .863,  .902,  .945} \textbf{XPOT~\cite{Lau:2016}} & \cellcolor[rgb]{ 1,  1,  1}2016 & \cellcolor[rgb]{ 1,  1,  1}Medium & \cellcolor[rgb]{ 1,  1,  1} \checkmark & \cellcolor[rgb]{ 1,  1,  1}Virtual & \cellcolor[rgb]{ 1,  1,  1}S7comm, SNMP& \cellcolor[rgb]{ 1,  1,  1}Server & \cellcolor[rgb]{ 1,  1,  1}ICS \\\hline
      
    \rowcolor[rgb]{ .863,  .902,  .945} \textbf{HosTaGe~\cite{Vasilomanolakis:2016}} & \cellcolor[rgb]{ 1,  1,  1}2016 & \cellcolor[rgb]{ 1,  1,  1}Low & \cellcolor[rgb]{ 1,  1,  1} \checkmark & \cellcolor[rgb]{ 1,  1,  1}Virtual & \cellcolor[rgb]{ 1,  1,  1}Modbus, S7comm, HTTPS, FTP, MySQL, SIP, SSH, SNMP, HTTP, Telnet, SMB and SMT& \cellcolor[rgb]{ 1,  1,  1}Server & \cellcolor[rgb]{ 1,  1,  1}ICS \\\hline
       
    \rowcolor[rgb]{ .863,  .902,  .945} \textbf{S7CommTrace~\cite{Xiao:2018}} & \cellcolor[rgb]{ 1,  1,  1}2018 & \cellcolor[rgb]{ 1,  1,  1}Medium & \cellcolor[rgb]{ 1,  1,  1} \checkmark & \cellcolor[rgb]{ 1,  1,  1}Virtual & \cellcolor[rgb]{ 1,  1,  1}S7comm& \cellcolor[rgb]{ 1,  1,  1}Server & \cellcolor[rgb]{ 1,  1,  1}ICS \\\hline 
    
    \rowcolor[rgb]{ .863,  .902,  .945} \textbf{Disso et al.~\cite{Disso:2013}} & \cellcolor[rgb]{ 1,  1,  1}2013 & \cellcolor[rgb]{ 1,  1,  1}Hybrid & \cellcolor[rgb]{ 1,  1,  1}Limited & \cellcolor[rgb]{ 1,  1,  1}Hybrid & \cellcolor[rgb]{ 1,  1,  1}N/A& \cellcolor[rgb]{ 1,  1,  1}Server & \cellcolor[rgb]{ 1,  1,  1}ICS \\\hline 
    
    \rowcolor[rgb]{ .863,  .902,  .945} \textbf{Honeyd+~\cite{Winn:2015}} & \cellcolor[rgb]{ 1,  1,  1}2015 & \cellcolor[rgb]{ 1,  1,  1}High & \cellcolor[rgb]{ 1,  1,  1}Limited & \cellcolor[rgb]{ 1,  1,  1}Hybrid & \cellcolor[rgb]{ 1,  1,  1}EtherNet/IP, HTTP& \cellcolor[rgb]{ 1,  1,  1}Server & \cellcolor[rgb]{ 1,  1,  1}ICS \\\hline 
    
    \rowcolor[rgb]{ .863,  .902,  .945} \textbf{Gallenstein \cite{Gallenstein:2017}} & \cellcolor[rgb]{ 1,  1,  1}2017 & \cellcolor[rgb]{ 1,  1,  1}Low & \cellcolor[rgb]{ 1,  1,  1}\checkmark & \cellcolor[rgb]{ 1,  1,  1}Virtual & \cellcolor[rgb]{ 1,  1,  1}EtherNet/IP. ISO-TSAP, HTTP& \cellcolor[rgb]{ 1,  1,  1}Server & \cellcolor[rgb]{ 1,  1,  1}ICS \\\hline 
    
    \rowcolor[rgb]{ .863,  .902,  .945} \textbf{Abe et al. \cite{ABE:2018}} & \cellcolor[rgb]{ 1,  1,  1}2018 & \cellcolor[rgb]{ 1,  1,  1}Low & \cellcolor[rgb]{ 1,  1,  1}\checkmark & \cellcolor[rgb]{ 1,  1,  1}Virtual & \cellcolor[rgb]{ 1,  1,  1}Modbus, S7comm, BACNet, IPMI, Guardian AST, HTTP, SNMP& \cellcolor[rgb]{ 1,  1,  1}Server & \cellcolor[rgb]{ 1,  1,  1}ICS \\\hline
    
    \rowcolor[rgb]{ .863,  .902,  .945} \textbf{Haney et al. \cite{Haney:2014}} & \cellcolor[rgb]{ 1,  1,  1}2014 & \cellcolor[rgb]{ 1,  1,  1}Low & \cellcolor[rgb]{ 1,  1,  1}\checkmark & \cellcolor[rgb]{ 1,  1,  1}Virtual & \cellcolor[rgb]{ 1,  1,  1}Modbus/TCP, Telnet, SSH, HTTP(S)& \cellcolor[rgb]{ 1,  1,  1}Server & \cellcolor[rgb]{ 1,  1,  1}ICS \\\hline

    \rowcolor[rgb]{ .863,  .902,  .945} \textbf{Kuman et al. \cite{Kuman:2017}} & \cellcolor[rgb]{ 1,  1,  1}2017 & \cellcolor[rgb]{ 1,  1,  1}Low & \cellcolor[rgb]{ 1,  1,  1}\checkmark & \cellcolor[rgb]{ 1,  1,  1}Virtual & \cellcolor[rgb]{ 1,  1,  1}Modbus/TCP& \cellcolor[rgb]{ 1,  1,  1}Server & \cellcolor[rgb]{ 1,  1,  1}ICS \\\hline
    
    \rowcolor[rgb]{ .863,  .902,  .945} \textbf{Ding et al. \cite{Ding:2018}} & \cellcolor[rgb]{ 1,  1,  1}2018 & \cellcolor[rgb]{ 1,  1,  1}Medium & \cellcolor[rgb]{ 1,  1,  1}\checkmark & \cellcolor[rgb]{ 1,  1,  1}Virtual & \cellcolor[rgb]{ 1,  1,  1}S7comm, SNMP& \cellcolor[rgb]{ 1,  1,  1}Server & \cellcolor[rgb]{ 1,  1,  1}ICS \\\hline 
     
    \rowcolor[rgb]{ .863,  .902,  .945} \textbf{Bodenheim \cite{Bodenheim:2014}} & \cellcolor[rgb]{ 1,  1,  1}2014 & \cellcolor[rgb]{ 1,  1,  1}High & \cellcolor[rgb]{ 1,  1,  1}Limited & \cellcolor[rgb]{ 1,  1,  1}Physical & \cellcolor[rgb]{ 1,  1,  1}HTTP, EtherNet/IP, SNMP& \cellcolor[rgb]{ 1,  1,  1}Server & \cellcolor[rgb]{ 1,  1,  1}ICS \\\hline 
    
    \rowcolor[rgb]{ .863,  .902,  .945} \textbf{Piggin et al. \cite{Piggin:2016}} & \cellcolor[rgb]{ 1,  1,  1}2016 & \cellcolor[rgb]{ 1,  1,  1}High & \cellcolor[rgb]{ 1,  1,  1}X & \cellcolor[rgb]{ 1,  1,  1}Physical & \cellcolor[rgb]{ 1,  1,  1}SSH, HTTP, RDP& \cellcolor[rgb]{ 1,  1,  1}Server  & \cellcolor[rgb]{ 1,  1,  1}ICS \\\hline 
    
    \rowcolor[rgb]{ .863,  .902,  .945} \textbf{Haney \cite{Haney:2019}} & \cellcolor[rgb]{ 1,  1,  1}2019 & \cellcolor[rgb]{ 1,  1,  1}High & \cellcolor[rgb]{ 1,  1,  1}\checkmark & \cellcolor[rgb]{ 1,  1,  1}Hybrid & \cellcolor[rgb]{ 1,  1,  1}Modbus/TCP, SSH, Telnet, SNMP, HTTP& \cellcolor[rgb]{ 1,  1,  1}Client, Server& \cellcolor[rgb]{ 1,  1,  1}ICS \\\hline 
    
    \rowcolor[rgb]{ .863,  .902,  .945} \textbf{Hilt et al. \cite{Hilt:2020}} & \cellcolor[rgb]{ 1,  1,  1}2020 & \cellcolor[rgb]{ 1,  1,  1}High & \cellcolor[rgb]{ 1,  1,  1}X & \cellcolor[rgb]{ 1,  1,  1}Hybrid & \cellcolor[rgb]{ 1,  1,  1}S7comm, Omron FINS, EtherNet/IP, VNC& \cellcolor[rgb]{ 1,  1,  1}Server  & \cellcolor[rgb]{ 1,  1,  1}ICS \\\hline 
    
    \rowcolor[rgb]{ .863,  .902,  .945} \textbf{Berman \cite{Berman:2012}} & \cellcolor[rgb]{ 1,  1,  1}2012 & \cellcolor[rgb]{ 1,  1,  1}Low & \cellcolor[rgb]{ 1,  1,  1}Limited & \cellcolor[rgb]{ 1,  1,  1}Virtual & \cellcolor[rgb]{ 1,  1,  1}Modbus/TCP& \cellcolor[rgb]{ 1,  1,  1}Server & \cellcolor[rgb]{ 1,  1,  1}ICS \\\hline 
    
    \rowcolor[rgb]{ .863,  .902,  .945} \textbf{Jaromin \cite{Jaromin:2013}} & \cellcolor[rgb]{ 1,  1,  1}2013 & \cellcolor[rgb]{ 1,  1,  1}Low & \cellcolor[rgb]{ 1,  1,  1}Limited & \cellcolor[rgb]{ 1,  1,  1}Virtual & \cellcolor[rgb]{ 1,  1,  1}Modbus/TCP, HAP, HTTP& \cellcolor[rgb]{ 1,  1,  1}Server & \cellcolor[rgb]{ 1,  1,  1}ICS \\\hline 
     
    
    \rowcolor[rgb]{ .863,  .902,  .945} \textbf{Holczer et al. \cite{Holczer:2015}} & \cellcolor[rgb]{ 1,  1,  1}2015 & \cellcolor[rgb]{ 1,  1,  1}High & \cellcolor[rgb]{ 1,  1,  1}\checkmark & \cellcolor[rgb]{ 1,  1,  1}Virtual & \cellcolor[rgb]{ 1,  1,  1}S7comm, SNMP, HTTP(S)& \cellcolor[rgb]{ 1,  1,  1}Server & \cellcolor[rgb]{ 1,  1,  1}ICS \\\hline 
    
    \rowcolor[rgb]{ .863,  .902,  .945} \textbf{Serbanescu \newline et al. \cite{SerbanescuHoneynet:2015}} & \cellcolor[rgb]{ 1,  1,  1}2015 & \cellcolor[rgb]{ 1,  1,  1}Low & \cellcolor[rgb]{ 1,  1,  1}\checkmark & \cellcolor[rgb]{ 1,  1,  1}Virtual & \cellcolor[rgb]{ 1,  1,  1}DNP3, IEC-104, Modbus, ICCP, SNMP, TFTP, XMPP& \cellcolor[rgb]{ 1,  1,  1}Server & \cellcolor[rgb]{ 1,  1,  1}ICS \\\hline 
    
    \rowcolor[rgb]{ .863,  .902,  .945} \textbf{Simões \cite{Simoes:2015}} & \cellcolor[rgb]{ 1,  1,  1}2015 & \cellcolor[rgb]{ 1,  1,  1}Low & \cellcolor[rgb]{ 1,  1,  1}\checkmark & \cellcolor[rgb]{ 1,  1,  1}Virtual & \cellcolor[rgb]{ 1,  1,  1}Modbus, SNMP, FTP& \cellcolor[rgb]{ 1,  1,  1}Server & \cellcolor[rgb]{ 1,  1,  1}ICS \\\hline 
    
    \rowcolor[rgb]{ .863,  .902,  .945} \textbf{Ahn et al. \cite{Ahn:2019}} & \cellcolor[rgb]{ 1,  1,  1}2019 & \cellcolor[rgb]{ 1,  1,  1}Low & \cellcolor[rgb]{ 1,  1,  1}\checkmark & \cellcolor[rgb]{ 1,  1,  1}Virtual & \cellcolor[rgb]{ 1,  1,  1}Modbus& \cellcolor[rgb]{ 1,  1,  1}Server & \cellcolor[rgb]{ 1,  1,  1}ICS \\\hline 
    
    \rowcolor[rgb]{ .863,  .902,  .945} \textbf{Belqruch et al. \cite{Belqruch:2019} } & \cellcolor[rgb]{ 1,  1,  1}2019 & \cellcolor[rgb]{ 1,  1,  1}Medium& \cellcolor[rgb]{ 1,  1,  1}\checkmark & \cellcolor[rgb]{ 1,  1,  1}Virtual & \cellcolor[rgb]{ 1,  1,  1}SSH& \cellcolor[rgb]{ 1,  1,  1}Server & \cellcolor[rgb]{ 1,  1,  1}ICS \\\hline 
    
    \rowcolor[rgb]{ .863,  .902,  .945} \textbf{SHaPe \cite{Koltys2015}} & \cellcolor[rgb]{ 1,  1,  1}2015 & \cellcolor[rgb]{ 1,  1,  1}Low& \cellcolor[rgb]{ 1,  1,  1}\checkmark & \cellcolor[rgb]{ 1,  1,  1}Virtual & \cellcolor[rgb]{ 1,  1,  1}IEC 61850 MMS, HTTP, FTP, SMB& \cellcolor[rgb]{ 1,  1,  1}Server & \cellcolor[rgb]{ 1,  1,  1}Smart Grid \\\hline 
    
    \rowcolor[rgb]{ .863,  .902,  .945} \textbf{GridPot \cite{Redwood:2015}} & \cellcolor[rgb]{ 1,  1,  1}2015 & \cellcolor[rgb]{ 1,  1,  1}Hybrid& \cellcolor[rgb]{ 1,  1,  1}\checkmark & \cellcolor[rgb]{ 1,  1,  1}Virtual & \cellcolor[rgb]{ 1,  1,  1}IEC 61850 GOOSE/MMS, Modbus, HTTP& \cellcolor[rgb]{ 1,  1,  1}Server & \cellcolor[rgb]{ 1,  1,  1}Smart Grid \\\hline 
    
    \rowcolor[rgb]{ .863,  .902,  .945} \textbf{Scott \cite{Scott:2014}} & \cellcolor[rgb]{ 1,  1,  1}2014 & \cellcolor[rgb]{ 1,  1,  1}Low& \cellcolor[rgb]{ 1,  1,  1}\checkmark & \cellcolor[rgb]{ 1,  1,  1}Virtual & \cellcolor[rgb]{ 1,  1,  1}Modbus/TCP, HTTP, SNMP& \cellcolor[rgb]{ 1,  1,  1}Server & \cellcolor[rgb]{ 1,  1,  1}Smart Grid \\\hline
    
    \rowcolor[rgb]{ .863,  .902,  .945} \textbf{Mashima et al. \cite{Mashima:2017}} & \cellcolor[rgb]{ 1,  1,  1}2017 & \cellcolor[rgb]{ 1,  1,  1}Medium / High& \cellcolor[rgb]{ 1,  1,  1}\checkmark & \cellcolor[rgb]{ 1,  1,  1}Virtual & \cellcolor[rgb]{ 1,  1,  1}IEC 60870-5-104, IEC 61850, SSH& \cellcolor[rgb]{ 1,  1,  1}Server & \cellcolor[rgb]{ 1,  1,  1}Smart Grid \\\hline
    
    \rowcolor[rgb]{ .863,  .902,  .945} \textbf{CryPLH~\cite{Buza:2014}} & \cellcolor[rgb]{ 1,  1,  1}2018 & \cellcolor[rgb]{ 1,  1,  1}Low & \cellcolor[rgb]{ 1,  1,  1}\checkmark & \cellcolor[rgb]{ 1,  1,  1}Virtual & \cellcolor[rgb]{ 1,  1,  1}HTTP(S), SNMP, Step7 ISO-TSAP& \cellcolor[rgb]{ 1,  1,  1}Server & \cellcolor[rgb]{ 1,  1,  1}Smart Grid \\\hline
     
    \rowcolor[rgb]{ .863,  .902,  .945} \textbf{Pliatsios et al.~\cite{Pliatsios:2019}} & \cellcolor[rgb]{ 1,  1,  1}2019 & \cellcolor[rgb]{ 1,  1,  1}Low & \cellcolor[rgb]{ 1,  1,  1} Limited & \cellcolor[rgb]{ 1,  1,  1}Hybrid & \cellcolor[rgb]{ 1,  1,  1}Modbus/TCP& \cellcolor[rgb]{ 1,  1,  1}Client, Server & \cellcolor[rgb]{ 1,  1,  1}Smart Grid \\\hline
    
    \rowcolor[rgb]{ .863,  .902,  .945} \textbf{Mashima et al. \cite{Mashima:2019}} & \cellcolor[rgb]{ 1,  1,  1}2019 & \cellcolor[rgb]{ 1,  1,  1}Low & \cellcolor[rgb]{ 1,  1,  1}\checkmark & \cellcolor[rgb]{ 1,  1,  1}Virtual & \cellcolor[rgb]{ 1,  1,  1}TCP port listener on IEC 61850 MMS, S7comm, Modbus/TCP, Niagara Fox, EtherNet/IP, IEC 60870-5-104, DNP3 and BACnet ports& \cellcolor[rgb]{ 1,  1,  1}Server & \cellcolor[rgb]{ 1,  1,  1}Smart Grid \\\hline
    
    \rowcolor[rgb]{ .863,  .902,  .945} \textbf{Murillo et al. \cite{Murillo:2018}} & \cellcolor[rgb]{ 1,  1,  1}2018 & \cellcolor[rgb]{ 1,  1,  1}Low& \cellcolor[rgb]{ 1,  1,  1}\checkmark & \cellcolor[rgb]{ 1,  1,  1}Virtual & \cellcolor[rgb]{ 1,  1,  1}EtherNet/IP& \cellcolor[rgb]{ 1,  1,  1}Server & \cellcolor[rgb]{ 1,  1,  1}Water System \\\hline
    
    \rowcolor[rgb]{ .863,  .902,  .945} \textbf{Petre et al. \cite{Petre:2019}} & \cellcolor[rgb]{ 1,  1,  1}2019 & \cellcolor[rgb]{ 1,  1,  1}Medium& \cellcolor[rgb]{ 1,  1,  1}\checkmark & \cellcolor[rgb]{ 1,  1,  1}Virtual & \cellcolor[rgb]{ 1,  1,  1}Modbus& \cellcolor[rgb]{ 1,  1,  1}Server & \cellcolor[rgb]{ 1,  1,  1}Water System \\\hline

    \rowcolor[rgb]{ .863,  .902,  .945} \textbf{Wilhoit~\cite{Wilhoit:2013-1}} & \cellcolor[rgb]{ 1,  1,  1}2013 & \cellcolor[rgb]{ 1,  1,  1}Hybrid & \cellcolor[rgb]{ 1,  1,  1}Limited & \cellcolor[rgb]{ 1,  1,  1}Hybrid & \cellcolor[rgb]{ 1,  1,  1}Modbus/TCP, HTTP, FTP& \cellcolor[rgb]{ 1,  1,  1}Server & \cellcolor[rgb]{ 1,  1,  1}Water System \\\hline 
    
    \rowcolor[rgb]{ .863,  .902,  .945} \textbf{Antonioli et al. \cite{Antonioli:2016}} & \cellcolor[rgb]{ 1,  1,  1}2016 & \cellcolor[rgb]{ 1,  1,  1}High & \cellcolor[rgb]{ 1,  1,  1}\checkmark & \cellcolor[rgb]{ 1,  1,  1}Virtual & \cellcolor[rgb]{ 1,  1,  1}EtherNet/IP, SSH, Telnet, VPN& \cellcolor[rgb]{ 1,  1,  1}Server & \cellcolor[rgb]{ 1,  1,  1}Water System \\\hline
     
    \rowcolor[rgb]{ .863,  .902,  .945} \textbf{MimePot~\cite{Bernieri:2019}} & \cellcolor[rgb]{ 1,  1,  1}2019 & \cellcolor[rgb]{ 1,  1,  1}High & \cellcolor[rgb]{ 1,  1,  1} \checkmark & \cellcolor[rgb]{ 1,  1,  1}Virtual & \cellcolor[rgb]{ 1,  1,  1}Modbus/TCP& \cellcolor[rgb]{ 1,  1,  1}Client, Server & \cellcolor[rgb]{ 1,  1,  1}Water System \\\hline
    
    \rowcolor[rgb]{ .863,  .902,  .945} \textbf{GasPot \cite{Wilhoit:2015}} & \cellcolor[rgb]{ 1,  1,  1}2015 & \cellcolor[rgb]{ 1,  1,  1}Low& \cellcolor[rgb]{ 1,  1,  1}\checkmark & \cellcolor[rgb]{ 1,  1,  1}Virtual & \cellcolor[rgb]{ 1,  1,  1}N/A& \cellcolor[rgb]{ 1,  1,  1}Server & \cellcolor[rgb]{ 1,  1,  1}Gas System \\\hline
    
    \rowcolor[rgb]{ .863,  .902,  .945} \textbf{Zamiri et al. \cite{Zamiri:2019}} & \cellcolor[rgb]{ 1,  1,  1}2019 & \cellcolor[rgb]{ 1,  1,  1}Medium& \cellcolor[rgb]{ 1,  1,  1}\checkmark & \cellcolor[rgb]{ 1,  1,  1}Virtual & \cellcolor[rgb]{ 1,  1,  1}Veeder-Root ATG& \cellcolor[rgb]{ 1,  1,  1}Server & \cellcolor[rgb]{ 1,  1,  1}Gas System \\\hline
    
    \rowcolor[rgb]{ .863,  .902,  .945} \textbf{HoneyPhy \cite{Litchfield:2016}} & \cellcolor[rgb]{ 1,  1,  1}2016 & \cellcolor[rgb]{ 1,  1,  1}Hybrid& \cellcolor[rgb]{ 1,  1,  1}X & \cellcolor[rgb]{ 1,  1,  1}Hybrid & \cellcolor[rgb]{ 1,  1,  1}DNP3& \cellcolor[rgb]{ 1,  1,  1}Server & \cellcolor[rgb]{ 1,  1,  1}Building Auto.\\\hline

    \rowcolor[rgb]{ .863,  .902,  .945} \textbf{HoneyIo3~\cite{Ammar:2018}} & \cellcolor[rgb]{ 1,  1,  1}2018 & \cellcolor[rgb]{ 1,  1,  1}Hybrid & \cellcolor[rgb]{ 1,  1,  1} \checkmark & \cellcolor[rgb]{ 1,  1,  1}Hybrid & \cellcolor[rgb]{ 1,  1,  1}IPMI, S7comm, Kamstrup, SNMP, HTTP(S), Ntopng, SSH & \cellcolor[rgb]{ 1,  1,  1}Server & \cellcolor[rgb]{ 1,  1,  1}IIoT \\\hline
     
    \rowcolor[rgb]{ .863,  .902,  .945} \textbf{Du and Wang~\cite{Du:2020}} & \cellcolor[rgb]{ 1,  1,  1}2020 & \cellcolor[rgb]{ 1,  1,  1}Hybrid & \cellcolor[rgb]{ 1,  1,  1} \checkmark & \cellcolor[rgb]{ 1,  1,  1}Virtual & \cellcolor[rgb]{ 1,  1,  1}Not identified & \cellcolor[rgb]{ 1,  1,  1}Server & \cellcolor[rgb]{ 1,  1,  1}IIoT \\\hline

  %
\end{longtable}%
\clearpage
\twocolumn


\clearpage
\onecolumn
  \LTcapwidth=\textwidth
  \renewcommand\arraystretch{1.5}
  \begin{longtable}
  {|>{\footnotesize}p{1.2cm}|>{\footnotesize}p{3.1cm}|>{\footnotesize}p{3.8cm}|>{\footnotesize}p{3cm}|>{\footnotesize}p{3.9cm}|>{\footnotesize}p{0.7cm}|}\caption{Summary of Tools, Implementation, and Attack Types of Honeypots and Honeynets for IIoT and CPS}\label{tab:iiot-cps-honeypots-tools} \\\hline

    \rowcolor[rgb]{ .141,  .251,  .384} \textcolor[rgb]{ 1,  1,  1}{\textbf{Work}} & \textcolor[rgb]{ 1,  1,  1}{\textbf{Tools}} & \textcolor[rgb]{ 1,  1,  1}{\textbf{Ports}} & \textcolor[rgb]{ 1,  1,  1}{\textbf{Attack Types}} & \textcolor[rgb]{ 1,  1,  1}{\textbf{Data Analyzed}} & \textcolor[rgb]{ 1,  1,  1}{\textbf{Length of the Study}} \\\endfirsthead
    
\multicolumn{6}{c}{\tablename\ \thetable: Summary of Tools, Implementation, and Attack Types of Honeypots and Honeynets for IIoT and CPS (Cont.)} \\
\hline
\rowcolor[rgb]{ .141,  .251,  .384} \textcolor[rgb]{ 1,  1,  1}{\textbf{Work}} & \textcolor[rgb]{ 1,  1,  1}{\textbf{Tools}} & \textcolor[rgb]{ 1,  1,  1}{\textbf{Ports}} & \textcolor[rgb]{ 1,  1,  1}{\textbf{Attack Types}} & \textcolor[rgb]{ 1,  1,  1}{\textbf{Data Analyzed}} & \textcolor[rgb]{ 1,  1,  1}{\textbf{Length of the Study}} \\\endhead

    \rowcolor[rgb]{ .863,  .902,  .945} \textbf{CISCO \cite{CISCO-SCADA-Honeynet}} & \cellcolor[rgb]{ 1,  1,  1}N/A & \cellcolor[rgb]{ 1,  1,  1}Modbus/TCP (502), Telnet (23), HTTP (80), FTP (21) & \cellcolor[rgb]{ 1,  1,  1} N/A & \cellcolor[rgb]{ 1,  1,  1}N/A & \cellcolor[rgb]{ 1,  1,  1}N/A\\\hline
    
    \rowcolor[rgb]{ .863,  .902,  .945} \textbf{Digital Bond \cite{DB-SCADA-Honeynet}}& \cellcolor[rgb]{ 1,  1,  1}Sebek, Argus, Walleye, Snort IDS & \cellcolor[rgb]{ 1,  1,  1}Modbus/TCP (502), SNMP (161), Telnet (23), HTTP (80), FTP (21) & \cellcolor[rgb]{ 1,  1,  1} N/A & \cellcolor[rgb]{ 1,  1,  1}N/A & \cellcolor[rgb]{ 1,  1,  1}N/A\\\hline
    
    \rowcolor[rgb]{ .863,  .902,  .945} \textbf{Conpot \cite{Conpot}}& \cellcolor[rgb]{ 1,  1,  1}N/A& \cellcolor[rgb]{ 1,  1,  1}IEC 60870-5-104 (2404), BACnet (47808), EtherNet/IP (44818), Guardian AST (10001), Kamstrup (1025, 50100), Modbus (502), S7comm (102), HTTP (80), FTP (21), SNMP (161), IPMI (623), TFTP (69)& \cellcolor[rgb]{ 1,  1,  1} N/A & \cellcolor[rgb]{ 1,  1,  1}N/A & \cellcolor[rgb]{ 1,  1,  1}N/A\\\hline
    
    \rowcolor[rgb]{ .863,  .902,  .945} \textbf{Zhao and Qin \cite{Zhao:2017}}& \cellcolor[rgb]{ 1,  1,  1}Flask framework, Wireshark& \cellcolor[rgb]{ 1,  1,  1}N/A& \cellcolor[rgb]{ 1,  1,  1}Traffic from 244 IP addresses from 34 countries & \cellcolor[rgb]{ 1,  1,  1}Types, sources, requests from IPs & \cellcolor[rgb]{ 1,  1,  1}43 days\\\hline
    
    \rowcolor[rgb]{ .863,  .902,  .945} \textbf{DiPot \cite{Cao:2018}}& \cellcolor[rgb]{ 1,  1,  1}N/A& \cellcolor[rgb]{ 1,  1,  1}N/A &\cellcolor[rgb]{ 1,  1,  1}Modbus and Kamstrup scan, Modbus over-length packets & \cellcolor[rgb]{ 1,  1,  1}Access sequences to protocols and their IPs & \cellcolor[rgb]{ 1,  1,  1}6 months\\\hline   
    
    \rowcolor[rgb]{ .863,  .902,  .945} \textbf{CamouflageNet \cite{Naruoka:2015}}& \cellcolor[rgb]{ 1,  1,  1}Nmap, Kali Linux& \cellcolor[rgb]{ 1,  1,  1}N/A &\cellcolor[rgb]{ 1,  1,  1}Scanning & \cellcolor[rgb]{ 1,  1,  1}N/A & \cellcolor[rgb]{ 1,  1,  1}N/A\\\hline 
   
    \rowcolor[rgb]{ .863,  .902,  .945} \textbf{XPOT \cite{Lau:2016}}& \cellcolor[rgb]{ 1,  1,  1}Nmap, nfqueue& \cellcolor[rgb]{ 1,  1,  1}N/A &\cellcolor[rgb]{ 1,  1,  1}N/A& \cellcolor[rgb]{ 1,  1,  1}S7comm handshakes and queries & \cellcolor[rgb]{ 1,  1,  1}1 month\\\hline 
   
    \rowcolor[rgb]{ .863,  .902,  .945} \textbf{HosTaGe \cite{Vasilomanolakis:2016}}& \cellcolor[rgb]{ 1,  1,  1}Wireshark, Bro IDS, snp4j& \cellcolor[rgb]{ 1,  1,  1}Modbus (502) &\cellcolor[rgb]{ 1,  1,  1}Multi-stage attacks consisting of different scanning and attack attempts& \cellcolor[rgb]{ 1,  1,  1}Attacks to Modbus, S7comm, HTTP, Telnet and IP addresses targeting HosTaGe and Conpot & \cellcolor[rgb]{ 1,  1,  1}12 weeks\\\hline
    
    \rowcolor[rgb]{ .863,  .902,  .945} \textbf{S7CommTrace \cite{Xiao:2018}}& \cellcolor[rgb]{ 1,  1,  1}N/A& \cellcolor[rgb]{ 1,  1,  1}S7comm (102) &\cellcolor[rgb]{ 1,  1,  1}N/A& \cellcolor[rgb]{ 1,  1,  1}Indexing in Shodan, valid and invalid requests, function coverage of S7comm, received IP address diversity & \cellcolor[rgb]{ 1,  1,  1}60 days\\\hline
   
    \rowcolor[rgb]{ .863,  .902,  .945} \textbf{Disso et al. \cite{Disso:2013}}& \cellcolor[rgb]{ 1,  1,  1}N/A& \cellcolor[rgb]{ 1,  1,  1}N/A &\cellcolor[rgb]{ 1,  1,  1}N/A& \cellcolor[rgb]{ 1,  1,  1}Link latency, network traffic counting and connection limiting, background network traffic& \cellcolor[rgb]{ 1,  1,  1}N/A\\\hline
    
    \rowcolor[rgb]{ .863,  .902,  .945} \textbf{Honeyd+ \cite{Winn:2015}}& \cellcolor[rgb]{ 1,  1,  1}Nmap, Zenmap, Wget& \cellcolor[rgb]{ 1,  1,  1}EtherNet/IP (44818, 2222), HTTP (80)&\cellcolor[rgb]{ 1,  1,  1}Scanning& \cellcolor[rgb]{ 1,  1,  1}Fingerprints of Honeyd+ hosts, error rates and protocol data rates& \cellcolor[rgb]{ 1,  1,  1}N/A\\\hline
    
    \rowcolor[rgb]{ .863,  .902,  .945} \textbf{Gallenstein \cite{Gallenstein:2017}}& \cellcolor[rgb]{ 1,  1,  1}Nmap, Shodan Honeyscore, RSLinx, STEP7, Wget, Wireshark& \cellcolor[rgb]{ 1,  1,  1}EtherNet/IP (44818), ISO-TSAP (102), HTTP (80)&\cellcolor[rgb]{ 1,  1,  1}Scanning& \cellcolor[rgb]{ 1,  1,  1}Nmap fingerprint similarity, Honeyscore performance, RSLinx and STEP7 PLC module discovery performance, comparison of responses to Wget requests& \cellcolor[rgb]{ 1,  1,  1}N/A\\\hline
    
    \rowcolor[rgb]{ .863,  .902,  .945} \textbf{Abe et al. \cite{ABE:2018}}& \cellcolor[rgb]{ 1,  1,  1}Nmap& \cellcolor[rgb]{ 1,  1,  1}Modbus (502), S7comm (102), BACNet (47808), IPMI (623), Guardian AST (10001), HTTP (80), SNMP (161)&\cellcolor[rgb]{ 1,  1,  1}Havex RAT, Modbus Stager, PLC blaster& \cellcolor[rgb]{ 1,  1,  1}Behavior against Havex RAT, Modbus Stager and PLC blaster attacks& \cellcolor[rgb]{ 1,  1,  1}N/A\\\hline
    
    \rowcolor[rgb]{ .863,  .902,  .945} \textbf{Haney et al. \cite{Haney:2014}}& \cellcolor[rgb]{ 1,  1,  1}IMUNES, JAMOD Library, Snort IDS, Snort daemon logger, Sebek, Honeywall& \cellcolor[rgb]{ 1,  1,  1}Modbus/TCP (502), HTTP (80), HTTPS (443), Telnet (23), SSH (22)&\cellcolor[rgb]{ 1,  1,  1}Network and port scan, Modbus packet capture, injection and out of band packets& \cellcolor[rgb]{ 1,  1,  1}N/A& \cellcolor[rgb]{ 1,  1,  1}N/A\\\hline

    \rowcolor[rgb]{ .863,  .902,  .945} \textbf{Kuman et al. \cite{Kuman:2017}}& \cellcolor[rgb]{ 1,  1,  1}OSSEC host-based IDS, PLCScan, Shodan, iptables& \cellcolor[rgb]{ 1,  1,  1}Modbus/TCP (502)&\cellcolor[rgb]{ 1,  1,  1}Port scans on Modbus and HTTP protocols& \cellcolor[rgb]{ 1,  1,  1}Conpot logs& \cellcolor[rgb]{ 1,  1,  1}2 weeks\\\hline
    
    \rowcolor[rgb]{ .863,  .902,  .945} \textbf{Ding et al. \cite{Ding:2018}}& \cellcolor[rgb]{ 1,  1,  1}Nmap, snmpwalk, STEP7 software, PLCscan& \cellcolor[rgb]{ 1,  1,  1}S7comm (102)&\cellcolor[rgb]{ 1,  1,  1}Scanning& \cellcolor[rgb]{ 1,  1,  1}Scanning result& \cellcolor[rgb]{ 1,  1,  1}N/A\\\hline
    
    \rowcolor[rgb]{ .863,  .902,  .945} \textbf{Bodenheim \cite{Bodenheim:2014}}& \cellcolor[rgb]{ 1,  1,  1}Nmap, TCPdump, SSH, Tshark, Wireshark, Shodan API, Security Onion Linux, Snort, netcat& \cellcolor[rgb]{ 1,  1,  1}EtherNet/IP (44818), HTTP (80) SNMP (161)&\cellcolor[rgb]{ 1,  1,  1}Scanning& \cellcolor[rgb]{ 1,  1,  1}Shodan's functionality and indexing, effect of being indexed on the received traffic, effect of modifying device service banners & \cellcolor[rgb]{ 1,  1,  1}55 days\\\hline
       
    \rowcolor[rgb]{ .863,  .902,  .945} \textbf{Piggin et al. \cite{Piggin:2016}}& \cellcolor[rgb]{ 1,  1,  1}Google Dorks& \cellcolor[rgb]{ 1,  1,  1}N/A&\cellcolor[rgb]{ 1,  1,  1}Scanning, password attack, execute malicious program, SSH brute-force, an attack originated from TOR network, DoS on the PLC& \cellcolor[rgb]{ 1,  1,  1}Origin and target protocols of the attacks & \cellcolor[rgb]{ 1,  1,  1}N/A\\\hline  
    
    \rowcolor[rgb]{ .863,  .902,  .945} \textbf{Haney \cite{Haney:2019}}& \cellcolor[rgb]{ 1,  1,  1}SecurityOnion, iptables, SnortSam, Sebekd, Argus, JAMOD, IMUNES, LabVIEW, Matlab Simulink& \cellcolor[rgb]{ 1,  1,  1}Modbus/TCP (502), HTTP (80), SSH (22), SNMP (161)&\cellcolor[rgb]{ 1,  1,  1}Modbus scanning via Shodan, brute-force login& \cellcolor[rgb]{ 1,  1,  1}The most common usernames and passwords used for attacks, attack origins & \cellcolor[rgb]{ 1,  1,  1}2 weeks\\\hline
    
    \rowcolor[rgb]{ .863,  .902,  .945} \textbf{Hilt et al. \cite{Hilt:2020}}& \cellcolor[rgb]{ 1,  1,  1}Tshark, Moloch, Chaosreader, VNCLogger, Suricata, Syslog& \cellcolor[rgb]{ 1,  1,  1}S7comm (102), Omron FINS (9600), EtherNet/IP (44818), VNC (5900, 5901)&\cellcolor[rgb]{ 1,  1,  1}Scanning, ransomware, malicious cryptomining, robotic workstation beaconing attempt& \cellcolor[rgb]{ 1,  1,  1}Unique IP addresses, amount of traffic, protocol-specific traffic and commands to PLCs, communication with scanners, VNC screen recording, attacker's downloads& \cellcolor[rgb]{ 1,  1,  1}7 months\\\hline 
    
    \rowcolor[rgb]{ .863,  .902,  .945} \textbf{Berman \cite{Berman:2012}}& \cellcolor[rgb]{ 1,  1,  1}Nmap, Wireshark, SSH, TCPDump, Syslog, Triangle MicroWorks Protocol Test Harness& \cellcolor[rgb]{ 1,  1,  1}Modbus/TCP (502)&\cellcolor[rgb]{ 1,  1,  1}Scanning, invalid ICS traffic& \cellcolor[rgb]{ 1,  1,  1}Modbus/TCP traffic tests, response statistics, fingerprint analysis,  response to invalid ICS traffic, logging capabilities& \cellcolor[rgb]{ 1,  1,  1}N/A\\\hline
    
    \rowcolor[rgb]{ .863,  .902,  .945} \textbf{Jaromin \cite{Jaromin:2013}}& \cellcolor[rgb]{ 1,  1,  1}Nmap, Metasploit, NetEdit3, DirectSOFT5, iptables and netfilter modules, libpcap library, HAP API, Syslog& \cellcolor[rgb]{ 1,  1,  1}Modbus/TCP (502), HAP (28784), HTTP (80)&\cellcolor[rgb]{ 1,  1,  1}Brute-force password guessing, fingerprinting& \cellcolor[rgb]{ 1,  1,  1}Packet level accuracy and logging capability, OS fingerpring accuracy, Metasploit attack performance, response timing& \cellcolor[rgb]{ 1,  1,  1}N/A\\\hline
    
    
    \rowcolor[rgb]{ .863,  .902,  .945} \textbf{Holczer et al. \cite{Holczer:2015}}& \cellcolor[rgb]{ 1,  1,  1}Step7, Szilu SSL, MiniWeb, iptables& \cellcolor[rgb]{ 1,  1,  1}S7comm (102), HTTP (80), HTTPS (443), SNMP (161)&\cellcolor[rgb]{ 1,  1,  1}Pings, port scans, SSH scans& \cellcolor[rgb]{ 1,  1,  1}Attack origins, logs& \cellcolor[rgb]{ 1,  1,  1}50 days\\\hline
    
    \rowcolor[rgb]{ .863,  .902,  .945} \textbf{Serbanescu et al. \cite{SerbanescuHoneynet:2015}}& \cellcolor[rgb]{ 1,  1,  1}Snort, Matlab, Amazon EC2 environment& \cellcolor[rgb]{ 1,  1,  1}N/A&\cellcolor[rgb]{ 1,  1,  1}Scanning& \cellcolor[rgb]{ 1,  1,  1}Modbus traffic (connections, requests, port scans, activity types, country of origin), impact of Shodan listing the devices, attractiveness of ICS protocols& \cellcolor[rgb]{ 1,  1,  1}28 days\\\hline
    
    \rowcolor[rgb]{ .863,  .902,  .945} \textbf{Simões \cite{Simoes:2015}}& \cellcolor[rgb]{ 1,  1,  1}Modbus-tk, Pymodbus and Libpcap libraries, NET-SNMP, VSFTPd& \cellcolor[rgb]{ 1,  1,  1}N/A&\cellcolor[rgb]{ 1,  1,  1}N/A& \cellcolor[rgb]{ 1,  1,  1}Resource usage of honeypot, response time, reliability& \cellcolor[rgb]{ 1,  1,  1}N/A\\\hline
    
    \rowcolor[rgb]{ .863,  .902,  .945} \textbf{Ahn et al. \cite{Ahn:2019}}& \cellcolor[rgb]{ 1,  1,  1}N/A& \cellcolor[rgb]{ 1,  1,  1}Modbus (502)&\cellcolor[rgb]{ 1,  1,  1}ARP poisoning& \cellcolor[rgb]{ 1,  1,  1}N/A& \cellcolor[rgb]{ 1,  1,  1}N/A\\\hline
    
    \rowcolor[rgb]{ .863,  .902,  .945} \textbf{Belqruch et al. \cite{Belqruch:2019}}& \cellcolor[rgb]{ 1,  1,  1}Kali Linux& \cellcolor[rgb]{ 1,  1,  1}SSH (22)&\cellcolor[rgb]{ 1,  1,  1}SSH brute force& \cellcolor[rgb]{ 1,  1,  1}Username-password  combinations, password attempts& \cellcolor[rgb]{ 1,  1,  1}N/A\\\hline
    
    \rowcolor[rgb]{ .863,  .902,  .945} \textbf{SHaPe \cite{Koltys2015}}& \cellcolor[rgb]{ 1,  1,  1}libiec61850& \cellcolor[rgb]{ 1,  1,  1}N/A&\cellcolor[rgb]{ 1,  1,  1}N/A& \cellcolor[rgb]{ 1,  1,  1}TCP connection information (connection ID, source and destination IPs and ports), Dionaea logs& \cellcolor[rgb]{ 1,  1,  1}N/A\\\hline
    
    \rowcolor[rgb]{ .863,  .902,  .945} \textbf{GridPot \cite{Redwood:2015}}& \cellcolor[rgb]{ 1,  1,  1}ETSY Skyline project anomaly detection modules, GridLab-D, hpfeeds logging& \cellcolor[rgb]{ 1,  1,  1}N/A&\cellcolor[rgb]{ 1,  1,  1}IED switching attack& \cellcolor[rgb]{ 1,  1,  1}Physics impact of the attack& \cellcolor[rgb]{ 1,  1,  1}N/A\\\hline
 
    \rowcolor[rgb]{ .863,  .902,  .945} \textbf{Scott \cite{Scott:2014}}& \cellcolor[rgb]{ 1,  1,  1}Tenable Nessus, Splunk Enterprise, Rsyslog& \cellcolor[rgb]{ 1,  1,  1}Modbus/TCP (502), HTTP (80), SNMP (161), Syslog (514), Splunk (8000), SMTP (25) &\cellcolor[rgb]{ 1,  1,  1}Scanning attack& \cellcolor[rgb]{ 1,  1,  1}Alerts generated by Splunk& \cellcolor[rgb]{ 1,  1,  1}N/A\\\hline
    
    \rowcolor[rgb]{ .863,  .902,  .945} \textbf{Mashima et al. \cite{Mashima:2017}}& \cellcolor[rgb]{ 1,  1,  1}VirtualBox, Mininet, SoftGrid and POWERWORLD simulators, SOCAT port forwarding,  OpenMUC& \cellcolor[rgb]{ 1,  1,  1}IEC 60870-5-104 (2404), IEC 61850 (102), SSH (22)&\cellcolor[rgb]{ 1,  1,  1}Nmap scan, Shodan& \cellcolor[rgb]{ 1,  1,  1}Fingerprinting, latency, scalability and cost analysis& \cellcolor[rgb]{ 1,  1,  1}N/A\\\hline
    
    \rowcolor[rgb]{ .863,  .902,  .945} \textbf{CryPLH \cite{Buza:2014}}& \cellcolor[rgb]{ 1,  1,  1}Nessus, Nmap, Backtrack Linux, Miniweb, NGINX, SNMPWalk& \cellcolor[rgb]{ 1,  1,  1}ISO-TSAP (102), HTTP (80), HTTPS (443), SNMP (161) &\cellcolor[rgb]{ 1,  1,  1}Attack tests with Backtrack Linux (Kali Linux), Nmap, nessus & \cellcolor[rgb]{ 1,  1,  1}Honeypot logs & \cellcolor[rgb]{ 1,  1,  1}38 days\\\hline 
    
    \rowcolor[rgb]{ .863,  .902,  .945} \textbf{Pliatsios et al.~\cite{Pliatsios:2019}}& \cellcolor[rgb]{ 1,  1,  1}Wireshark, Tshark& \cellcolor[rgb]{ 1,  1,  1}N/A &\cellcolor[rgb]{ 1,  1,  1}N/A & \cellcolor[rgb]{ 1,  1,  1}N/A& \cellcolor[rgb]{ 1,  1,  1}N/A\\\hline 
    
    \rowcolor[rgb]{ .863,  .902,  .945} \textbf{Mashima et al. \cite{Mashima:2019}}& \cellcolor[rgb]{ 1,  1,  1}Wireshark, ELK stack, Amazon Cloud& \cellcolor[rgb]{ 1,  1,  1}IEC 61850 MMS and S7comm (102), Modbus/TCP (502), Niagara Fox (1911, 4911), EtherNet/IP (ENIP) (2222, 44818), IEC 60870-5-104 (2404), DNP3 (19999, 20000), BACnet (47808)&\cellcolor[rgb]{ 1,  1,  1}SYN-flooding DoS, scanning& \cellcolor[rgb]{ 1,  1,  1}Access trends, protocol specific attempts, correlation of honeypots' data, attack origin dynamics& \cellcolor[rgb]{ 1,  1,  1}6 months\\\hline
    
    \rowcolor[rgb]{ .863,  .902,  .945} \textbf{Murillo et al. \cite{Murillo:2018}}& \cellcolor[rgb]{ 1,  1,  1}Mininet, MiniCPS, Odeint solver& \cellcolor[rgb]{ 1,  1,  1}N/A&\cellcolor[rgb]{ 1,  1,  1}Bias injection attack& \cellcolor[rgb]{ 1,  1,  1}Tank levels and plant behavior without attack, with attack and defense& \cellcolor[rgb]{ 1,  1,  1}N/A\\\hline
    
    \rowcolor[rgb]{ .863,  .902,  .945} \textbf{Petre et al. \cite{Petre:2019}}& \cellcolor[rgb]{ 1,  1,  1}Node-RED, Softing OPC UA Client, SQLite& \cellcolor[rgb]{ 1,  1,  1}N/A&\cellcolor[rgb]{ 1,  1,  1}Unauthorized access& \cellcolor[rgb]{ 1,  1,  1}Database entries& \cellcolor[rgb]{ 1,  1,  1}N/A\\\hline
    
    \rowcolor[rgb]{ .863,  .902,  .945} \textbf{Wilhoit~\cite{Wilhoit:2013-1,Wilhoit:2013-2}}& \cellcolor[rgb]{ 1,  1,  1}Snort, tcpdump, Pastebin, Amazon EC2& \cellcolor[rgb]{ 1,  1,  1}Modbus/TCP (502), HTTP (80), FTP (21)&\cellcolor[rgb]{ 1,  1,  1}Scanning, spearphishing, unauthorized access and modification, Modbus traffic modification, CPU fan speed modification on the water pump, malware exploitation& \cellcolor[rgb]{ 1,  1,  1}Attack types and origins& \cellcolor[rgb]{ 1,  1,  1}28 days\\\hline
    
    \rowcolor[rgb]{ .863,  .902,  .945} \textbf{Antonioli et al. \cite{Antonioli:2016}}& \cellcolor[rgb]{ 1,  1,  1}Mininet, MiniCPS, ocserv VPN, sshd, telnetd, tc link shaping, cpppo EtherNet/IP emulation& \cellcolor[rgb]{ 1,  1,  1}Ethernet/IP (44818), HTTP (80), SSH (22), Telnet (23)&\cellcolor[rgb]{ 1,  1,  1}DoS, Man in the Middle, port scan, service enumeration, physical process attacks (i.e., tank overflow)& \cellcolor[rgb]{ 1,  1,  1}Network metrics (address, packet loss, delay, bandwidth, topology, protocols, etc.) and physical metrics (realistic mathematical model, sensor and actuator operations, etc.)& \cellcolor[rgb]{ 1,  1,  1}Capture the Flag Competition\\\hline
    
    \rowcolor[rgb]{ .863,  .902,  .945} \textbf{MimePot~\cite{Bernieri:2019}}& \cellcolor[rgb]{ 1,  1,  1}Mininet, Scapy& \cellcolor[rgb]{ 1,  1,  1}N/A &\cellcolor[rgb]{ 1,  1,  1}Man in the Middle and integrity attack& \cellcolor[rgb]{ 1,  1,  1} Tank water levels, Mime Estimation and Control status by time, water pump status, flows between tanks& \cellcolor[rgb]{ 1,  1,  1}N/A\\\hline
    
    \rowcolor[rgb]{ .863,  .902,  .945} \textbf{GasPot \cite{Wilhoit:2015}}& \cellcolor[rgb]{ 1,  1,  1}N/A& \cellcolor[rgb]{ 1,  1,  1}N/A&\cellcolor[rgb]{ 1,  1,  1}Reconnaissance, DDoS& \cellcolor[rgb]{ 1,  1,  1}Connection attempts, commands, attack origins& \cellcolor[rgb]{ 1,  1,  1}N/A\\\hline
    
    \rowcolor[rgb]{ .863,  .902,  .945} \textbf{Zamiri et al. \cite{Zamiri:2019}}& \cellcolor[rgb]{ 1,  1,  1}Nmap& \cellcolor[rgb]{ 1,  1,  1}Veeder-Root ATG (10001)&\cellcolor[rgb]{ 1,  1,  1}N/A& \cellcolor[rgb]{ 1,  1,  1}N/A& \cellcolor[rgb]{ 1,  1,  1}N/A\\\hline
    \rowcolor[rgb]{ .863,  .902,  .945} \textbf{HoneyPhy \cite{Litchfield:2016}}& \cellcolor[rgb]{ 1,  1,  1}OpenDNP3 library, LabVIEW& \cellcolor[rgb]{ 1,  1,  1}N/A&\cellcolor[rgb]{ 1,  1,  1}N/A& \cellcolor[rgb]{ 1,  1,  1}Heating and cooling curve from both physical system and the process model& \cellcolor[rgb]{ 1,  1,  1}N/A\\\hline

   \rowcolor[rgb]{ .863,  .902,  .945} \textbf{Du and Wang \cite{Du:2020}} & \cellcolor[rgb]{ 1,  1,  1}SDN testbed & \cellcolor[rgb]{ 1,  1,  1}N/A & \cellcolor[rgb]{ 1,  1,  1}DDoS attacks, SYN Flood attack, FTP flow & \cellcolor[rgb]{ 1,  1,  1}Protocols, packets per port, packet characteristics & \cellcolor[rgb]{ 1,  1,  1}N/A\\\hline

   \rowcolor[rgb]{ .863,  .902,  .945} \textbf{HoneyIo3 \cite{Ammar:2018}} & \cellcolor[rgb]{ 1,  1,  1}Shodan, Nmap & \cellcolor[rgb]{ 1,  1,  1}IPMI (623), S7comm (102), Kamstrup (1025), SNMP (161), HTTP (80), HTTPS (443), ntopng (3000), SSH (9002) & \cellcolor[rgb]{ 1,  1,  1}Reconnaissance attacks & \cellcolor[rgb]{ 1,  1,  1} Protocols, packets per port, packets characteristics & \cellcolor[rgb]{ 1,  1,  1}N/A\\\hline

\end{longtable}%
\clearpage
\twocolumn




\subsection{Simulated services}
Honeypots and honeynets for IIoT and CPS support a wide variety of protocols and services that are both specific and not specific to industrial environments. The protocols and services supported by the honeypots and honeynets are shown in Table~\ref{tab:iiot-cps-honeypots} while ports that are exposed for such protocols in the honeypots are outlined in Table~\ref{tab:iiot-cps-honeypots-tools}. When we consider the protocols, we can see that Modbus, HTTP, SNMP, and S7comm are the most popular protocols among the studies. Our findings are also validated by a number of researchers \cite{Haney:2014, SerbanescuHoneynet:2015, Alves:2016, Humayed:2017, Almulla:2018, Petre:2019, Pliatsios:2019} who cite Modbus as the most widely used industrial protocol. Popularity of industrial protocols along with number of honeypots supporting them can be expressed as follows: Modbus (22), S7comm (12), EtherNet/IP (8), IEC 60870-5-104 (4), BACnet (4), Kamstrup (4), DNP3 (3), Guardian AST (3), IEC 51850 (3), and ISOTSAP (2). The popularity of non-industrial protocols, HTTP and SNMP are very reasonable. HTTP is used as the interface of HMIs of industrial systems~\cite{Antonioli:2016, Scott:2014} and also it enables the remote configuration of industrial components such as PLCs~\cite{Holczer:2015, Vasilomanolakis:2016}. For these reasons, it is stated as the target of scanning activities performed by malicious entities~\cite{Kendrick:2019}. SNMP on the other hand is used for monitoring and management purposes in industrial environments~\cite{Holczer:2015, Winn:2015}.

\subsection{Availability of open-source honeypot and honeynet solutions}
There exist eight honeypot and honeynet studies that provide their implementation openly. In this respect, CISCO SCADA HoneyNet~\cite{CISCO-SCADA-Honeynet} source code is still available. However, the last shared version was in 2015. Unfortunately, Digital Bond's SCADA Honeynet~\cite{DB-SCADA-Honeynet} is not reachable right now. Conpot on the other hand, is open-source and is still being actively maintained. Considering the rest of the honeypot and honeynet studies, only the honeypot of Zamiri et al.~\cite{Zamiri:2019} is actively maintained. However, their study was performed in 2019 and it is not known if they will continue to actively maintain it. The implementations of GridPot~\cite{Redwood:2015} and SHaPe~\cite{Koltys2015} are still available, but their last update was in 2015. The last update for  GasPot~\cite{Wilhoit:2015} was in 2016, and honeypot-like testbed of Murillo et al.~\cite{Murillo:2018} was maintained in 2018.

\subsection{Most commonly used tools}
The most commonly used tool for IIoT and CPS honeypot and honeynet studies is Nmap, which is followed by Wireshark, Snort IDS, Shodan tools, Mininet, iptables, tshark, TCPDUMP, and syslog. Researchers used Nmap to obtain fingerprints of their honeypots and to indetify the exposed ports. Wireshark was used for traffic capture and analysis. Snort IDS is used for attacker control attempts especially in honeywall configurations. Shodan tools were used to find out indexing information, honeypot's fingerprint from Shodan's point of view, and also to find out if Shodan detects the decoy system as a honeypot or not.

\subsection{Most Common Attacks}
\revision{
The most commonly detected/tested attacks in IIoT and CPS honeypots/honeynets are scanning attacks. Majority of the studies detected/tested scanning attacks to the IIoT and CPS environments. In addition to DoS/DDoS, SSH, brute-force, and Man-in-the-Middle attacks were also detected/tested in the proposed honeypots and honeynets. Although less common than the mentioned attacks, ransomware, malicious cryptocurrency mining, malware and ICS specific attacks such as HAVEX RAT, PLC Blaster, and tank overflow attacks were also detected/tested in the proposed systems.
}

\section{LESSONS LEARNED AND OPEN ISSUES}\label{sec:openissues}
Considering the honeypot and honeynets for IoT, IIoT, and CPS environments, we believe that it is crucial to stress the importance of key points. This is valuable to interpret the state-of-the-art and to motivate for further research and practice.

\subsection{Lessons Learned}
Any honeypot/honeynet developer and researcher for IoT, IIoT, and CPS needs to consider a few key factors at the very beginning of his/her work. The key factors that should be taken into account are target application area, purpose of the honeypot/honeynet, \revision{cost}, deployment location, 
intended level of interaction with the attacker, resource level, services that will be provided, simulated, or emulated, and their realistic service to the attackers, tools that will be used, fingerprintability and indexing, and liability issues that may come up. 

\noindent\textbf{Target Application Area Selection: }
 IIoT and CPS environments have their own characteristics which may affect the entire honeypot/honeynet design. Devices, communication channel characteristics, protocols, traffic rates,  application QoS requirements, and many other factors can be different for each unique application. CPS and IIoT devices have quite different characteristics from regular IoT devices. In addition, they work with industrial protocols which are not used in traditional ICT or IoT environments. \revision{Such industrial devices have life-times in the order of decades and work with real-time constraints which strictly require them to work without interruptions~\cite{Simoes:2013, Holczer:2015}.} Critical infrastructures of nations are controlled by such industrial devices. While typical IoT applications do not have any physical processes to be continuously monitored and controlled, it is very common for IIoT and CPS applications. For these reasons, it is extremely important to determine the target application and its characteristics.    

\noindent\textbf{Purpose of the Honeypot/Honeynet: }
\revision {The purpose of a honeypot or honeynet significantly affects the measures that need to be taken to ensure that attacks on the honeynet do not compromise the infrastructure on which it is implemented. In a research environment, this can be done by isolating the honeynet system. For example, by implementing it in a DMZ.} However, if production honeypots are to be deployed in IIoT and CPS environments where industrial devices monitor and control critical plant processes, then extra care has to be given to the decoy system design. Such production honeypots in industrial environments need to  ensure that they cannot be compromised by attackers, as well as ensure that they do not interfere with the communication and control processes (i.e., operational resources) of the existing industrial devices. In addition, one has to note that honeypots and honeynets do not stop attacks~\cite{Scott:2014}. For this reason, the alerts or logs created by them have to be considered by administrators. 

\noindent\textbf{Deployment Location: }
 \revision{While deployment location can have an important effect on honeypot activity, only twelve of the reviewed studies stated their deployment locations. Two CPS studies \cite{Bodenheim:2014, Holczer:2015} deployed their honeypots within the IP range of universities,  which may call the attention of attackers who check the IP address spaces of their targets. Another two CPS studies \cite{Serbanescu:2015, Mashima:2019} and three IoT studies \cite{Zhang:2019, Dang:2019, Vetterl:2019} chose cloud environments as deployment targets. Such an approach would provide a global view of attacks to honeypot/honeynet owners and also may be more attractive to attackers than the university option. However, attackers can still find out that the target system operates within the IP range of a cloud provider. 
 Additionally, two CPS studies \cite{Hilt:2020, Wilhoit:2015} and three IoT studies \cite{AHA:2018, Guarnizo:2017, Tambe:2019} use public IP addresses which is the better option.} In addition to this, Guarnizo \cite{Guarnizo:2017} identified that geographical location selection, in terms of country or city of deployment, or at least the location shown to attackers, is an important consideration. This is because attackers might seek to attack devices in certain cities if they are looking for a point to start targeted attacks or if they have an interest in reselling IPs after they are infected.
 
\noindent\textbf{Cost:} \revision{Cost is a crucial consideration in developing honeypots and honeynets. Setting up a honeypot or honeynet can be very expensive if physical resources and closed source tools are used instead of virtual resources and open source tools. Also, it is important to note that the PLCs, IEDs, RTUs, and RIOs used in industrial applications are considerably more expensive than Commercial of-the-shelf (COTS) IoT devices. In addition, complexity of a honeypot, especially a honeynet, can be another contributing factor for the cost of the system. Complexity is directly proportional to the level of interaction provided and also the number of services/protocols supported. As the interaction level and number of supported services increase for honeypots and honeynets, higher fidelity data in high volume is collected, which requires more resources to store and process. Moreover, deployment locations can have an effect on the cost of the system. To be more specific, although deployment of a honeypot or a honeynet in a university IP address space can be cost efficient for research, it can easily call the attention of adversaries. Honeypot/honeynet deployments in cloud environments would be significantly more costly compared to university environments. However, attackers can still determine that the IP addresses are in the cloud provider space. The third option would be renting private IP addresses to avoid suspicion by attackers, but such an option can be more costly than the cloud option. For these reasons, honeypot/honeynet developers and researchers need to consider how resource and interaction levels as well as deployment environment and complexity affect the cost.}


\noindent\textbf{Level of Interaction Considerations: }
\revision {The level of interaction of a honeypot/honeynet affects many different aspects, as explained in Section~\ref{sec:methodology}.} Considering the existing honeypots and honeynets for IoT, IIoT, and CPS, almost every possible level of interaction choice can be seen \revision{ as reviewed in Section~\ref{sec:taxonomy} and Section~\ref{sec:taxonomycps}. However, high interaction is needed in order to identify complex attacks that may target IoT, IIoT, and CPS devices and understand possible effects on industrial processes and critical infrastructures. Although COTS} IoT devices are more affordable, industrial devices in the order of thousands of dollars can be a significant issue to consider. Therefore, resource level choice and realistic simulation/emulation become important considerations. These are further discussed in the following categories. 

\noindent\textbf{Resource Level Selection: }The question of whether real, simulated, or both types of devices are to be used in honeypots/honeynets for IoT, IIoT, and CPS is quite a vital one. Real devices can act as high-interaction honeypots and provide high fidelity information. In addition, they would be  almost impossible to be detected as a honeypot by outsiders. However, as explained earlier, costs of real devices can change based on the target application area and constructing a realistic honeynet with a realistic number of industrial devices may cost a fortune. These important factors motivated researchers and developers to design honeypots/honeynets with virtual components. 
Virtualization enables scalability, heterogeneity, easy maintenance and cost-effective deployment of IoT, IIoT, and CPS honeypots. In this respect, Dang et al.~\cite{Dang:2019} found that approximately 92.1\% of malware-based attacks target multiple IoT device architectures and emphasized the need for a virtual IoT honeypot solution. At the same time, they identified that virtual honeypots attracted 37\% fewer suspicious connections and 39\% fewer attacks than physical honeypots. Also, the variety of attacks virtual honeypots captured were more than with physical honeypots. Dang et al.~\cite{Dang:2019} also pointed out that a virtual honeypot costs 12.5x less to maintain than a physical honeypot. These factors should be weighed in considering honeypot/honeynet design. Balancing the benefits of both physical and virtual resources in a hybrid solution is an important consideration. In addition to this, the choice of which model of devices to select, either real or simulated, can play a factor in attracting attackers. Guarnizo~\cite{Guarnizo:2017} identified that models with known vulnerabilities tend to be attacked more frequently.

\noindent\textbf{Choice of Services to Provide/Simulate and Realism: }
Choice of services to provide or simulate, and ensuring realism in such services are very critical factors in honeypot/honeynet design. These considerations get even more important for IIoT and CPS systems. Which services will be provided? Is it logical to support all of the protocols and services in the target application area? If not, how to choose among the set of protocols/services? Scott~\cite{Scott:2014} pointed out that honeypots and honeynets should simulate only the services that the mimicked device would usually accommodate. If the mimicked device does not have a certain service or does not support it, but the honeypot does, then attackers may realise that they are interacting with a decoy system. After determining the services/protocols to be supported, then comes another important aspect: realism. 
 
One of the principal considerations when deploying a honeypot or honeynet system for IoT, IIoT, or CPS is how to simulate a real system effectively in order to avoid hackers and search engines from identifying that they are interacting with a decoy system. This is vital for the honeypot system to be able to attract attackers and to gather as much information as possible from their interactions. In order to avoid detection more effectively for a honeynet deployed in an IoT environment, Surnin et al.~\cite{Surnin:2019} recommended the following: a limited number of services should be run to simulate a more realistic environment, ping command host requests should yield an existing host, files created by attackers should not be deleted, commands for utilities should return a list of running processes, no hardcoded values should be used, simulated Linux utilities should have full functionality from the origin, and attacker file requests should be sent to a sandbox with a specified delay before checking them on external services such as VirusTotal. Zamiri-Gourabi et al.~\cite{Zamiri:2019} pointed out the fact that default hardcoded configurations, missing features of the simulated services or protocols, unusual or unrealistic behaviors, fingerprintability of the hosting platform and response times can be the possible fingerprints of honeypots and honeynets. Simulations of plant processes in a realistic way comes to the scene for IIoT and CPS honeypots/honeynets. Unfortunately, only a small portion of honeypots/honeynets considered this vital issue with IIoT and CPS honeypots.With IoT honeypots, this factor was considered by various studies. In fact, the most commonly used tools for IoT honeypot and honeynet research were all tools which were used to check the available services, realism in responses, including response times, and other factors that affect fingerprintability, which will be discussed in the following sections.

\noindent\textbf{Choice of Tools: }
A honeypot or honeynet designer should consider the deployment area or target application area characteristics when he/she is choosing the tools such as scanners. Not every tool may support all of the IoT, IIoT, and CPS applications, their corresponding protocols and services. In addition, tools that also support vulnerability checks should be considered to be employed~\cite{Scott:2014}. A designer should also consider how to pair their honeypot or honeynet with tools that will best complement the honeynet for effective attack mitigation. \revision{ While medium and high interaction honeypots enable more interactivity for attackers, attackers may have tools to check whether they are interacting with a virtual environment and whether their activities are being recorded/logged.} Tools such as Sebek are used by researchers in order to seamlessly log the activities of the attackers.  

\noindent\textbf{Appearance on the Search Engines and Fingerprintability: }
One of the most important factors in honeypot/honeynet design is ensuring appearance on the search engines while not being fingerprinted as a decoy system. For this reason, honeypot/honeynet owners have to monitor IoT search engines which  identify and detect devices and honeypots on the Internet, such as Shodan. Different views exist in the literature whether being indexed by such search engines has an effect on the attacks to be received. For example, Guarnizo \cite{Guarnizo:2017} identified that the number of attacks on a device increase significantly in the first few weeks after they are listed on Shodan. Nevertheless, such indexing services can make the jobs of attackers easier by pointing out Internet-connected ready-to-attack targets. Being indexed by such search engines verifies the accessibility of the honeypot/honeynet system. Being listed as a real system rather than a honeypot/honeynet is an achievement that helps honeypot owners to reach their ultimate goal. 

\noindent\textbf{\revision{Comparison of IoT, IIoT, and CPS Honeypots/Honeynets: }}\revision{
Honeypot and honeynet research for IoT, IIoT, and CPS environments is an important research area. Although we summarized the studies and provided taxonomies in the previous sections, comparison of the decoy systems for IoT, IIoT, and CPS, and highlighting their similarities and differences can be very crucial. The first significant difference arises from the supported services. While the IoT decoys considered mostly support Telnet, SSH, and HTTP which are not IoT-specific, the CPS decoys considered mostly support industrial protocols such as Modbus, S7comm, EtherNet/IP, and non industry-specific protocols such HTTP and SNMP. Since there are only two decoys for IIoT and only one of them is disclosing its services, we can see that IIoT decoys stay in the intermediary position in this regard, supporting both industrial and non-industrial protocols. The second difference arises from the process simulations. While some CPS decoys employ simulations of industrial processes for ICS plants, water management, electrical grid, and building HVAC systems, we do not see such process simulations in the proposed IoT decoys. The third difference arises from the interaction level of proposed honeypots and honeynets. While the majority of the decoys proposed for IoT are medium interaction decoys (10 studies), the majority of the decoys for CPS are low interaction (16 studies). The cost of physical ICS devices and difficulty of realistic process simulations play an important role in the interaction level choice of CPS honeypots and honeynets. Considering the similarities, we see that decoys with virtual resources and server roles are common between IoT, IIoT, and CPS environments.    
}
\vspace{-0.5em}

\noindent\textbf{Control and Liability: }
When deploying a honeypot or honeynet for IoT, IIoT, and CPS environments, control and liability issues are the aspects that are greatly overlooked, but designers should always consider. The greater the level of interaction a honeypot allows, the greater the risk that it could be compromised and used by attackers for harming other systems in the network or even launching attacks on other networks. Scott~\cite{Scott:2014} advised to be familiar with laws before deployment of honeypots since honeypots are interpreted as entrapment by jurisdictions in some places. Haney~\cite{Haney:2019} emphasized the importance of taking liability and legal issues into account and putting data control as a first priority, even if this means data capture may be affected. Haney proposed setting up both automated as well as manual data control mechanisms, with at least two protection mechanisms to always have a second option if one data control method fails. Sokol~\cite{Sokol:2015} highlighted that a honeynet should contain the following parts in order to address security, data control, and liability issues: a firewall with only the necessary network ports opened, a dynamic (re)connection mechanism to  determine  if  a  connection  is 
trusted and can be allowed, a testbed for analysis, an emulated private virtual network to restrict attackers, and a control center to monitor connections and  respond to issues quickly.

\noindent\textbf{Improving Security of IoT, IIoT, and CPS Devices: }
The information gathered from research with honeypots and honeynets can lead to innovative ways to improve the security of IoT devices despite their constraints. One example of this is the proposal by Dang et al. \cite{Dang:2019} of a series of measures called IoTCheck to increase the security of IoT devices, which include asking whether the IoT device has a unique strong password, whether the default system user is a non-root user, and whether there are unnecessary components on the devices which can be eliminated. The same authors also suggest for manufacturers to disable shell commands that are enabled by default on Linux-based IoT devices but are not necessary, as these are used for attacks.

\subsection{Open Issues}
Honeypots and honeynets for IoT, IIoT, and CPS have been a very active field of research during the last decade. We studied 79 honeypots/honeynets in greater detail in this study. However, there are still open issues which need to be addressed by researchers.



\noindent\textbf{\revision{Emerging Technologies/Domains:}} In terms of decoy systems for IoT, we see that there are honeypots/honeynets for Smart Home, but not for emerging domains or technologies such as wearable devices, medical devices, and smart city. In terms of decoy systems for IIoT and CPS, we see that there are honeypots/honeynets for general ICS, smart grid, water, gas and building automation systems. However, we do not see such decoys for other IIoT and CPS applications such as smart city, transportation, nuclear plants, medical devices. \revision{As smart medical devices in modern healthcare applications are becoming more prevalent and are threatened by various attacks~\cite{newaz2020survey, Heka, HealthGuard}, decoy systems for modern healthcare applications are needed. In addition, to the best of our knowledge, there is only one honeypot system for building automation systems. Considering the rapid increase of notorious ransomware attacks~\cite{Harun:2021}, cryptocurrency mining attacks~\cite{Minos,tekiner2021sok}, and attacks to enterprise IoT systems~\cite{HDMI, hdmiwatch, PoisonIvy, LuisEIoT:2021}, we believe that further research is needed which may enable us to protect smart buildings from ransomware attacks. We would like to note that building honeypots and honeynets for the unexplored IoT, IIoT and CPS applications may require realistic process models (e.g., patient vitals models, vehicle operation models, nuclear process models, etc.) in case virtual or hybrid decoy systems are targeted.}

\noindent\textbf{\revision{Unexplored Protocols: }}Existing IoT, IIoT, and CPS honeypots/honeynets support a wide range of ICT, IoT, and industrial protocols. Various IoT honeypots emulate full devices. However, one cannot claim that the state-of-the-art honeypot/honeynet research considered every protocol or service. In addition to this, very few current studies focus on IoT specific protocols. There are also protocols and services that still need to be addressed by honeypot research. For instance, we did not find any study that supports \revision{Highway Addressable Remote Transducer} (HART) and WirelessHART~\cite{hart} industrial protocols. \revision{In addition, Enterprise IoT environments can employ various proprietary communication protocols that rely on security through obscurity~\cite{LuisEIoT:2021}. For this reason, decoy designs for such type of proprietary solutions are needed.} Researching unexplored protocols and services may provide valuable information for honeypot/honeynet research and practice. \revision{A potential solution to unexplored protocols for IoT, IIoT, and CPS honeypots/honeynets could be extending open source honeypots and honeynets such as Conpot, Honeyd, Dionaea, Kippo, etc. for the unexplored protocols. Although open source libraries for the unexplored well-known protocols can be found, researchers would have to perform reverse engineering for the proprietary communication protocols.}

\noindent\textbf{\revision{Emerging Platforms: }}\revision{In the recent years, several platforms were proposed/developed by both researchers and vendors for the management of the IoT devices~\cite{iotPlatforms}. In this regard, platforms such as openHAB, Samsung SmartThings, thingworx, Amazon AWS IoT, IBM Watson IoT, Apple HomeKit, etc. emerged for IoT applications. Such platforms have different characteristics in terms of supported IoT devices, communication protocols and network topologies, data processing and event handling approaches, and security. Although there exist decoy systems for generic IoT applications, one does not see any studies focusing on honeypot and honeynet design for the mentioned emerging IoT platforms. Since popularity of such platforms is increasing in recent years, IoT applications that are built on top of such platforms can be sweet spots for the adversaries. Therefore, there is a need for honeypot/honeynet research for the emerging IoT platforms. In order to propose novel decoy systems for the emerging platforms, researchers can benefit from the existing IoT honeypot/honeynet research and extend the open source IoT decoys.}

\noindent\textbf{\revision{Optimized Deployment Location: }}\revision{Honeypots and honeynets proposed for IoT, IIoT, and CPS employed various deployment locations (i.e., university, cloud, private locations) as explained in the previous sections. Each deployment location option has its own benefits and pitfalls in terms of fingerprintability~\cite{Babun:2020, Aksu:2018},  suitability for IoT, IIoT, or CPS application, complexity, and cost. Although a few studies investigated how a limited set of deploy locations attract attackers, one does  not see any study in the literature that aims to optimize the deployment location for the decoy system with respect to a set of constraints. We believe that, this is an important gap in the honeypot/honeynet research and there is a need for extensive analysis and novel frameworks in order to optimize deployment location decisions. Although this problem is hard to tackle, researchers can employ relaxation strategies in order to approximate the optimal deployment location solution for the IoT, IIoT, and CPS decoys.}

\noindent\textbf{\revision{Remote Management: }}\revision{Several tools can be utilized to manage honeypots/honeynets locally or remotely. While the decoys with virtual resources can be managed locally or remotely without much efforts, the decoys with physical resources may require researchers to be physically present in such locations for maintenance purposes. However, the Covid-19 pandemic caused lockdowns all around the world which forced researchers to perform their tasks remotely. Extraordinary times like the current pandemic, natural hazards, etc. can cause similar situations that can force people to remotely manage their decoy systems. We believe that researchers have to consider such conditions while designing and deploying their decoy systems for IoT, IIoT, and CPS. Remote management of decoy systems require employment of secure tools and secure configurations. However, vulnerabilities of such tools that are considered to be secure can be found, as in the case of SolarWinds ~\cite{Solarwinds}, which require continuous efforts to check for vulnerabilities and patch.}

\noindent\textbf{\revision{Anti-Detection Mechanisms: }} Honeypots and honeynets that are using virtual resources have been widely used in IoT, IIoT, and CPS environments. Such an approach has several advantages as discussed in the previous section. However, from the malware research domain we know that virtual environment detection techniques are frequently used by malicious software developers. When we checked the honeypot/honeynet studies for IoT, IIoT, and CPS that use virtual resources, we did not see any study which considers this important issue. In addition, the analysis parts of the studies did not mention detecting an attacker which uses such techniques. Although research did not observe the existence of a sample case, we think that attackers will be using such methods in the near future. For this reason, future honeypots and honeynets for IoT, IIoT, and CPS should consider to employ anti-detection mechanisms in their medium/high interaction virtual decoy systems. \revision{Researchers in this regard can benefit from existing anti-detection research from the malware analysis domain such as hiding the artifacts regarding the analysis environment, moving analysis logic to lower levels such as hypervisors or bare-metal, etc.~\cite{Afianian:2019}.}

\noindent\textbf{\revision{Vulnerabilities of Industrial Devices: }} IoT, IIoT, and CPS environments consist of several devices produced by different vendors. Vulnerabilities with device firmware, OS and other software are often found and listed in vulnerability databases such as Common Vulnerabilities and Exposures (CVE)~\cite{cve}. As explained earlier, devices with such vulnerabilities attract the attackers and stand as vulnerable targets to compromise. Considering the honeypots and honeynets, we see that there exist studies which take such vulnerabilities into account when designing honeypots. However, we did not encounter any proposal for IIoT and CPS that considers vulnerabilities of industrial devices. We believe that a research gap exists in the literature in regards to whether attackers really pay attention to industrial device vulnerabilities or not when choosing targets. \revision{A potential way to address this open problem could be deploying honeypots for IIoT and CPS environments that advertise both vulnerable and patched versions of ICS device firmware or management software. In this way, it would be possible to understand if adversaries pay attention to disclosed vulnerabilities when choosing their targets.}

\noindent\textbf{\revision{Insider Attacks: }}The target users for IoT, IIoT, and CPS are very diverse and have very different skill levels for deploying honeypot/honeynet systems. However, none of the IoT research studies consider how the systems being proposed could be implemented on a wider scale in the future, taking into account the need for simple deployment. In addition to this, none of the current research places focus on attacks initiated and carried out from inside the network. These types of attacks could be carried out by disgruntled employees or for corporate espionage. \revision{However, researchers may not deploy physical or virtual honeypots on a network in a straightforward way since insiders may have a chance to reach the decoys physically or virtually. We believe that virtualization technologies such as Network Function Virtualization (NFV) and containers, and SDN technologies can be utilized to develop moving target defense-like honeypot solutions for insider attackers.} 

\noindent\textbf{\revision{Machine Learning: }}Another open issue 
is the employment of ML and AI techniques for honeypot design. Considering the studies, we see that ML techniques have been employed \revision{by a limited number of honeypot/honeynet works for configuration and data analysis purposes. Although eight studies (\cite{Wagener:2011, Pauna:2014, Pauna:2018, Luo:2017, Shrivastava:2019, Pauna:2019, Lingenfelter:2020, Wang:2018}) employed ML for IoT honeypots/honeynets,  we see that only one study~\cite{Cao:2018} used ML techniques for IIoT and CPS honeypots. 
We believe that future IoT, IIoT, and CPS honeypots and honeynets can benefit from ML techniques to propose smarter decoy systems that can i) adapt themselves based on the actions of attackers, ii) discriminate known attacks from new attacks thus enable researchers to focus more on novel threats, and iii) increase the efficiency and prevalence of honeypots and honeynets.}

\noindent\textbf{\revision{Discrimination of Benign Decoy Traffic: }}\revision{Honeypots and honeynets are traditionally assumed to receive only malicious traffic which are in fact helpful for the existing IDS and IPS elements in the network to increase their true positive rates. However, IoT honeypots and honeynets employing physical IoT devices can receive benign traffic from vendors. For instance smart home devices provided by Google, Apple, Samsung, and Amazon can receive benign traffic from their vendors with application-specific motivations (e.g., cloud connectivity, health check, updates, etc.). Such benign traffic originating from device vendors targeting the decoy system, as well as the traffic generated by benign bots such as Shodan and Censys to index the Internet-connected devices, break the aforementioned assumption of incoming traffic to decoy systems. For this reason, researchers have to take such benign traffic into account while analyzing the decoy traffic. We believe that IP address lookup for the traffic sources can provide information on the benign origins of the decoy traffic. In addition, analysis of Ferretti et al.~\cite{Ferretti:2019} on the scanning patterns of legitimate scanners such as Shodan can give clues to researchers on discriminating legitimate traffic.}

\noindent\textbf{\revision{Production Decoys: }}\revision{Considering the reviewed honeypots and honeynets for IoT, IIoT, and CPS environments, we see that the majority of the reviewed works are research honeypots. Although research honeypots are important to understand the attacks and new tactics of attackers, they do not actively participate in securing an IoT, IIoT, or CPS environment. For this reason, more production honeypots are needed that can actively participate in securing IoT, IIoT, and CPS networks. Efforts in combining honeypots/honeynets with IDS solutions are noteworthy in this regard. Researchers can employ open source IDS solutions such as Snort, Zeek, Suricata etc., malware analysis platforms such as Cuckoo, and next generation networking technologies such as SDN to propose novel decoy solutions for IoT, IIoT, and CPS environments.}

\section{Conclusion}\label{sec:conclusion}
In this paper, we provided a comprehensive survey of honeypots and honeynets for IoT, IIoT, and CPS environments. 
We provided a taxonomy of honeypots and honeynets based on purpose, role, level of interaction, scalability, resource level, availability of source code and target IoT, IIoT, or CPS application. In addition, we analyzed the existing honeypots and honeynets extensively and extracted the common characteristics of state-of-the-art honeypots and honeynets for IoT, IIoT, and CPS. Moreover, we outlined and discussed the key design factors for honeypots and honeynets for IoT, IIoT, and CPS applications. We also summarized the open research problems that can be addressed by future honeypot and honeynet studies. As future work, we are planning to propose novel honeypot/honeynet systems for IoT and CPS environments that build upon this survey.

\section*{ACKNOWLEDGEMENTS}
This work is partially supported by the US National Science Foundation Awards: NSF-CAREER-CNS-1453647 and NSF-1663051. The views expressed are those of the authors only, not of the funding agencies.


\bibliographystyle{IEEEtran}

\bibliography{IEEEabrv,honeypot.bib}
\end{document}